\documentclass[12pt]{article}
\usepackage[dvips]{graphicx}
\usepackage{epsfig,cite}
\usepackage {amsmath}
\usepackage{amssymb}
\usepackage{slashed}
\newcount\timecount
\newcount\hours \newcount\minutes  \newcount\temp \newcount\pmhours
\hours = \time
\divide\hours by 60
\temp = \hours
\multiply\temp by 60
\minutes = \time
\advance\minutes by -\temp
\def\hour{\the\hours}
\def\minute{\ifnum\minutes<10 0\the\minutes
            \else\the\minutes\fi}
\def\clock{
\ifnum\hours=0 12:\minute\ AM
\else\ifnum\hours<12 \hour:\minute\ AM
      \else\ifnum\hours=12 12:\minute\ PM
            \else\ifnum\hours>12
                 \pmhours=\hours
                 \advance\pmhours by -12
                 \the\pmhours:\minute\ PM
                 \fi
            \fi
      \fi
\fi
}

\def\monthname{\relax\ifcase\month 0/\or January\or February\or
   March\or April\or May\or June\or July\or August\or September\or
   October\or November\or December\else\number\month/\fi}

\def\bold#1{\setbox0=\hbox{$#1$}%
     \kern-.025em\copy0\kern-\wd0
     \kern.05em\copy0\kern-\wd0
     \kern-.025em\raise.0433em\box0 }

\def\beq{\begin{equation}}
\def\eeq{\end{equation}}

\def\ga{\mathrel{\raise.3ex\hbox{$>$\kern-.75em\lower1ex\hbox{$\sim$}}}}
\def\la{\mathrel{\raise.3ex\hbox{$<$\kern-.75em\lower1ex\hbox{$\sim$}}}}
\def\gev{{\rm \, Ge\kern-0.125em V}}
\def\tev{{\rm \, Te\kern-0.125em V}}
\def\gyr{{\rm \, G\kern-0.125em yr}}

\def\tbt{\tan \beta}
\def\tanb{\tbt}

\def\gappeq{\mathrel{\rlap {\raise.5ex\hbox{$>$}}
{\lower.5ex\hbox{$\sim$}}}}
\def\lappeq{\mathrel{\rlap{\raise.5ex\hbox{$<$}}
{\lower.5ex\hbox{$\sim$}}}}
\def\Toprel#1\over#2{\mathrel{\mathop{#2}\limits^{#1}}}

\def\stau{\widetilde \tau}
\def\stop{\widetilde t}
\def\sbot{\widetilde b}

\def\mpl{M_{\rm Pl}}
\def\mchi{m_{\tilde \chi}}

\def\m12{m_{1\!/2}}

\newcommand\iso[2]{\mbox{${}^{#2}${\rm #1}}}
\def\he#1{\iso{He}{#1}}
\def\be#1{\iso{Be}{#1}}
\def\li#1{\iso{Li}{#1}}

\def\stau{\tilde{\tau}}

\def\mgrav{m_{3/2}}
\def\grav{\widetilde{G}}
\def\qbar{\overline{q}}
\def\mpl{M_{P}}
\def\mchi{m_{\chi}}
\def\cosw{\cos \theta_W}
\def\sinw{\sin \theta_W}

\def\bea{\begin{eqnarray}}
\def\eea{\end{eqnarray}}

\def\yx{y_X}
\def\nrg{\epsilon}
\def\pd{\partial}
\def\ra{\rightarrow}

\def\omp{\omega_{\rm p}}

\def\pref#1{(\ref{#1})}
\def\pfrac#1#2{\left(\frac{#1}{#2}\right)}
\def\avg#1{\langle #1 \rangle}
\def\beqar{\begin{eqnarray}}
\def\eeqar{\end{eqnarray}}

\begin{document}

\begin{titlepage}
\pagestyle{empty}
\baselineskip=21pt
%\rightline{\tt astro-ph/yymmnnn}
\rightline{CERN-PH-TH/2009-108}
\rightline{UMN--TH--2803/09}
\rightline{FTPI--MINN--09/23}
\vskip +0.4in
\begin{center}
{\large {\bf Nucleosynthesis Constraints on a Massive Gravitino in Neutralino Dark Matter Scenarios}}

\end{center}
\begin{center}
\vskip +0.4in
{\bf Richard~H.~Cyburt}$^{1}$, {\bf John~Ellis}$^2$,
{\bf Brian~D.~Fields}$^{3}$, \\
{\bf Feng Luo}$^{4}$, {\bf Keith~A.~Olive}$^{4,5}$,
and {\bf Vassilis~C.~Spanos}$^{6}$
\vskip 0.6in
{\small {\it
$^1${Joint Institute for Nuclear Astrophysics(JINA), National Superconducting
Cyclotron Laboratory(NSCL), Michigan State University, East Lansing,
MI 48824, USA}\\ 
$^2${TH Division, Physics Department, CERN, CH-1211 Geneva 23, Switzerland}\\
$^3${Center for Theoretical Astrophysics,
Departments of Astronomy and of Physics, \\ University of Illinois, Urbana, IL 61801, USA}\\
$^4${School of Physics and Astronomy, \\
University of Minnesota, Minneapolis, MN 55455, USA}\\
$^5${William I. Fine Theoretical Physics Institute, \\
University of Minnesota, Minneapolis, MN 55455, USA}\\
$^6${Department of Physics, University of Patras, GR-26500 Patras, Greece}} \\
}
%\vskip 0.1in
\clearpage
{\bf Abstract}
\end{center}
%\baselineskip=18pt \noindent
%%%%%%%%%%%%%%%%%%%%%%%%%%%%%%%%%%%%%%%%%%%%%%%%%
{\small
The decays of massive gravitinos into neutralino dark matter
particles and Standard Model secondaries during
or after Big-Bang nucleosynthesis (BBN) may alter the
primordial light-element abundances.  We present here details of a new suite of
codes for evaluating such effects, including a new treatment based on
{\tt PYTHIA} of the evolution of showers induced by hadronic decays of
massive, unstable particles such as a gravitino.  We present several sets of results
obtained using these codes, including general constraints on the
possible lifetime and abundance of an unstable particle decaying into
neutralino dark matter under
various hypotheses for its decay mechanism.
We also develop an analytical treatment of non-thermal hadron
propagation in the early universe, and use this to
derive analytical estimates for light-element production
and in turn on decaying particle lifetimes and abundances, which confirm our numerical
results and illuminate the underlying physics.
We then consider specifically
the case of an unstable
massive gravitino within the
constrained minimal supersymmetric extension of the Standard Model
(CMSSM). We present upper limits
on its possible primordial abundance before decay for different
possible gravitino masses, with CMSSM parameters along strips where the lightest neutralino
provides all the astrophysical cold dark matter density.
We do not find any CMSSM solution to the cosmological \li7 problem for small $m_{3/2}$.
Discounting this, for $m_{1/2} \sim 500$~GeV and $\tanb = 10$ the other light-element
abundances impose an upper limit $m_{3/2} n_{3/2}/n_\gamma \la 3 \times 10^{-12} \gev$ to
$\la 2 \times 10^{-13} \gev$ for $m_{3/2} = 250$~GeV to 1~TeV, which is similar
in both the coannihilation and focus-point strips and
somewhat weaker for $\tanb = 50$, particularly for larger $m_{1/2}$.
The constraints also weaken in general for larger $m_{3/2}$, and for
$m_{3/2} > 3$~TeV we find a narrow range of $m_{3/2} n_{3/2}/n_\gamma$,
at values which increase with $m_{3/2}$,
where the \li7 abundance is marginally compatible with the other
light-element abundances.}

%%%%%%%%%%%%%%%%%%%%%%%%%%%%%%%%%%%%%%%%%%%%%%%%

\vfill
\leftline{CERN-PH-TH/2009-108}
\leftline{July 2009}
\end{titlepage}
\baselineskip=18pt
%%%%%%%%%%%%%%%%%%%%%%%%%%%%%%%%%%%%%%%%%%%%%%%%%%

\section{Introduction}

Many extensions of the Standard Model of particle physics predict the
existence of massive, unstable or metastable particles. If these have
lifetimes longer than about a second, their decay products may affect
the abundances of light nuclei, such as D, \he4 and \li7, produced
during Big-Bang nucleosynthesis (BBN).  If the particles have
lifetimes exceeding about 100 seconds, they decay after BBN, but their
decay products may re-process the abundances produced during
BBN. Both of these possibilities are strongly constrained by the degree
of concordance between abundance predictions of D and \he4 and the
observational determination of these abundances,
if the baryon density is within the range determined
from measurements of fluctuations in the cosmic microwave background
radiation (CMB)~\cite{wmap}.
This concordance is not only an important vindication
of BBN calculations~\cite{cfo1,cfo2,bbn2,cyburt,cfo5},
but also sets severe limits on the possible
abundance of any massive, metastable particle as a function of its
lifetime.  These limits have been the subject of many previous studies
that have modeled both the electromagnetic and hadronic components of
the showers induced by the decays of such heavy
particles~\cite{Lindley:1984bg}-\cite{stef}.

We revisit here these constraints using a new suite of codes to model
the decay spectra, the evolution of the showers that they induce in
the electromagnetic plasma that filled the universe during the epoch
between BBN and the decoupling of the CMB, and their potential for
either destroying or creating light nuclear species~\cite{cefos}.
Studying the decays of heavy particles during BBN requires a code that calculates
the direct impact of electromagnetic and/or hadronic showers during
nucleosynthesis, rather than just the effects on abundances that had
previously been fixed independently by BBN.
Our earlier code~\cite{cefo}
considered only electromagnetic decays and did not provide
constraints for particle lifetimes $\la 10^4$s. Here we describe
a new suite of codes that includes hadronic decays and is applicable also
to particles with lifetimes $\la 10^4$s.

The effects included in this suite of codes are potentially relevant
to any extension of the Standard Model that predicts the existence
of massive unstable or metastable particles. One example is
supersymmetry (SUSY). Many supersymmetric models incorporate a discrete
$Z_2$ symmetry known as $R$ parity. If $R$ parity is not maintained,
the lightest supersymmetric particle (LSP) is unstable, and the
consistency of BBN with astrophysical observations and the density
of baryons inferred from the CMB imposes important constraints on the
mechanism and amount of $R$ violation \cite{Bouquet:1986mq}. On the other hand, if $R$
parity is conserved, the LSP is stable, but there is in general a
heavier metastable supersymmetric particle. The possibility studied
most frequently has been that the LSP is the lightest neutralino
$\chi$ \cite{EHNOS}. In this case, the gravitino would be
metastable, and the BBN constraints on its decays impose important
limits on the primordial production of gravitinos that are
potentially problematic for some inflationary cosmological scenarios
\cite{Nanopoulos:1983up} - \cite{pradler}. However, the neutralino
is not the only candidate for the LSP, and another generic
possibility is the gravitino itself. In this case, the gravitino
itself would be stable, but the next-to-lightest supersymmetric
particle (NLSP) would in general be metastable, and its decays would
be subject to the BBN constraints \cite{SFT,0404231,eoss5}.

Our goals in this paper are threefold. The first is to document our
new interlocking suite of codes for decay showers in the early universe
and their impact on  BBN;
our codes include the effects of both hadronic and
electromagnetic showers, and are applicable to a large range
of unstable particle lifetimes. The second goal is
to set out the constraints obtained using this code on the lifetime and
abundance of a generic unstable or metastable particle, under various
plausible hypotheses for its dominant decay modes. The third goal is to
explore in more detail the consequences of the shower
effects for supersymmetric models with a neutralino LSP.  In the context of such
models, we demonstrate how our new BBN codes may be used to obtain constraints on the gravitino
abundance as a function of its mass.

In previous analyses of the 
constrained minimal supersymmetric extension of the Standard Model
(CMSSM) with a neutralino LSP, cosmological constraints on
the decays of the metastable gravitino have usually been ignored. Essentially, it has been assumed
implicitly that the gravitino is so heavy, and hence its lifetime is so short,
that its decays take place so early in the history of the universe that they do not
affect light-element abundances. However, from a theoretical point of view, a
large hierarchy between the gravitino and LSP masses may appear unlikely.
Indeed, in minimal supergravity (mSUGRA) and related 
models~\cite{vcmssm}, it is usually the case that the gravitino is not
much heavier than the LSP, if it is not the LSP itself.
Therefore, it is important to analyze the case when the decays of gravitinos {\it do}
affect the light-element abundances, and evaluate the resulting constraint on
the primordial gravitino abundance as a function of its mass. 

We exemplify these constraints in the specific case of the
CMSSM, in which supersymmetry-breaking gaugino and scalar masses
$m_{1/2}$ and $m_0$ are each assumed to be universal \cite{eoss,cmssmwmap}.
In this case, when the LSP is the lightest neutralino, its relic density 
$0.0975 < \Omega_{\rm CDM} h^2 < 0.1223$ as inferred from astrophysical and
cosmological measurements~\cite{wmap} constrains $(m_{1/2}, m_0)$ to lie along narrow
``WMAP strips" for given fixed values of the other CMSSM parameters \cite{eoss,bench}. Moreover,
the gravitino lifetime and decay modes are known as functions of its mass
and the CMSSM parameters. Thus, we are able to set firm upper limits on
the possible abundance of the gravitino as a function of its mass and $m_{1/2}$. This
upper limit on the gravitino abundance may, in some
circumstances, constrain severely the maximum temperature reached in
the universe, e.g., following an early inflationary epoch \cite{Nanopoulos:1983up} - \cite{pradler}.

The layout of this paper is as follows. In Section \ref{sect:data} we summarize the
cosmological data that we use in our analysis. Then, in Section \ref{sect:nuke} we explain our
treatment of the evolution of the hadronic and
electromagnetic showers initiated by the decay of a generic metastable
particle into the cosmic plasma, and their effects on the primordial light-element abundances.
Appendix~\ref{app:prop} summarizes our
treatment of hadronic spectra, and describes the propagation of
non-thermalized particles in the cosmic plasma. 
In Section \ref{sect:decays} we turn to the specific case in which the
gravitino is the metastable particle,
relegating some technical details to Appendix~\ref{app:gravdec}. 
In Section \ref{sect:constraints} we discuss the
resulting constraints on the abundance of a metastable
neutral particle, under plausible assumptions on its dominant decay
modes. 
We consider constraints  as a function of a generic particle
lifetime, as well as constraints 
on the gravitino abundance in specific CMSSM scenarios with a neutralino LSP.
For $m_{3/2} \le 1$~TeV, we find no parameter choices where the \li7
abundance can be reconciled with the other light-element abundances.
However, for $m_{3/2} > 3$~TeV we find narrow ranges of the gravitino
abundance that may reconcile marginally the \li7 and other light-element
abundances. Finally, in Section \ref{sect:conclude}
we summarize our conclusions and present some prospects for future
work.

\section{Cosmological Data}

\label{sect:data}

\subsection{Light-Element Abundances}
\label{sect:abs}

The abundances of the light elements D and \he4 predicted by BBN theory agree quite well
with the values determined by observation, if the baryon-to-photon ratio
$\eta$ is that inferred from CMB measurements~\cite{wmap}. This concordance
provides the basis for the constraints on metastable particles to be discussed
in this paper. However, there is known to be an issue regarding the abundance
of \li7, which we discuss below.

Deuterium provides a powerful constraint,
as it is very sensitive to the baryon content in the universe,
and thus offers by itself a measure of $\eta$.
Local deuterium that is measured in the solar neighborhood in the
interstellar medium provides a strong lower limit on the cosmological
abundance, given that astrophysical effects destroy more
deuterium than they produce~\cite{els76}.
Recent observations by FUSE show a wide dispersion in the
deuterium abundance in local gas seen via its absorption,
$({\rm D/H})_{\rm local~gas} = (0.5-2.2)\times10^{-5}$\cite{fuse}.
This surprisingly large spread, taken together with the positive correlation
of D/H with temperature and metallicity along various sightlines,
led \cite{fuse} to suggest that deuterium may suffer significant
and preferential depletion onto dust grains.  In this case
the true local interstellar D/H value would lie at the upper limit
of the observed values, giving
${(\rm D/H)}_{\rm ISM} \ga (2.31 \pm 0.24) \times 10^{-5}$.
However, extracting a primordial deuterium value requires a Galactic
chemical evolution model (e.g., \cite{galdeut}), whose model dependences yield
uncertainties in the determination of the primordial deuterium abundance.

A high-redshift, metal-poor system is free of many such model
uncertainties.  In particular, observing the absorption features of quasar light due to a
dense high-redshift cloud can be used to determine the primordial deuterium
abundance.
Several studies yield
results~\cite{bt98a,bt98b,omeara,pettini,kirkman,omeara2,pettini2}
that are broadly consistent, but with a reduced $\chi^2_\nu = 2.95$,
suggesting that the uncertainties have been underestimated.
Taking a weighted mean, we find a
``world average" range:
\beq
\label{eq:D}
\left( \frac{\rm D}{\rm H} \right)_p = \left( 2.82 \pm 0.21 \right)
  \times10^{-5},
\eeq
where the error bar has been inflated by a factor of $\sqrt{\chi^2_\nu}$.
Because it is likely that systematic errors dominate,
an even more conservative approach would be to use the
sample variance of the D/H data\cite{cfo1}; this would give
$\sigma({\rm D/H}) = ( \pm 0.53) \times 10^{-5}$,
i.e., a significantly higher error.
In this study we adopt $3.2 \times 10^{-5}$
as our fiducial upper limit on the D/H abundance,
but we also illustrate the effect of significant variations in D/H
around this value.

Since \he3 is also quite sensitive to the baryon density, one might hope
that it too could be used as a baryometer.  Observations of H{\sc II}
regions in our own Galaxy yield values of the \he3/H ratio that are compatible with
calculations of the primordial value~\cite{bania,vofc}.
However, the extrapolation from the
observations to a primordial abundance is complicated by the unknown chemical
evolution of \he3.  Indeed, one does not even know whether \he3/H is
increasing or decreasing with cosmic time.  Thus, a primordial
extrapolation yields only an order-of-magnitude range of allowable
values of \he3/H~\cite{foscv}.
However, if we use the results of~\cite{els76,sigl} that the deuterium abundance is always
decreasing with time, and assume
that \he3 changes relatively slowly, we can adopt
their ratio:
\beq
\left( \frac{\he3}{\rm D} \right)_p < 1.0
\eeq
as a conservative constraint on the primordial \he3/D
ratio.

H{\sc II} regions also yield observations of \he4.  However,
measurements of extra-galactic and metal-poor sources yield {\it a priori}
more reliable estimates of the primordial \he4 abundance.  Various
evaluations of the same data yield differing results ranging from $Y_p =
0.234$~\cite{peim} to  $Y_p= 0.244$~\cite{izotov}.  These differences point
to the dominance of systematic uncertainties~\cite{OSk2001} over statistical errors.
The allowed ranges of systematic shifts are discussed in~\cite{IzotovThuan2004}, but
their analysis does not provide a full estimate for the systematic error.
Subsequently, another analysis of the data was
performed~\cite{OSk2004}, making more realistic assumptions than those
used in previous analyses.  This analysis found a systematically higher
$Y_p$, with significantly increased errors,
suggesting that previous analyses had underestimated their
systematics:
\beq
Y_p = 0.249\pm0.009.
\eeq
The inclusion of one of these systematics has been explored in~\cite{Fukugita}, which finds
results similar to~\cite{OSk2004}. In our subsequent analysis, we adopt the lower
limit $Y_p > 0.240$.

Lithium is seen in the atmospheres of the most primitive, metal-poor
stars in the stellar halo of our Galaxy (extreme Population II).
Some $\ga 100$ such
stars show a plateau\cite{spite}
in (elemental) lithium versus metallicity,
with a small scatter consistent with observational uncertainties.
This insensitivity of Li/H to (proto)-Galactic metal content and thus
stellar nucleosynthesis indicates that lithium in these stars
has a primordial or at least
pre-Galactic origin.
An analysis \cite{rbofn}
of field halo stars gives a plateau abundance of
\beq
\label{eq:li_halo}
\left( \frac{\rm Li}{{\rm H}} \right)_{\rm halo \star}
 = (1.23^{+0.68}_{-0.32}) \times 10^{-10} ,
\eeq
where the errors represent a 95\% CL and include
both statistical and systematic uncertainties.
On the other hand, studies of globular cluster stars
generally find higher lithium abundances.
The measurement \cite{liglob} 
of
\beq
\label{eq:li_gc}
\left( \frac{\rm Li}{{\rm H}} \right)_{\rm gc}
 = (2.19 \pm 0.28) \times 10^{-10} ,
\eeq
is consistent with results for other systems:
${\rm Li/H} = (1.91 \pm 0.44) \times 10^{-10}$ \cite{pasquini};
${\rm Li/H} = (1.69 \pm 0.27) \times 10^{-10}$ \cite{thevenin};
and ${\rm Li/H} = (2.29 \pm 0.94) \times 10^{-10}$ \cite{li_m92}.
Both sets of abundances
are substantially lower than the standard BBN prediction of \li7
based on the
WMAP estimate of $\eta$~\cite{wmap},
by factors of $\sim 2.4-4.3$, or $4-5 \sigma$ \cite{cfo5}.
This discrepancy is known as the ``lithium problem''
or more specifically the ``\li7 problem.'' In our later analysis,
we adopt $\li7/{\rm H} < 2.75 \times 10^{-10}$ as our
fiducial upper limit on the cosmological \li7 abundance.

The existence of the \li7 problem, and the nature of its solution,
both have a direct impact on our analysis.
It is possible that the problem points to new physics, in particular if the
observations and standard theory are both correct with accurate
error estimates.  If SUSY were to lead to a solution of the
\li7 problem, this would tie together a wide array of 
particle astrophysics experiments and observations.
Indeed, it has been suggested \cite{jed,kkm2}
that the \li7 problem could be solved by hadronic decays of a metastable
neutral particle $X$
with lifetime $\tau_X \sim 10^3$ sec.  We will directly address
and update this issue below.
On the other hand, it remains possible that the standard BBN
light-element abundance predictions remain correct, i.e., unperturbed
by any dark matter interactions and/or decays.  Rather, it could be that
the \li7 problem
instead reflects astronomical observational systematics in recovering a
Li/H abundance from stellar spectra,
though a recent study \cite{hosford} of the effective temperature of metal poor
stars confirmed the use of relatively low temperatures and a
Li abundance in the range ${\rm Li/H} = (1.3 - 1.4 \pm 0.2)
\times 10^{-10}$.  The \li7 problem may also
reflect astrophysical systematics due to Li depletion 
via circulation and nuclear burning over the $> 10$ Gyr lifetime
of metal-poor stars \cite{depletion}, though
the lack of star-to-star scatter in Li/H suggests
the depletion could be negligibly small~\cite{rbofn}.
If there is depletion, for our problem the correct procedure would be to use
the initial, undepleted Li/H abundances.  However, since these cannot be determined
accurately, we do not adopt a quantitative
constraint for this possibility.  Instead, we
consider the limits on unstable particles when the \li7
is assumed to be solved and thus the \li7 constraints are ignored.

Recent work has claimed to resolve lithium isotope information
in halo stars \cite{li6obs,asp}, and indicates the presence of \li6.
In fact, the isotopic ratio has a roughly constant value of
\beq
\label{eq:li67}
\frac{\li6}{\li7} \approx 0.05 .
\eeq
This has several immediate implications.  First,
we see that most of the lithium is in the from of \li7, so that
the total elemental lithium abundance (as in eqs.~(\ref{eq:li_halo})
and (\ref{eq:li_gc}))
is mainly \li7: ${\rm Li/H \approx \li7/H}$.
Secondly, given that Li/H in these stars has a plateau,
the constancy of the isotopic ratio implies that both
isotopes have a plateau.
However, the inferred \li6 plateau abundance
is about 1000 times the \li6 abundance predicted by standard BBN~\cite{bbnli6,van}.

Extracting isotopic information from thermally broadened absorption lines
is extremely challenging, and the existence of a \li6 plateau
has been questioned \cite{cay}.
If real, the \li6  plateau
abundance cannot be explained by
conventional Galactic cosmic-ray nucleosynthesis \cite{gcr,fo,van}.
Therefore, one is led to think that there is a ``\li6 problem'' in addition to that for \li7.

It is possible to arrange decaying-particle scenarios
in which \li6 is produced at the plateau level with some destruction of  \li7,
thus fixing both lithium problems simultaneously~\cite{Jed2,cefos,grant,jed3,cumber,bailly},
a point to which we will return in detail below.
However, even granting the existence of the \li6 plateau,
the \li6 problem really only requires a pre-Galactic source,
which need not arise prior to
the first star formation.
For example, the \li6 problem
might be explained by cosmological cosmic-ray nucleosynthesis due to
cosmic rays produced at the epoch of structure formation \cite{ccr}.
For our purposes, therefore, we simply interpret eq.~(\ref{eq:li67})
as a firm upper limit on any primordial \li6 \cite{asp,cay}.

\subsection{Baryon Density}

The baryon-to-photon ratio $\eta \equiv n_{\rm B}/n_\gamma \equiv 10^{-10} \eta_{10}$ is
related to the present baryon density parameter $\Omega_{\rm B} = \rho_{\rm B}/\rho_{\rm crit}$
by $\Omega_{\rm B}h^2 = \eta_{10}/274$.
Adopting the value of $\Omega_{\rm B}h^2$
indicated by the WMAP 5-year CMB data \cite{wmap}
gives the following estimate of the baryon-to-photon ratio:
\beq
\eta_{10} = 6.23 \pm 0.17.
\eeq
This is the default value we assume later in our BBN analysis.
The central value and error range here correspond to
WMAP data combined with large-scale structure information,
in a framework for which the primordial power
spectrum is a simple power law with fixed index.
Similar but slightly different values would result from, e.g., a
running spectral index.

\section{Hadronic Decays during Primordial Nucleosynthesis}

\label{sect:nuke}

Particle decays during BBN generally have
two main effects~\cite{EHNOS,Lindley:1984bg}.
First, they change the cosmic expansion
rate due to the injection of additional
relativistic species.
This effect is model-independent, in that it is insensitive to
the details of the decays,
beyond the assumption that the daughter particles are relativistic~\cite{Scherrer:1987rr}.
Secondly, they introduce new, non-thermal
decay products--possibly electromagnetic and/or hadronic--which
can interact with the background thermal nuclei
and change the final light-element abundances.
The branching ratios and spectra of hadronic and electromagnetic
decay particles and energies
are model-dependent, and are further
discussed below in Section~\ref{sect:decays}.
While both effects occur in principle, the expansion effects are negligible
for the decay parameters of practical interest to us,
and the non-thermal interaction effects are the dominant
perturbations to the abundances.
Consequently, we focus on these effects,
and henceforth neglect the perturbation to the
expansion rate.
If the decaying particles are electrically charged and thus
can form bound states with nuclei, additional effects
arise~\cite{maxim,kota,kapling,cefos,grant,hama,jittoh,jed3,cumber,pradler2,a=8,pps,bailly,stop}.
This is not the case in the
scenarios considered in the present paper, but
we will revisit this subject in future work.

To fix notation:  we consider decays of some generic heavy particle $X$,
which we will eventually specialize to the case of the gravitino.
The particle has lifetime $\tau_X$ and decay rate $\Gamma_X = \tau_X^{-1}$.
We can quantify the $X$ abundance
either as
the number of decaying particles per background baryon,
\beq
Y_X \equiv \frac{n_X}{n_{\rm B}} ,
\label{ydef}
\eeq
or as the number of particles per
unit entropy
\beq
\yx = \frac{n_X}{s} = Y_X \frac{n_{\rm B}}{s} \approx \frac{1}{7} \eta Y_X ,
\eeq
where $s \simeq 7 n_\gamma$ is the entropy
density, and the factor of $\simeq 7$ is appropriate after $e^\pm$
annihilation.
We note that, in the absence of $X$ decays,
conservation of baryon number guarantees
the constancy of $Y_X$, while
$\yx$ changes during BBN due to the
usual photon heating from $e^\pm$ annihilation.
Due to decays, we have $Y_X(t) = Y_X^{\rm init} e^{-t/\tau_X}$;
our constraints will be on the initial, pre-decay abundance
$Y_X^{\rm init}$, and this should be understood hereafter whenever
we write $Y_X$. The only exception in our notation is that we 
write $Y_p$ for the \he4 mass fraction and this should not be
confused with the ratio of number densities as defined in eq. (\ref{ydef}).

We are interested in both hadronic and electromagnetic decays of the
heavy particles $X$.
We denote the
electromagnetic branching ratio of $X$ by
$B_{\rm EM} = \Gamma_{X\rightarrow {\rm EM}}/\Gamma_X$.
The abundance perturbations from electromagnetic
decays, and thus the constraints on such decay modes,
scale with the product of $B_{\rm EM}$
and the decay energy release per photon:
\beq
\label{eq:zeta}
\zeta_X \equiv \frac{m_X n_X}{n_\gamma}
  = m_X Y_X \eta ,
\eeq
where the $X$ abundance is evaluated prior to decay.
As emphasized particularly by~\cite{kkm2},
electromagnetic decays inevitably accompany hadronic decays,
and so both sets of decay products and interactions need to be included.
Electromagnetic decays were discussed in detail
in \cite{cefos}; we include those processes here,
incorporating the treatment described in~\cite{cefos}.
However, the analysis there assumed that
the electromagnetic decays occurred entirely after
the usual BBN thermonuclear reactions have run to completion,
i.e., the decays were treated as a``post-processing''
modification after the usual light-element abundances had been
established.
Here, we supplement the previous treatment by including electromagnetic decay effects
consistently throughout BBN.

We quantify generic decays $X \rightarrow h$ of a particle $X$
into a hadronic species $h$ as follows.
We write
the $h$ production rate per unit volume,  and daughter kinetic energy $\nrg$
as $q_h(\nrg)$.
The total $X$
decay rate per baryon is $\Gamma_X Y_X$
and the total $X$ decay rate per volume is
$q_{X,\rm tot} = \Gamma_X Y_X n_{\rm B}$.
It will be convenient to isolate the $h$ production
per $X$ decay as
$Q_h(\nrg) = q_h(\nrg)/q_{X,\rm tot}$, so that we have
\beq
q_h(\nrg) \equiv \Gamma_X Y_X n_{\rm B} Q_h(\nrg) .
\eeq
We refer to $Q_h(\nrg)$ as the spectrum of $X$ decays,
which gives the number of $h$ particles produced per
$X$ decay per energy interval.  The integral
$B_h \equiv \int Q_h(\nrg) \ d\nrg$ gives
the total number of $h$ per $X$ decay, and thus
represents a multiplicity-weighted
hadronic branching ratio.
Gravitino decays are described in detail in the next
Section.  

\subsection{Interactions with Background Nuclides}

The dominant effect of hadronic decays on BBN
is the addition of new interactions between
hadronic shower particles and background nuclides.
These act to alter the evolution and final values of
the light-element abundances, as follows.
For each light nuclide $\ell$,
the abundance per background baryon, $Y_\ell \equiv n_\ell/n_{\rm B}$,
changes according to
\beq
\label{eq:bbnrate}
\pd_t Y_\ell =  \left( \pd_t Y_\ell \right)_{\rm SBBN}
+ \left( \pd_t Y_\ell \right)_{\rm EM}
+ \left( \pd_t Y_\ell \right)_{\rm had} ,
\eeq
where $( \pd_t Y_\ell )_{\rm SBBN}$ denotes the usual
rate of change of $\ell$ in standard
BBN due to thermonuclear reactions.
The second term on the right-hand side of \pref{eq:bbnrate} accounts for
non-thermal
electromagnetic interactions due to $X$ decays, either directly from the
decays of $X$ to photons or leptons, or through electromagnetic
secondaries in the hadronic showers.
These are treated as in~\cite{cefo},
but are not dominant when hadronic branchings are significant.
The final term on the right-hand side of \pref{eq:bbnrate}
represents the non-thermal hadronic interactions,
which are a major focus of this paper.

The inclusion of hadronic decays in BBN thus
requires a detailed evaluation of
the rates for such interactions.
For each background light nuclide species $\ell$,
we can write the hadronic contributions to $\pd_t Y_\ell$ as
\beq
\label{eq:dYdthad}
\left( \pd_t Y_\ell \right)_{\rm had}
= - \Gamma_{\rm \ell \rightarrow {\rm inel}} Y_\ell
  + \sum_{hb} \Gamma_{hb \rightarrow \ell} Y_b .
\eeq
The first term
accounts for $\ell$ destruction by hadro-dissociation,
where $\Gamma_{\rm \ell \rightarrow {\rm inel}}$ is the
total rate (per unit of the species $\ell$) of all inelastic interactions of shower
particles with $\ell$.
The second term accounts for
production due to hadro-dissociation of
heavier background species (e.g., deuteron production via
helium erosion
$p_{\rm shower} \alpha_{\rm bg} \rightarrow d+\cdots$).
The sum of inelastic rates $\Gamma_{hb \rightarrow \ell}$
producing $\ell$ runs over shower species $h$
and background targets $b$.
In the particular case of lithium isotopes, production
occurs via the interaction of energetic (i.e., non-thermal) mass-3 dissociation
products with background \he4,
e.g., $\he3 + \alpha \rightarrow \li{6,7} + \cdots$.

The non-thermal reaction rates on light elements
are themselves set by the abundance and
energy distribution of non-thermal particles, which we quantify as follows.
Consider a hadronic (non-thermal projectile) species $h$,
with an energy spectrum $\frac{dn_{h}}{d\nrg}(\nrg,t)$
(particle number per unit volume per unit energy interval)
and total number density $n_h = \int \frac{dn_{h}}{d\nrg} d\nrg$.
It is convenient to define the energy spectrum of non-thermal $h$
as $N_h \equiv \frac{dn_{h}}{d\nrg}/n_X$,
which is normalized such that
$\int N_h(\nrg) \, d\nrg = n_h/n_X = n_h/Y_X n_{\rm B}$
measures the number of propagated non-thermal $h$  particles
per $X$.

The rate for non-thermal production of species $\ell$
due to $hb \rightarrow \ell$
is
\beq
\label{eq:non-thermal}
\Gamma_{hb \rightarrow \ell}
   =  Y_X  n_{\rm B}
  \int \ N_{h}(\nrg,t) \ \sigma_{hb \rightarrow \ell}(\nrg) \ v_{\rm rel}(\nrg)
  \   d\nrg
  \approx  Y_X n_{\rm B}
\int \ N_{h}(\nrg,t) \ \left[ \sigma_{hb \rightarrow \ell} \; v \right] (\nrg) \ d\nrg ,
\eeq
where $\sigma_{hb \rightarrow \ell}(\nrg)$ is the
cross section for this process.
Here $v_{\rm rel}(\nrg) \approx v_h(\nrg)$ is the relative velocity
between the projectile and the background target,
which in practice amounts to the projectile velocity.
In a similar way, the inelastic-loss rate is just
a sum over all inelastic channels removing $\ell$, i.e.,
where the final state $f$ does not include $\ell$:
\beq
\Gamma_{\ell \rightarrow {\rm inel}}
   \approx  Y_X n_{\rm B} \sum_{f} \int \ N_{h}(\nrg,t) \
  \left[ \sigma_{h\ell \rightarrow f} \; v \right] (\nrg)
  \   d\nrg .
\eeq
The rates $\Gamma_\ell$ thus depend on
the shower development
and evolution of $N_{h}(\nrg,t)$
in the background environment.
A major effort of this paper
is to calculate these spectra and their evolution.

\subsection{Hadronic Showers}
\label{sect:showers}

We wish to follow the evolution of the non-thermal
spectra $N_{h}$
over the multiple generations produced by the hadronic $X$
decays.
In the context of BBN,
this problem of shower development has
been approached via Monte Carlo computation of
the multiple generations of shower particles \cite{kohri,kkm,kkm2,jed}.
The final particle spectrum is obtained
via iterating an initial decay spectrum,
accounting for both energy losses as well as
the energy distribution of collision products.

Here we introduce an alternative, but equivalent,
approach based on a ``cascade equation" treatment, that
follows the well-studied treatment of the development of hadronic
showers due to cosmic-ray interactions
in the atmosphere.
As we show, this approach
offers new physical insight as well as computational
advantages.

Once a hadron $h$ is produced by $X$ decay,
it loses energy, interacts with background nuclei,
and possibly decays if it is unstable (e.g., $n,\pi^\pm$).
The spectrum of $h$ evolves according to
the energy-space ``propagation equation":
\beq
\label{eq:cascade}
\pd_t N_{h}(\epsilon)
  =  J_{h}(\epsilon) - \Gamma_{h}(\epsilon)N_{h}(\epsilon)-\pd_\nrg\left[b_{h}(\epsilon) N_{h}(\epsilon)\right] ,
\eeq
where $J_{h}$ is the sum of all source terms, $\Gamma_{h}$ is the sum
of all sink terms, and $b_h$ is the energy-loss rate of particle-conserving processes.
Note that inelastic interactions lead to
secondary production of $h$, sometimes by other
non-thermal species $h^\prime$.  For this reason,
eq.~\pref{eq:cascade} ultimately represents
a set of coupled, integro-differential equations.

Details of eq.~\pref{eq:cascade}
and its numerical solution appear in Appendix \ref{app:prop}.
In the next section, we summarize the key physics
via an analytical description informed by the full
numerical results.

\subsection{Non-Thermalized Particle Spectra and Interactions:
Analytical Model}
\label{sect:analytic}

To develop a physical intuition for the
full numerical results used later, we first present
a simplified analytical treatment
of the propagation and interactions of non-thermalized hadrons.

A crucial feature of
eq.~\pref{eq:cascade}
is that it contains both source and sink terms for
$h$.
The rates for energy loss in the cosmic plasma,
and for scattering on the background nuclei, are
always faster than the cosmic expansion rate $H$.
Thus, redshift and expansion effects may safely be
ignored in non-thermal particle propagation, and have
been neglected in eq.~\pref{eq:cascade}.
Moreover, the energy-loss and scattering rates
are generally faster than the non-thermal source rate $\Gamma_X = 1/\tau_X$ set
by the $X$ decay time.
Thus eq.~\pref{eq:cascade} is {\em self-regulating}, with
$N_h$ tracking a quasi-static equilibrium set by the balance of source
and sink terms.
That is, at any cosmic epoch
we have $\pd_t N_{h}(\epsilon) \approx 0$.
This provides a crucial simplification
and reduces the problem to a series of ordinary
integro-differential equations, which
one may solve at each epoch.

We thus begin with a simplified version of the full
propagation equation, which
neglects elastic and inelastic source terms:
\beq
\label{eq:simpleprop}
0 = -\pd_\nrg (b_h N_h)
  - (\Gamma_{h,\rm sc} + \Gamma_\beta) N_h + \Gamma_X Q_h ,
\eeq
where $b_h = - d\nrg_h/dt$ is the total energy-loss
rate,
and $\Gamma_{h,\rm sc}$ is the total rate for inelastic and
elastic scattering.
For neutrons, $\Gamma_\beta = 1/\gamma_n \tau_n$ is the rate for the
beta decay of a free neutron with Lorentz factor $\gamma_n$;
for protons $\Gamma_\beta = 0$.

Equation \pref{eq:simpleprop} is formally identical
to the ``leaky box'' equation for the propagation
of cosmic rays in our Galaxy.  Indeed,
our problem can be regarded as the injection of
cosmic rays in the early universe plasma.
The solutions to eq.~\pref{eq:simpleprop}
have been well-studied for the
case of Galactic cosmic rays
\cite{leakybox},
and we adapt the solutions for our problem.
We focus on the propagation of the
primary species, i.e.,
the  $p$ and $n$, which have sources in the $X$ decays
themselves; in this case the decay term dominates the sources.
Our full numerical treatment also includes
primary $\pi^\pm$, but these are important only for short
$\tau_X \la 1$ sec.

On physical grounds, we expect that the hadronic
spectra $N_h$ at any given time should scale linearly with
the $X$ decay rate $\Gamma_X$.
One can infer this formally from eq.~\pref{eq:cascade},
as outlined in Appendix \ref{app:prop}.
Thus we have the scaling $N_h \propto \Gamma_X$,
and so the ratio $N_h/\Gamma_X$
is independent of the decay rate $\Gamma_X$.

If we ignore secondary production of $h$, the
exact solution
to eq.~\pref{eq:simpleprop} is known
and appears in Appendix \ref{app:prop}.
However, for our purposes it is useful to focus on
two limiting cases.  Generally,
one of the sink terms is dominant and
controls the resulting equilibrium.
If the scattering term dominates the energy-loss term,
$\Gamma N \gg \pd_\nrg (bN)$ we refer to this as
the {\em thin-target} limit.  Physically,
$h$ particles interact catastrophically
before they lose energy, i.e., the cosmic plasma
is ``thin'' against stopping.
The reverse case, $\Gamma N \ll \pd_\nrg (bN)$,
constitutes the {\em thick-target} limit, where
particles lose energy before interacting.
We can express the ratio of the two terms
as $\pd_\nrg (bN)/\Gamma N = (b/\nrg \Gamma) \pd \ln(bN)/\pd \ln \nrg$.
We see that, as long as the logarithmic term is slowly varying,
the dominant loss term is set
by a comparison of scattering timescale $\Gamma^{-1}$
with energy-loss timescale $\nrg/b$:
\beqar
\Gamma_{\rm sc} \gg \frac{b}{\nrg} & \Rightarrow & \mbox{thin target}, \\
\Gamma_{\rm sc} \ll \frac{b}{\nrg} & \Rightarrow & \mbox{thick target}.
\eeqar
In the thin-target limit, scattering losses dominate,
so that eq.~\pref{eq:simpleprop}
reduces to $- \Gamma_{h, \rm sc} N_h + \Gamma_X Q_h \simeq 0$ and the solution
becomes algebraic
\beq
\label{eq:specthin}
N_h(\nrg) \simeq \Gamma_X \frac{Q_h(\nrg)}{\Gamma_{h,\rm sc}(\nrg)}
 = \Gamma_X \frac{Q_h(\nrg)}{n_{\rm B} Y_b v \sigma_{hb \ra \rm inel}(\nrg)} .
\eeq
Here the reaction on background species $b$ dominates the scattering
losses.
Note the inverse scaling with baryon density $n_{\rm B}$.
Note also that we have implicitly assumed $\Gamma_\beta \ll \Gamma_{\rm sc}$.
In the case of neutrons, once $\Gamma_\beta > \Gamma_{\rm inel}$
(which is indeed well into the thin-target regime),
then neutrons decay before they interact.
This occurs when
$\Gamma_{{\rm inel},n} \sim n_{\rm B} v_n \sigma_{\rm inel} < \Gamma_\beta$,
which corresponds to $T \la 0.4 \ {\rm keV} \ \gamma_n^{-1/3} (v_n/0.1 c)^{-1/3}$
and $t \ga 8 \times 10^{6} \ {\rm s} \ \gamma_n^{2/3} (v_n/0.1 c)^{2/3}$.
At longer times, non-thermal protons and, more importantly,
electromagnetic cascades, dominate light-element production.

Turning now to the thick-target limit,
when energy losses dominate the sinks
we have $0 \simeq -\pd_\nrg (bN_h)  +  \Gamma_X Q_h$,
and thus
\beq
\label{eq:specthick}
N_h(\nrg) \simeq \Gamma_X \frac{Q_h(>\nrg)}{b(\nrg)} ,
\eeq
where $Q_h(>\nrg) = \int_\nrg Q_h(\nrg^\prime) \, d\nrg^\prime$
is the integral source spectrum above $\nrg$.
Taking $b = b_{\rm coul}$ for Coulomb losses, we have
$b \sim 4\pi \alpha^2 \hbar^2 c^2 Y_{e^\pm} n_{\rm B}/m_e v$,
or
\begin{equation}
d\nrg/dR = b/\rho_{\rm B} v \sim 4\pi \alpha^2 \hbar^2 c^2 Y_{e^\pm} /m_em_pv^2
\sim 1.5 {\rm MeV/(g cm^{-2})} \, Y_{e^\pm} (100 {\rm MeV}/\nrg).
\end{equation}
Crucially, the electron/positron abundance $Y_{e^\pm}$ per baryon
is enormous before pair annihilation, so
that $Y_{e^\pm} \sim 1/\eta \sim 10^9$ at $T \ga 0.2 \ {\rm MeV}$,
and even afterwards the pairs dominate the baryons
until $T \sim m_e/25 \sim 0.02 \ \rm MeV$.
Thus we write
\beq
N_h(\nrg) \simeq \Gamma_X
   \frac{Q_h(>\nrg)}{m_p n_{\rm B} v(\nrg) \, d\nrg_h/dR(\nrg)} ,
\eeq
where
we use the energy loss per grammage $d\nrg/dR = b/\rho_{\rm B} v$,
which is independent of background density.
Here again we see that the spectrum scales inversely with
background density, but also has an inverse dependence
on the strongly-varying electron/positron abundance
$Y_{e^\pm}$, contained in $d\nrg/dR$.

\subsection{Non-thermal Reaction Rates}

Consider a reaction $hb \rightarrow \ell$
on a light background species $b$
which produces  a daughter species
$\ell$, e.g., $n \alpha \ra d + \cdots$.
The rate for this reaction is
\beq
\label{eq:reactionrate}
\Gamma_{hb \ra \ell} =
 \int \phi_h(\nrg) \ \sigma_{hb \ra \ell}(\nrg) \ d\nrg
\eeq
per background target $b$.
Here the (angle-integrated) flux of non-thermal $h \in n,p$ is
$\phi_h(\nrg) = n_X N_h(\nrg) v(\nrg) = Y_X n_{\rm B} N_h(\nrg) v(\nrg)$.
Recalling that $N_h \propto 1/n_{\rm B}$,
we see that the non-thermal flux $\phi_h \propto n_{\rm B} N_h$ and thus
the reaction rate $\Gamma_{hb \ra \ell}$ are both
independent of the background baryon density.
Also, since $N_h \propto \Gamma_X$, we
have $\Gamma_{hb \ra \ell} \propto \Gamma_X Y_X$.

We now examine the reaction rate
for the two limiting cases of propagation.
Using $N_h$ in the two limits, we have
\beq
\Gamma_{hb \ra \ell} =
\left\{
\begin{array}{cl}
  \Gamma_X \frac{Y_X}{Y_k^{\rm bg}} \int_{\nrg_{\rm th}} Q_h(\nrg) \
     \frac{\sigma_{hb \ra \ell}(\nrg)}
          {\sigma_{hk \ra \rm inel}}
     \ d\nrg  &
     \mbox{thin target} , \\
  \Gamma_X Y_X \int_{\nrg_{\rm th}} Q_h(> \nrg)
     \ \frac{\sigma_{hb \ra \ell}(\nrg)/m_p}{d\nrg_h/dR}
     d\nrg &
     \mbox{thick target} ,
\end{array}
\right.
\label{eq:rxnlite}
\eeq
where the integral begins at the threshold energy $\nrg_{\rm th}$ (if any).
Here we take the interaction $hk \rightarrow {\rm inelastic}$
of the non-thermal particle $h$ with background nuclide $k$
to dominate the inelastic losses.

With these reaction rates in hand, we illustrate their importance by
estimating below the perturbation on
light-element abundances that arises in the CMSSM with a massive gravitino.
Before doing this, we must first have an understanding
of the gravitino decays and the resulting hadronic spectra.

\section{Gravitino  Decays}
\label{sect:decays}

We work in the context of the CMSSM~\cite{cmssmwmap} which is described by
four parameters: universal gaugino, scalar, and trilinear masses, $m_{1/2}, m_0, A_0$,
and the ratio of the two Higgs vacuum expectation values, $\tan \beta$, along with
the sign of the Higgs mixing parameter, $\mu$. 
Motivated by $g_\mu - 2$ and $b \to s \gamma$, 
we restrict our attention to $\mu > 0$ and, for simplicity, we
consider only $A_0 = 0$.
We consider scenarios in which the lightest neutralino is the LSP,
but the gravitino is not assumed necessarily to be the
NLSP.
For fixed values of $\tan \beta$, the regions of parameter space for which the
relic density of neutralinos is computed to lie within the range 
$0.0975 < \Omega_{\rm CDM} h^2 < 0.1223$ determined
by WMAP \cite{wmap} and other observations for cold dark matter form narrow strips that foliate
the $(m_{1/2},m_0)$ plane~\cite{eoss,bench}.
These strips correspond to regions where there are enhanced annihilation 
cross sections that reduce the neutralino relic density to acceptable values.
These strips occur when the neutralino is nearly degenerate with some other
supersymmetric particle such as the partner of the tau lepton (coannihilation strip), 
when the neutrino is close to half the mass of the heavy Higgs scalar and pseudoscalar
boson so that rapid s-channel annihilation occurs (the funnel region), or 
at large values of $m_0$ when the renormalization-group evolution 
drives the value of $\mu$ to low values so
that the neutralino acquires a significant Higgsino component and new 
final-state channels become important (the focus-point strip).
Our analysis is based on the WMAP strips
for two representative values of $\tanb=10, 50$.
For $\tanb=10$ one strip follows  the coannihilation corridor, and
for $\tanb=50$ this strip also includes the funnel at larger values of $m_{1/2}$.
We also consider the focus-point strips for both values of $\tan \beta$. 

\begin{figure}[ht!]
\begin{center}
\epsfig{file=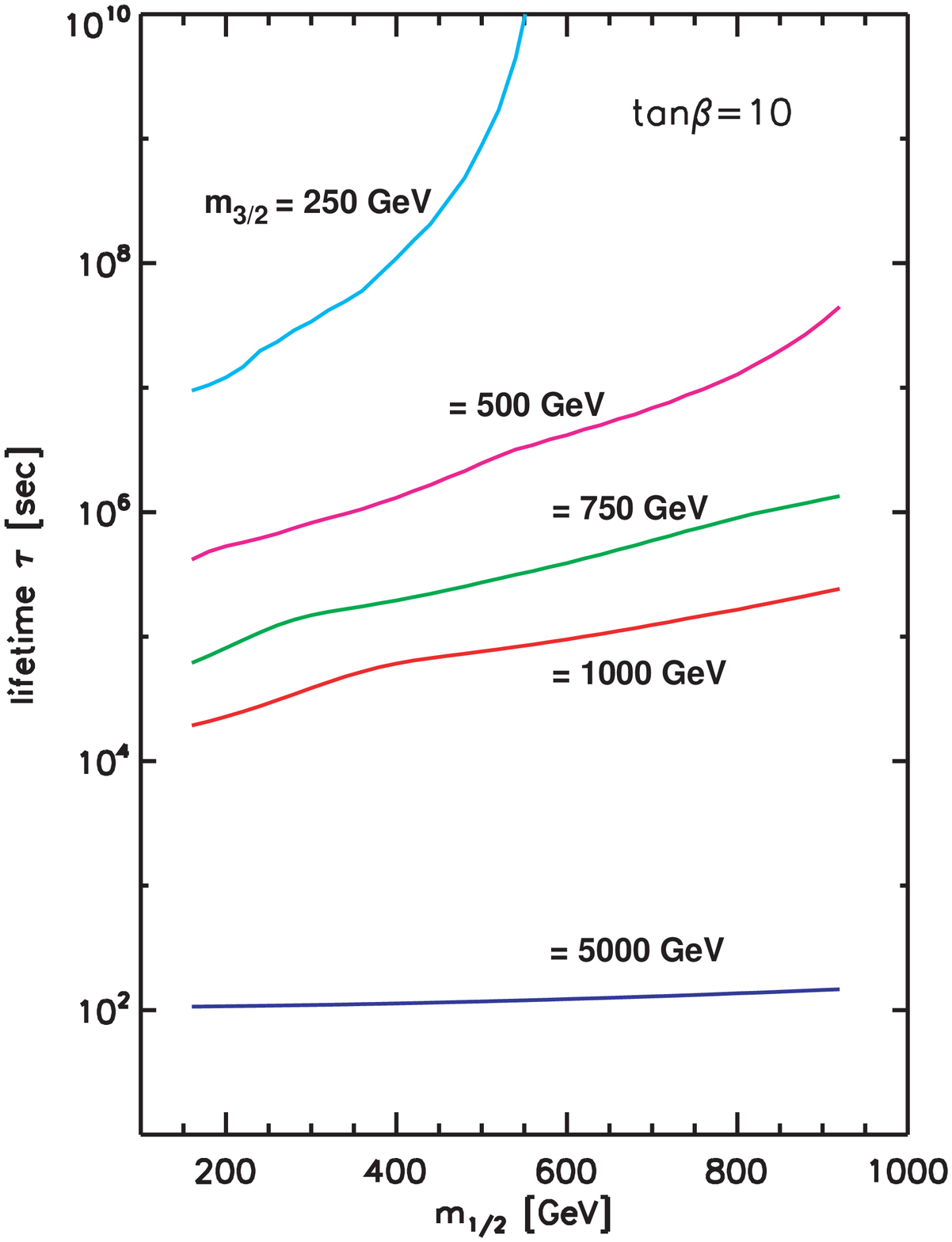,width=0.40\textwidth}
\epsfig{file=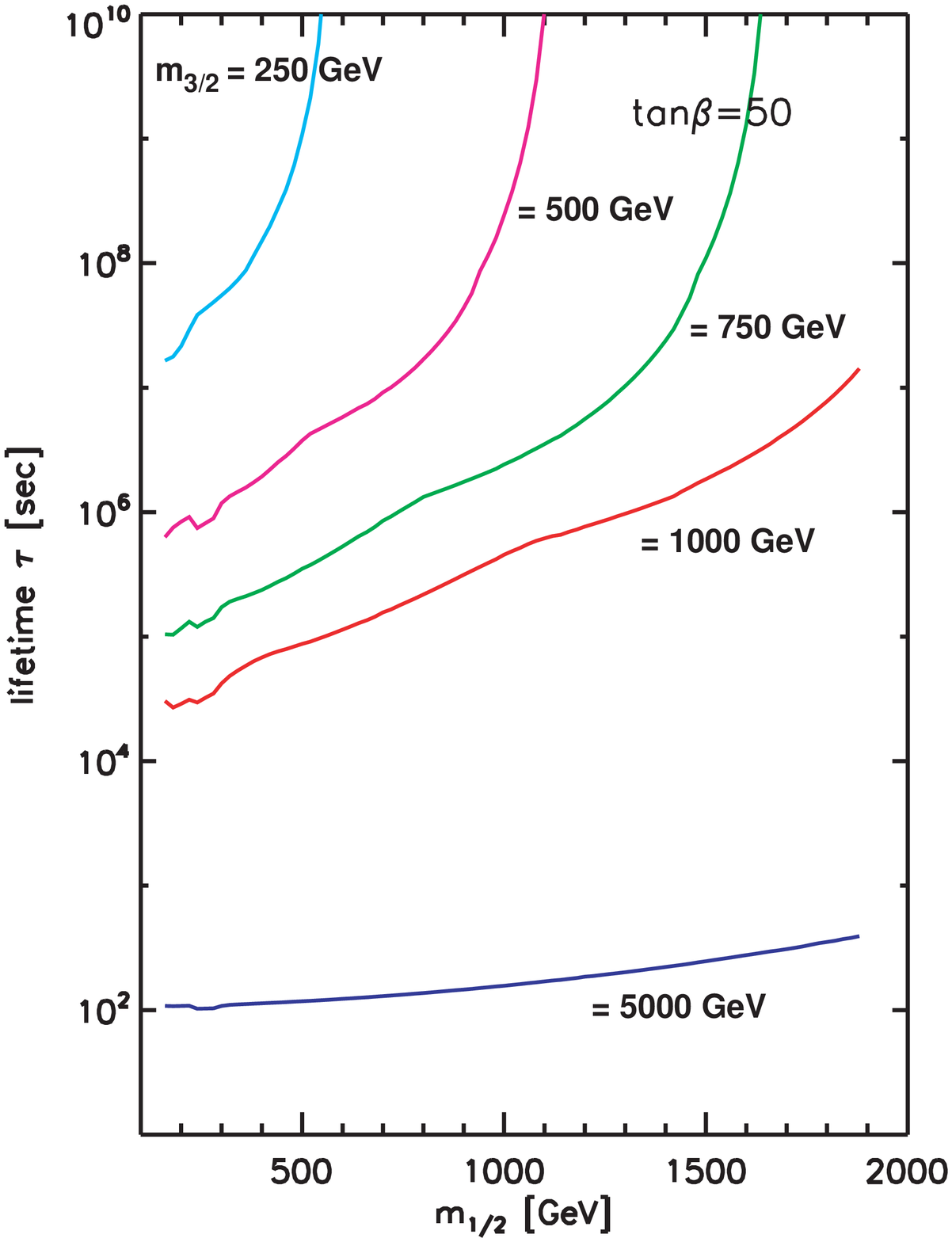,width=0.40\textwidth} \\
\epsfig{file=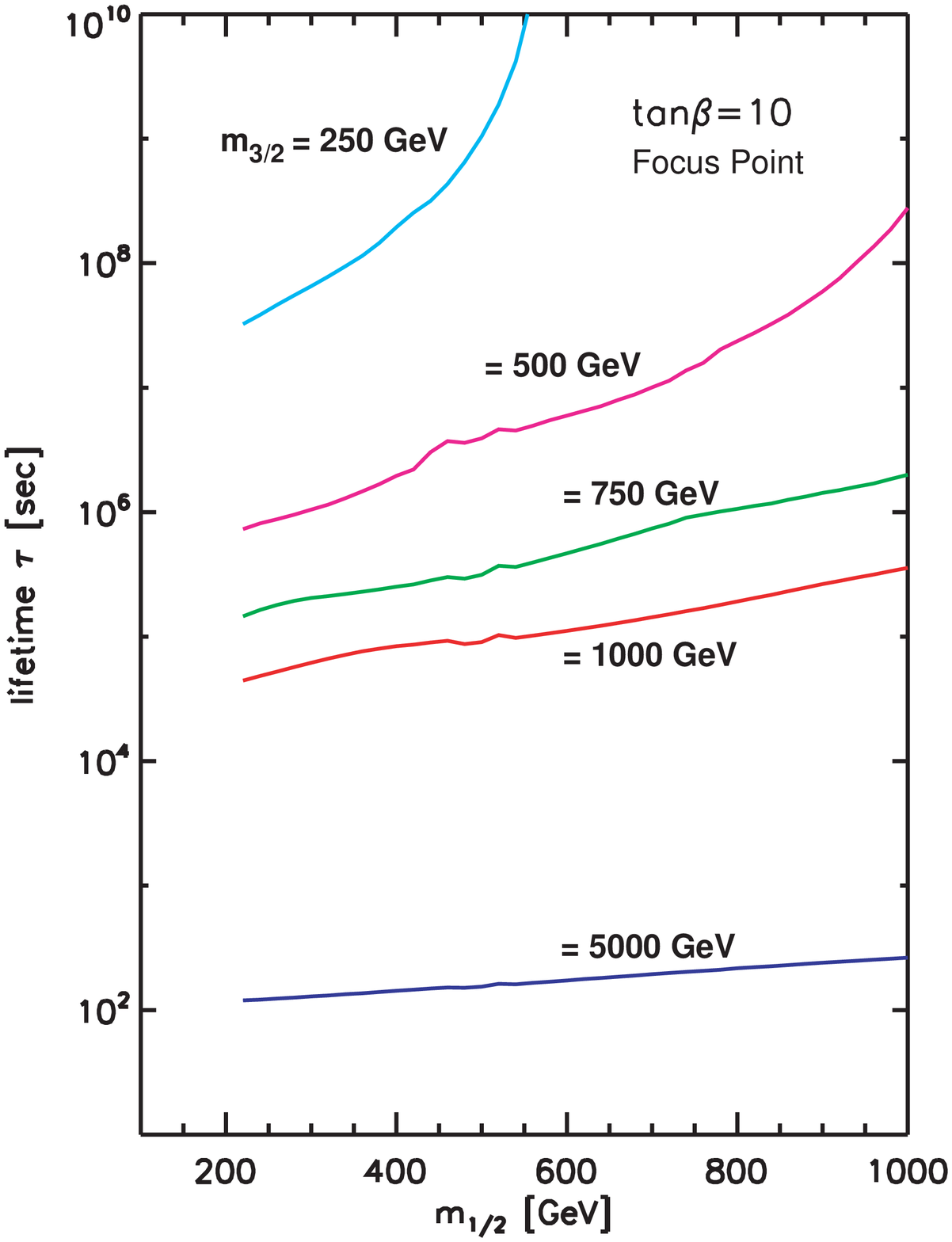,width=0.40\textwidth}
\epsfig{file=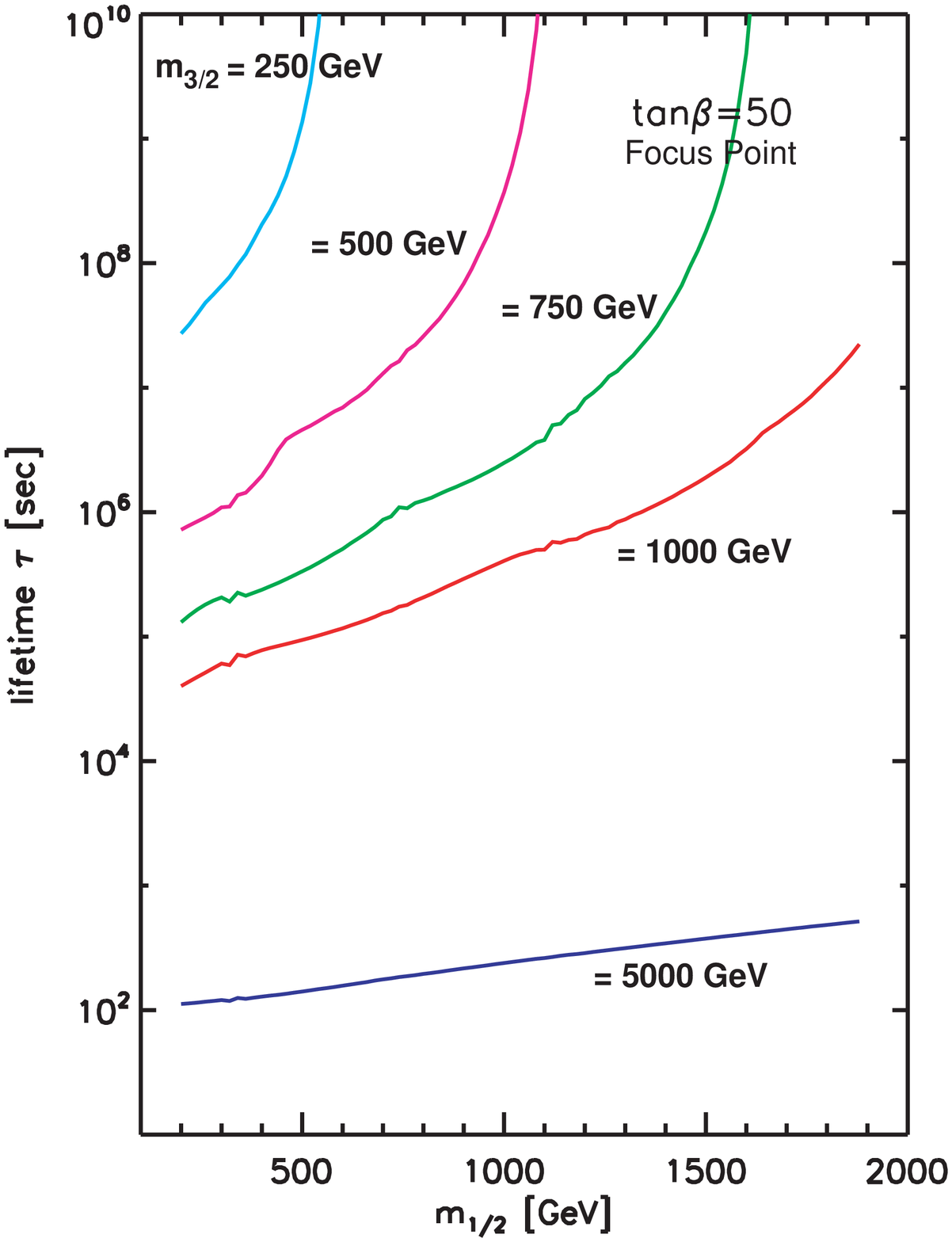,width=0.40\textwidth}
\end{center}
\caption{
\it The gravitino lifetime for representative values of $m_{3/2}$
as a function of $m_{1/2}$  
for $\tanb = 10$ (left) and $\tanb=50$ (right) along WMAP strips, 
in the coannihilation and funnel region (top) and the focus-point region (below).}
\label{fig:lifetimes}
%\end{center}
\end{figure}

In the scenario under study the gravitino  decays
into lighter sparticles, including the neutralino LSP but also others in general.
We take into account all the dominant decay channels
of the gravitino into (s)particles, including the complete set of two-body decays of
the gravitino into Standard Model particles and their spartners.
These fall into the following main categories: $\grav \to \tilde{f}\,f $,
$\grav \to \tilde{\chi}^+ \, W^- (H^-)$,
$\grav \to \tilde{\chi}^0_i \, \gamma (Z)$,
$\grav \to \tilde{\chi}^0_i \, H^0_i$ and $\grav \to \tilde{g} \, g$.
Analytical expressions for these amplitudes can be found
in Appendix~\ref{app:gravdec}.
In addition, we include the dominant
three-body decays  $\grav \to  \tilde{\chi}^0_i \, \gamma^*  \to \tilde{\chi}^0_i \, q  \qbar$,
 $\grav \to  \tilde{\chi}^0_i  \, W^+ W^-$, which are also discussed in Appendix~\ref{app:gravdec}.
In principle, one
should also include $q \qbar$ pair production through the virtual $Z$-boson
channel $\grav \to  \tilde{\chi}^0_i  \, Z^*  \to  \tilde{\chi}^0_i  \, q  \qbar$~\cite{kmy}
and the corresponding interference term. However,
this process is suppressed by a factor of $M_Z^4$ with respect to
$\grav \to  \tilde{\chi}^0_i  \, \gamma^*  \to  \tilde{\chi}^0_i  \, q  \qbar$, and the interference term is also
suppressed by $M_Z^2$. These contributions are therefore not very important,
and  we drop these amplitudes in our calculation.
We also drop the corresponding amplitudes for Higgs and squark exchange.
We calculate the lifetime of the gravitino by first
calculating the partial widths
of its dominant relevant decay channels, and then summing them.
Typical results are shown in Fig.~\ref{fig:lifetimes}, where we see that the
lifetimes for $\tanb = 50$ (right panels) are typically longer than those for
$\tanb = 10$ (left panels), particularly for larger $m_{1/2}$. On the other
hand, the lifetimes in the coannihilation/funnel regions (top panels) are
quite similar to those in the focus-point regions (bottom panels).

In order to calculate the resulting electromagnetic (EM) and
hadronic (HD) spectra, we first calculate the EM and HD decay
spectra of the different gravitino decay modes, and then weight them
by the corresponding branching ratios. The decay products that yield
EM energy obviously include directly-produced photons, but also
indirectly-produced photons, charged leptons (electrons and muons)
and neutral pions ($\pi^0$), which are produced via the secondary
decays of unstable heavy particles such as gauge and Higgs bosons.
Hadrons (nucleons and  mesons such as the $K_L^0$, $K^\pm$ and
$\pi^\pm $) are also usually produced through the secondary decays
of heavy particles, as well as (for the mesons) via the decays of
the heavy $\tau$ lepton. It is important to note that strange mesons and 
neutral pions decay
before interacting with the hadronic
background~\cite{kohri,SFT,0404231}. Hence they are relevant to
BBN processes and to our analysis only via their decays into
photons and charged leptons, which contribute to the EM component
of the decay showers. Therefore, the HD injections that
concern us are those that produce nucleons, via the decays of heavy
particles such as gauge and Higgs bosons and quark-antiquark pairs
and to a lesser extent charged pions.

\begin{figure}[!ht]
\begin{center}
\epsfig{file=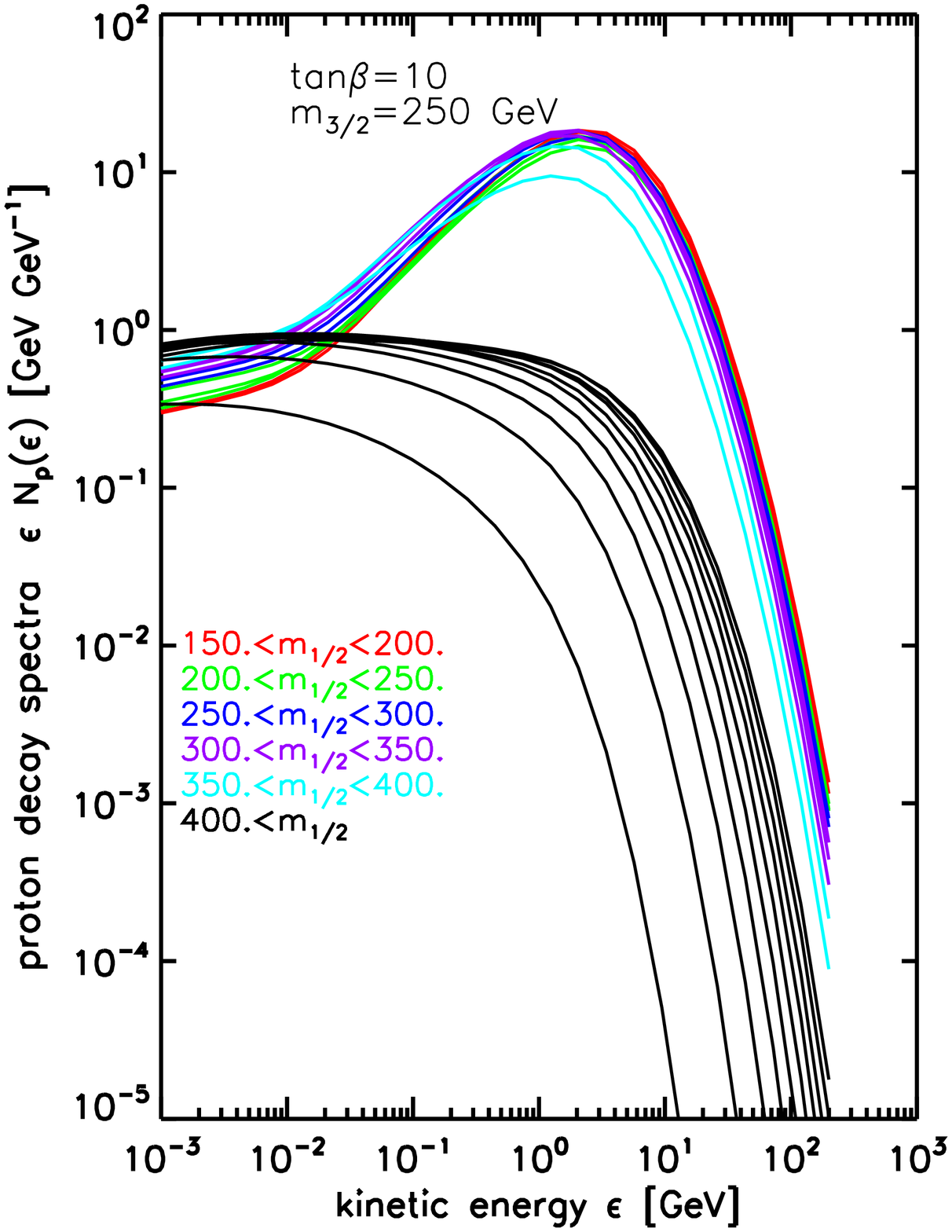,width=0.475\textwidth}
\epsfig{file=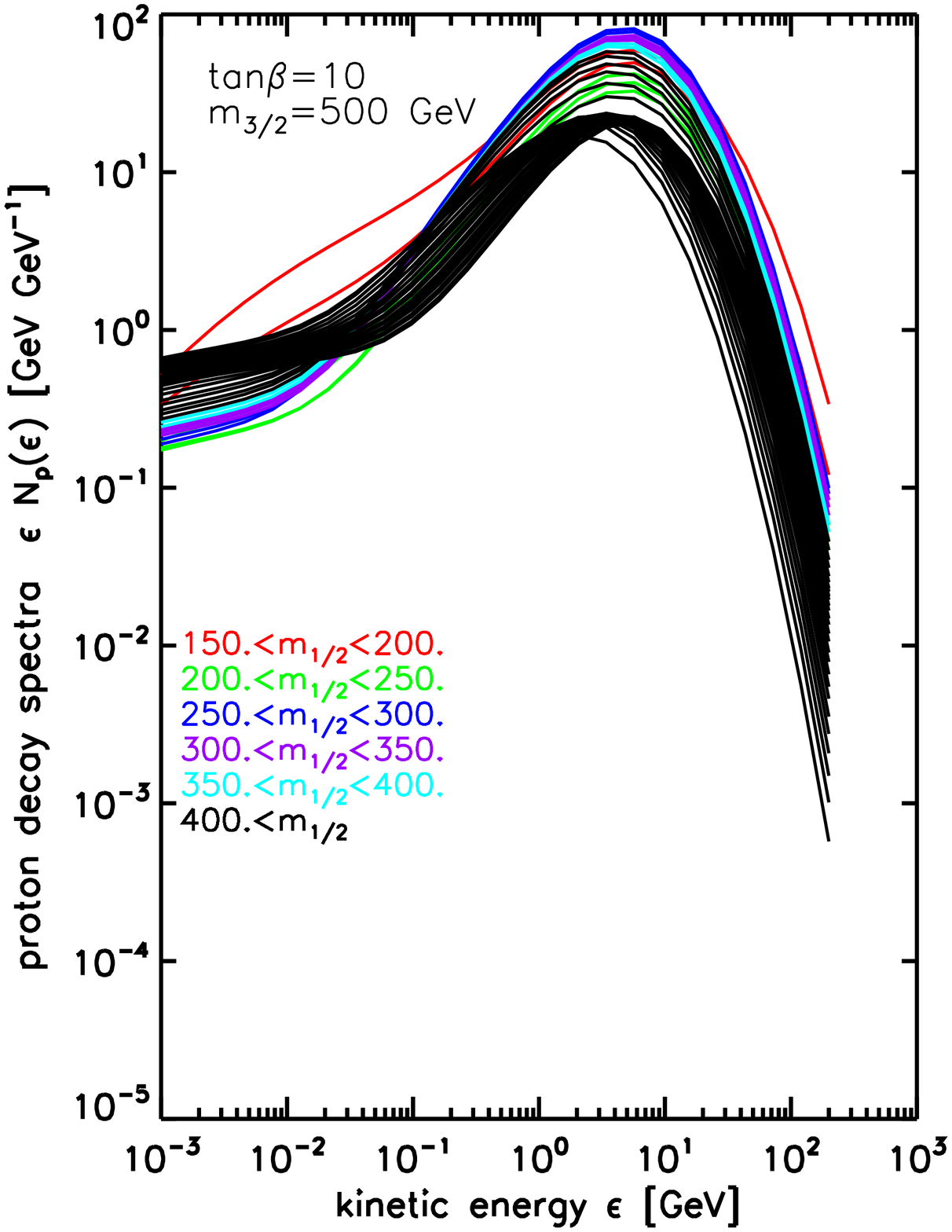,width=0.475\textwidth}
\caption{
\it Sample spectra for gravitino decays into protons.
We plot values of the combination $\nrg \ N_p(\nrg)$, which gives
the particle number per logarithmic energy range, for different
representative values of the gaugino mass
parameter $m_{1/2}$ along the coannihilation strip for $\tanb = 10$,
assuming the indicated values of the gravitino mass $m_{3/2}$.
\label{fig:decayspectra}
}
\end{center}
\end{figure}

\begin{figure}[ht!]
\begin{center}
\epsfig{file=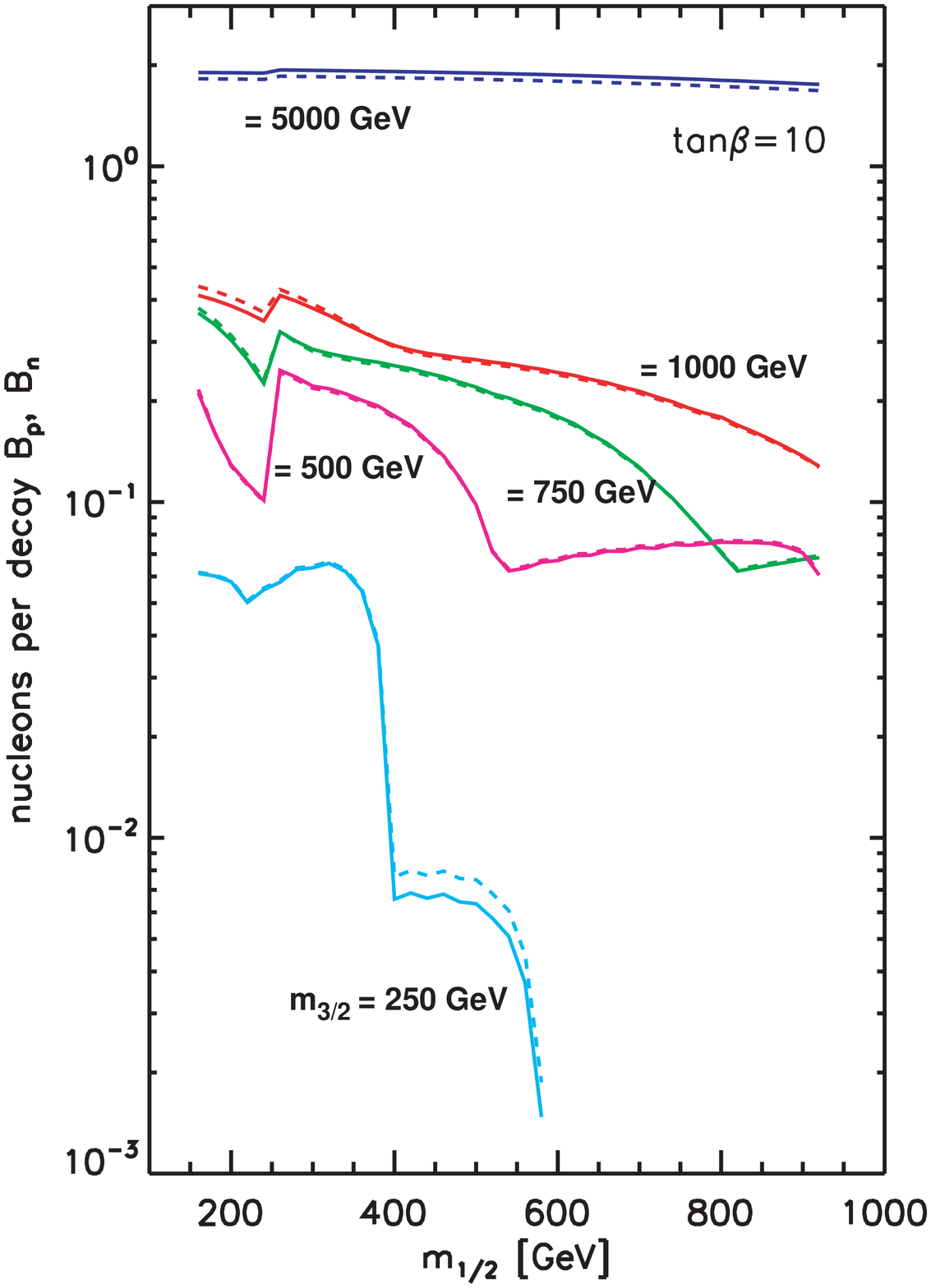,width=0.38\textwidth}
\epsfig{file=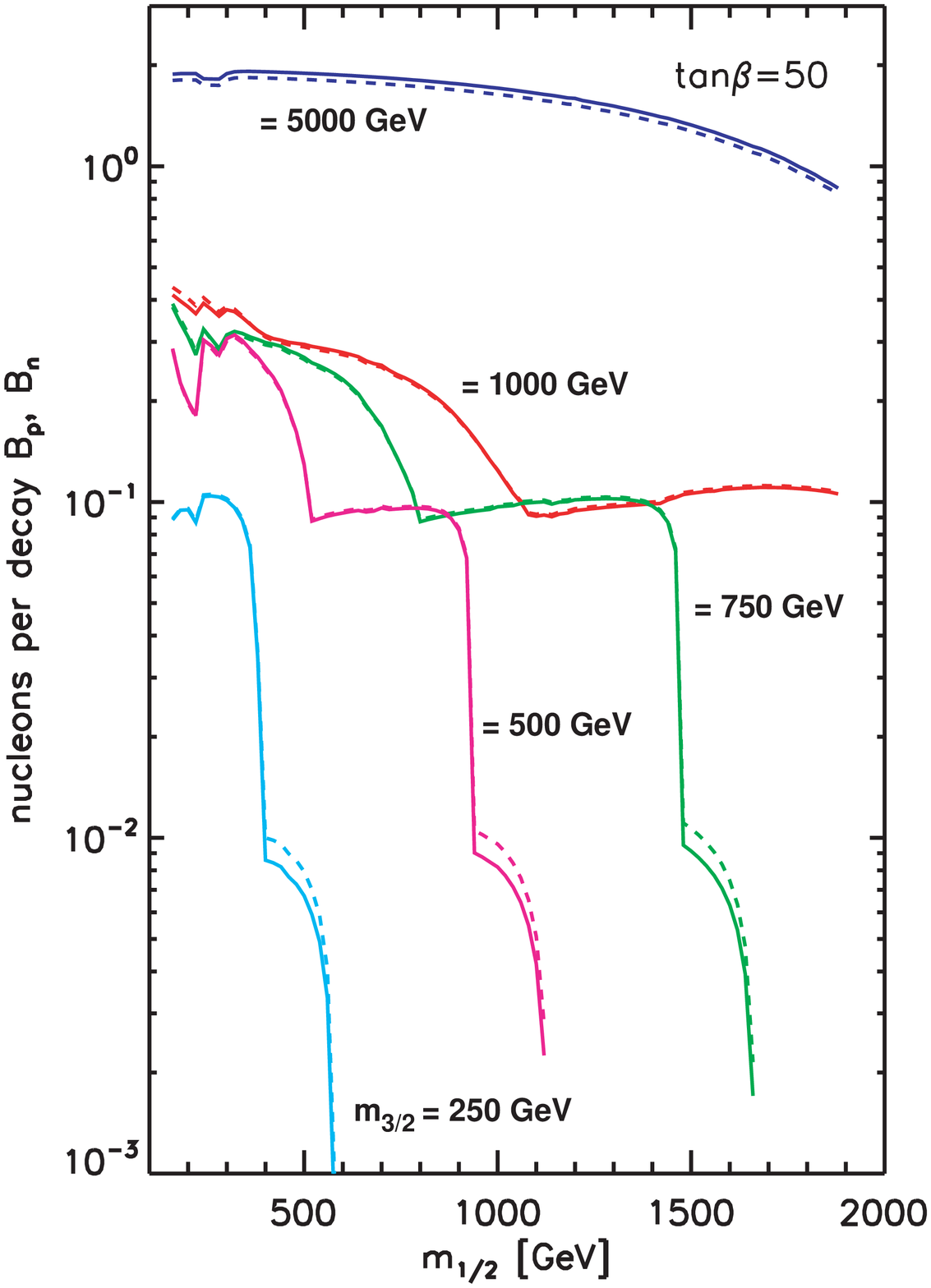,width=0.38\textwidth} \\
\epsfig{file=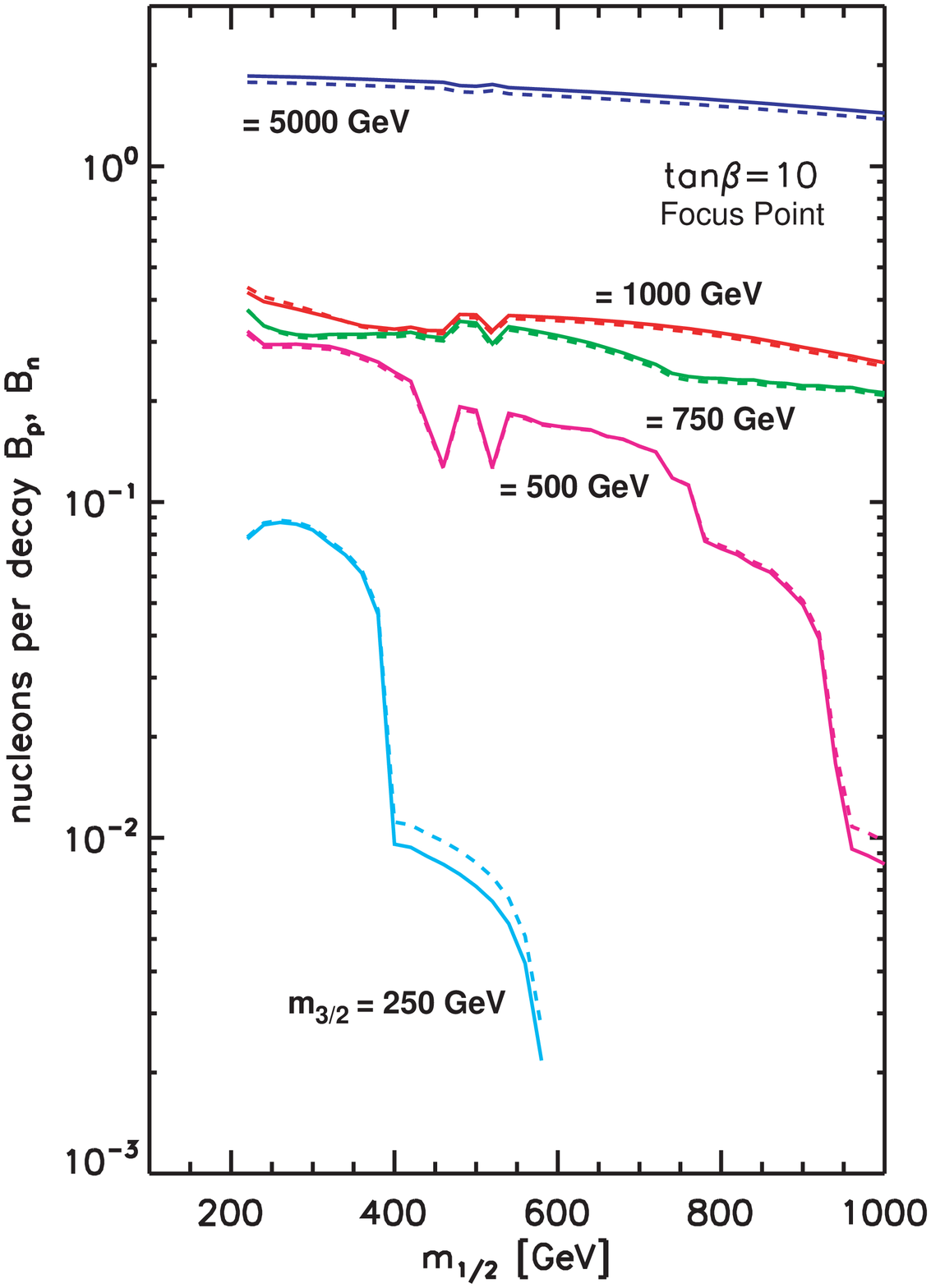,width=0.38\textwidth}
\epsfig{file=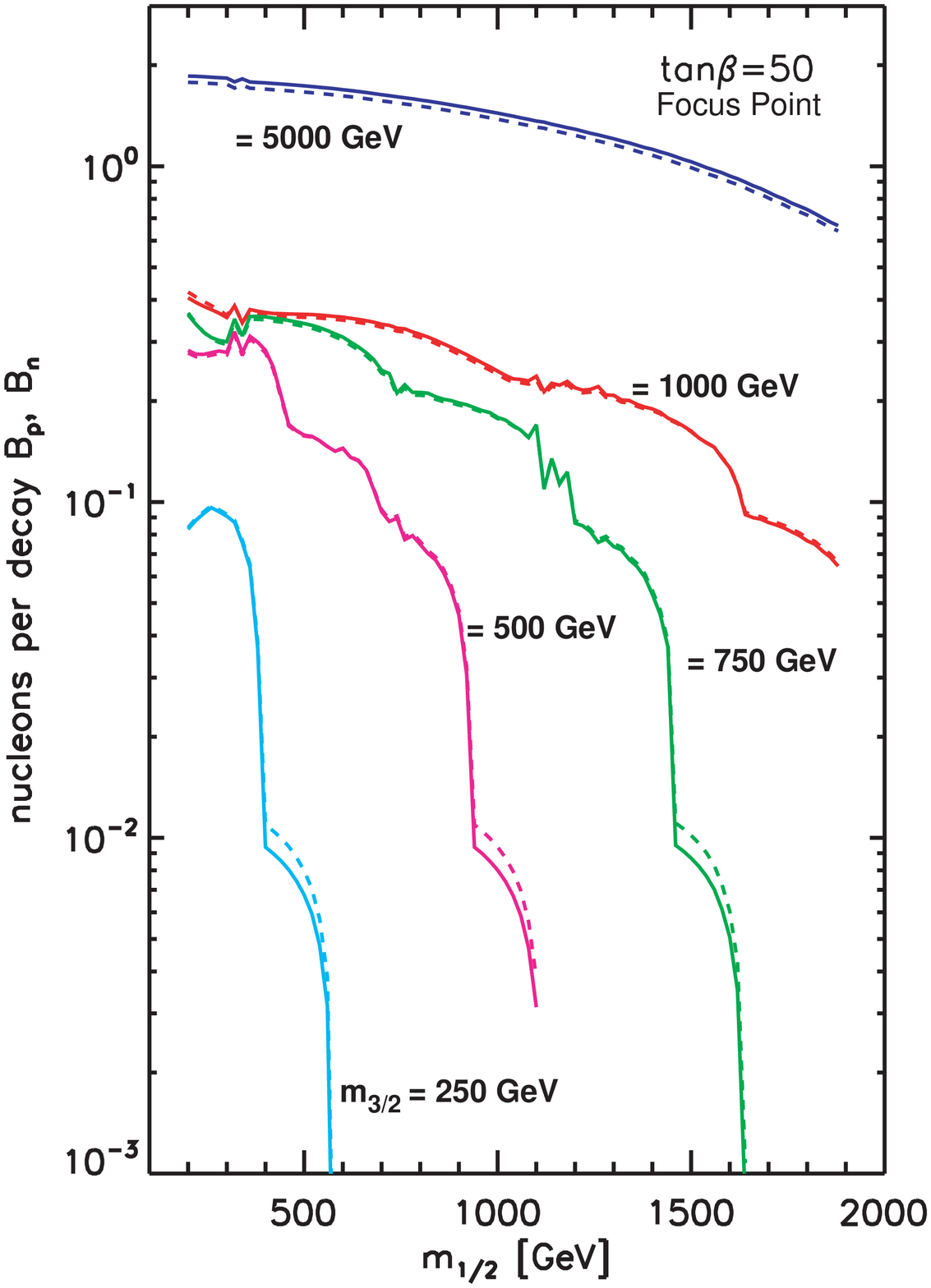,width=0.38\textwidth}
\end{center}
\caption{
\it The number of nucleons
per gravitino decay, as a function of $m_{1/2}$
 for $\tanb = 10$ (left) and $\tanb=50$ (right).
The upper panels are for WMAP strips in the coannihilation and rapid-annihilation regions,
and the lower panels are for WMAP strips in the focus-point regions.
Solid curves:  number $B_p$ of protons per decay.
Broken curves:  number $B_n$ of neutrons per decay.
We see that generally $B_p \approx B_n$ to a good approximation.
\label{fig:branching}
}
%\end{center}
\end{figure}

After calculating the partial decay widths and branching ratios, we then employ
the {\tt PYTHIA} event generator~\cite{pythia} to model both the EM and
the HD decays of the direct products of the gravitino decays. We first
generate a sufficient number of spectra for the secondary decays of the
gauge and Higgs bosons and the quark pairs.  Then, we perform fits to
obtain the relation between the injected  energy of the decaying particle, 
that we call $E_{inj}$, and the
quantity that characterizes the hadronic spectrum, namely $Q_h$, the
number of produced nucleons as a function of the nucleon energy, for various 
values of the $E_{inj}$.
We have performed fits that cover the range $200 \gev < E_{inj} < 20 \tev$.
In Table~\ref{tbl:pythia_fit} we present a representative set of fitting parameters 
for $E_{inj}=1000\gev$.
Then, using the equation 
\beq
\label{eq:spect}
Q_h(\nrg_h)=\frac{\nrg_h}{ E_{\rm inj}^2} \,  \frac {1}{x} \, \tilde{Q_h} (x) \,,
\eeq
 the spectrum  distribution $Q_h(\nrg_h)$ is computed using the function
$ \tilde{Q_h} (x)=Q_h(\nrg_h) d\nrg_h/dx$, with $x= \sqrt{\nrg_h^2 -m_h^2}/E_{inj}$.
 Some typical spectra for gravitino decays into protons are shown in Fig.~\ref{fig:decayspectra}.
These spectra and the fraction of the energy of the decaying particle that is
injected as EM energy are then used to calculate the light-element
abundances.
We stress that this procedure is repeated separately for each point sampled in the
supersymmetric parameter space. That is, given a set of parameters
$m_0, m_{1/2}, A_0, \tan \beta$, sgn$(\mu)$, and $m_{3/2}$, after
determining the sparticle spectrum, all of the relevant branching fractions
are computed, and the hadronic spectra and the injected EM energy
determined case by case.
Thus, in our analysis, $Q_p$ ($Q_n$) and hence the  total number of
protons (neutrons) per gravitino decay, $B_p$ ($B_n$),
varies between different points
in the parameter space, and Fig.~\ref{fig:branching} illustrates
this variation in  $B_p$  (and $B_n$) across the supersymmetric
parameter space we sample.

\begin{table}[ht]
\begin{center}
\caption{\it The values of the parameters $A$, $B$,
$C$, $D$ and $E$ used in fitting {\tt PYTHIA} hadronic decay spectra.
We use the function $Y \equiv \log_{10}(  \tilde{Q} (x) )$ parametrized via
$Y=A \, X^4+B \, X^3+C \, X^2+D \, X + E$,
where $X\equiv  \log_{10}(x)$, and the scaling parameter $x$ is defined as
$x=\sqrt{\nrg_h^2 -m_h^2}/E_{inj}$. The values of this table correspond to $E_{inj}=1\tev$.}

\vspace*{3mm}
\label{tbl:pythia_fit}
\begin{tabular}{|c|c|c|c|c|c|c|}
\hline \hline
decaying particles& inj. species  & $A$   &  $B$ & $C$ & $D$ & $E$  \\
\hline
$Z$& $p$          & 0.000   & 0.000   & -1.230   & -4.823  &  -3.401 \\
$$ & $n$          & 0.000   & 0.000   &  -1.281   & -4.917   & -3.418  \\
\hline
\hline
$h$ & $p$          & 0.000   & 0.000   & -1.011    & -4.302   &  -3.071  \\
$$ & $n$          & 0.000   & 0.000   &  -0.850    & -3.628    & -2.499   \\
\hline
\hline
$H$&$p$          & 0.000   & 0.000   & -0.647    & -3.398   &  -2.154  \\
$$& $n$          & 0.000   & 0.000   &  -0.644  & -3.414    & -2.230   \\
\hline
\hline
$A$&$p$          & 0.000   & 0.000   & -0.633    & -3.356   &  -2.133  \\
$$& $n$          & 0.000   & 0.000   &  -0.637   & -3.377    & -2.179   \\
\hline
\hline
$q \bar{q}$&$p$          & -0.065   & -0.551   & -2.007   & -4.326  &  -2.602 \\
$$&     $n$          & -0.033   &  -0.195    &  -0.603   & -2.039    & -1.327  \\
\hline
\hline
$W^+ \, W^-$&$p$          &  0.000   &    0.000   & -1.146   & -5.180  &  -4.036 \\
$$&$n$          &  0.000   &     0.000    &  -1.107  & -4.936    & -3.742  \\
\hline
\hline
\end{tabular}
\end{center}
\end{table}

The relative sizes of the  gravitino partial widths and
the locations of the various particle thresholds  help us to understand the lifetime and the
hadronic spectra curves.
In general, the two-body channels $\grav \to  \tilde{\chi}^0_i  \, \gamma$;
$ \tilde{f}\,f $; $ \tilde{g} \, g$; $ \tilde{\chi}^+ \, W^- $
 dominate the  $\grav$  decays. In particular,  the decay to $\tilde{\chi}^0_i  \, \gamma$
yields the bulk of the injected EM energy.
However, the decays to sfermions, gluinos and charginos become of the same order of magnitude
as the $\grav \to  \tilde{\chi}^0_i  \, \gamma$ channel whenever they are kinematically possible.
Also, when the $\grav$ is heavy enough to decay into a real $Z$ boson, the
channel $\grav \to  \tilde{\chi}^0_i  \, Z$ is  the dominant channel for producing HD
injections. 
When kinematically allowed, $\grav \to  \tilde{\chi}^+ \, W^- $
and $\grav \to  \tilde{g} \, g$ are also important in producing HD injections.
The Higgs boson channels are smaller by a few orders
of magnitude (due to couplings and kinematics)
and, in particular, in the large-$m_{1/2}$ region, decays to heavy Higgs bosons ($H,A$)
become kinematically accessible only for heavy $\grav$ 
and are unimportant otherwise. 

Turning to the three-body channels, the decay through the virtual photon to a $q \qbar$ pair
can become comparable to the channel $\grav \to  \tilde{\chi}^0_i  \, Z$,
injecting nucleons even in the kinematical region $\mgrav < \mchi + M_Z$,
where direct on-shell $Z$-boson production is not possible.
Finally, we note that the partial width of the
three-body decay  $\grav \to  \tilde{\chi}^0_i  \, W^+ W^-$   is usually
smaller by at least  an order of magnitude relative to the dominant two-body decays,
except when this three-body decay exhibits resonant behavior. For example,
the subprocess  $\grav \to  {\tilde{\chi}^{+*}} W^- \to   \tilde{\chi}^0_i  \, W^+ W^-$
can lift the contribution of the  $\tilde{\chi}^0_i  \, W^+ W^-$ channel to the level of the
dominant two-body decays  in the threshold region where the chargino can be produced on-shell.

With these observations in mind,  one can understand the gravitino lifetime curves
along the  WMAP strips for $\tanb=10$ and $50$ in Fig.~\ref{fig:lifetimes}.
We recall that the funnel region is only present in the CMSSM for large $\tan \beta$.
Since the relation $m_\chi \approx m_A/2$ is realized at large $m_{1/2}$, the
WMAP strip extends to significantly higher values of $m_{1/2}$ for $\tan \beta = 50$
than for $\tan \beta = 10$. In the latter case (upper left panel) , the WMAP strip shown
consists only of a coannihilation region, which terminates around $m_{1/2} = 900$ GeV.
We notice that the gravitino lifetime is longer for $\tanb=50$
than for  $\tanb=10$, for the same values of $m_{3/2}$ and $m_{1/2}$.
Especially for  the two upper panels,
 this is because for $\tanb=50$ the WMAP strip, in the coannihilation region and (particularly) in
the Higgs rapid-annihilation funnel  region, occurs at larger $m_0$ and hence
heavier Êsquark and slepton masses, than in the $\tanb=10$ case.
This implies that the dominant  two-body channels $\grav \to \tilde{f}\,f $
are very suppressed or even closed for $\tanb=50$. Thus, the total gravitino decay width
is smaller (and hence the lifetime longer) for $\tanb=50$  than for $\tanb=10$.

The lower panels in  Fig.~\ref{fig:lifetimes} correspond to the focus-point region. 
It is worth noticing that, unlike the coannihilation 
or the rapid-annihilation region, the focus-point strip extends 
to remarkably  high values of $m_0$ and $m_{1/2}$.  For example,
for $\tanb=50$ (lower right panel) 
$m_0\sim 3\tev$ ($5\tev$) at $m_{1/2}=1000 \gev$ ($2000 \gev$). 
As a result, the sfermion masses in this region are 
much larger than in the funnel or the coannihilation strip.  Hence, for  gravitino masses 
up to $1\tev$ all the  fermion-sfermion decay  channels $\grav \to \tilde{f}\,f $ are closed,
and the lifetime is larger than the upper panels. 
On the other hand, when $m_{3/2}=5\tev$ the dominant decay channels  $ \tilde{g} \, g$,  
$ \tilde{f}\,f $ and  $ \tilde{\chi}^+ \, W^- $
are open along the WMAP strips also in the focus-point region, 
resulting in relatively flat lifetime curves as functions of $m_{1/2}$.

We note that for $m_{3/2} = 5$~TeV, the largest value shown, the
gravitino lifetime $\sim$ few $\times 10^2$~s in all the cases
shown in in  Fig.~\ref{fig:lifetimes}, which is comparable with the
duration of BBN. Lighter gravitinos would decay after BBN is completed.

As $m_{1/2}$ increases for fixed $m_{3/2}$, the gaugino masses increase. Therefore,
one by one the various gravitino decay channels are closed. The last to be closed are the
two-body decay to $ \tilde{\chi}^0_1  \, \gamma$ and the three-body decay to light quark pairs
$\tilde{\chi}^0_1  \, q  \qbar$. Eventually all the channels are closed when $\mgrav < \mchi$.
For $m_{3/2}=250 \gev$  this occurs for $m_{1/2}\sim 580 \gev$, as can be seen in
Fig.~\ref{fig:lifetimes}.
When the dominant channel for hadron production, $ \tilde{\chi}^0_1  \, Z$, closes we
observe a significant decline in the nucleon spectra. This is the reason that in
Fig.~\ref{fig:branching}  in all the panels, for
$\mgrav=250 \gev$ the value of $B_p$ becomes smaller than $10^{-2}$
at  $m_{1/2}=400\gev $, and the same at  $m_{1/2}=940\gev $
when $\tanb=50 $ and $\mgrav=500\gev$ (right panels). Above these values of $m_{1/2}$,
the only channel that produces some hadrons is $\grav \to \tilde{\chi}^0_1  \, q  \qbar$.

The  importance of the
channel $ \grav \to \tilde{\chi}^0_i  \, Z$ for producing nucleons can be seen
in Fig.~\ref{fig:decayspectra}. There we plot the quantity $\nrg \ N_p(\nrg)$ for the case of protons
for $\tanb=10$ and $m_{3/2}=250\gev \, (500 \gev)$ in the left (right) panel. The  curves in these figures
correspond to the various $m_{1/2}$ we sample along the  WMAP coannihilation strip.
The reason for the peaks in the $\gev$ region  is that
the protons that are produced by on-shell hadronic decays of the $Z$ boson have
typical energies of a few $\gev$.
As discussed earlier, for $\tanb=10$ and $m_{3/2}=250\gev$ the $ \tilde{\chi}^0_i  \, Z$
channel closes above $m_{1/2}=400\gev$. Therefore, we observe two kinds of curves
in Fig.~\ref{fig:decayspectra} (left).
These that peak in the $\gev$ region are fed by the $ \tilde{\chi}^0_1  \, Z$ channel, whereas
these without the peak  originate from  the three-body channel $\tilde{\chi}^0_1  \, q  \qbar$.
For  $\tanb=10$ and $m_{3/2}=500 \gev$,
the decay $ \grav \to \tilde{\chi}^0_1  \, Z$ is not closed anywhere along the WMAP strip,
so all the curves
in Fig.~\ref{fig:decayspectra} (right) exhibit the $Z$-boson peak.

There are a few other features in Fig.~\ref{fig:branching}
to be discussed. For  $m_{1/2} \ga 240 \gev$ in the upper panels the lightest chargino can decay
on-shell to $ \tilde{\chi}^0_i  \, W^+$.  The same happens in the lower left (right) panel for 
$m_{1/2}=520\gev$ ($m_{1/2}=360\gev$).
This causes a sudden increase in the number of the produced protons,
that is more noticeable for $\tanb=10$ (upper left panel), 
since there the relative importance of the decay channel
$\grav \to  {\tilde{\chi}_1^{+*}} \, W^- \to \tilde{\chi}^0_i  \, W^+ W^-$ is greater.
This channel is closed for larger values of $m_{1/2}$ as the chargino becomes heavier,
resulting in the shoulders we observe  for $\tanb=10$ (upper left panel)  in the $B_p$ curves at
$m_{1/2}=520\gev$ ($\mgrav=500\gev$)
and $m_{1/2}=800\gev$ ($\mgrav=750\gev$). The corresponding features appear
also at the same points for $\tanb=50$ in Fig.~\ref{fig:branching} (upper right panel). 
In the same figure, for $m_{3/2} \ge 500$ GeV, 
the additional wiggle seen at  $m_{1/2} \sim 280$ GeV is due to the
closing of the  $\grav \to {\tilde{\chi}}^+_2  W^-$ channel. 
In the focus-point  figures (lower  panels), at 
$m_{1/2}=400\gev$ and $m_{3/2}=250\gev$ the $ \tilde{\chi}^0_1  \, Z$
channel closes. The same occurs at $m_{1/2}=960\gev$ ($m_{1/2}=1460\gev$) 
for  $m_{3/2}=500\gev$ ($m_{3/2}=750\gev$). After this point, $B_p$ diminishes. 
Along the focus-point strip, we also can see  in the $B_p$ curves the effect of the closing of the
$\tilde{\chi}_{1,2}^+ \, W^-$ channels. In particular, the $\tilde{\chi}_1^+ \, W^-$ channel closes 
 at $m_{1/2}=780\gev$ for $\mgrav=500\gev$,  $\tanb=10$,  at $m_{1/2}=1200\gev$
for $\mgrav=750\gev$, $\tanb=50$, and  at $m_{1/2}=1640\gev$ 
for $\mgrav=1000\gev$, $\tanb=50$. 
For   $\tanb=50$, the $\tilde{\chi}_2^+ \, W^-$ channel closes at 
$m_{1/2}=440\gev$ for $\mgrav=500\gev$,  at $m_{1/2}=760\gev$
for $\mgrav=750\gev$, and  at $m_{1/2}=1040\gev$ for $\mgrav=1000\gev$.
%For   $\tanb=10$ and $\mgrav=750\gev$ closes again at $m_{1/2}=760\gev$.
These thresholds produce distinctive features in the corresponding curves.
For very heavy gravitino masses, such as the case $m_{3/2}= 5\tev$
considered here, the dominant two-body decay channels
 $ \tilde{g} \, g$,  $ \tilde{t}\,t $, $\tilde{\chi}^0_i  \, Z$ and $ \tilde{\chi}^+ \, W^- $
are  kinematically  available, even in the focus-point  region, so no specific features are observed,
and the nucleon fractions are similar in all the scenarios studied. 

Finally, we note that various other channels close as $m_{1/2}$ increases, without producing any 
significant features in the $B_p$ curves. For instance,  for $\tanb=10$ 
(upper left panel in  Fig.~\ref{fig:branching}), 
for $\mgrav=250 \gev$, the $\tilde{\chi}^0_1 h $ channel closes at $m_{1/2}=340\gev$. 
The same occurs  for the channels  $\tilde{g} g  $ and $ \tilde{\chi}_1^+  H^-$ when $\mgrav=500 \gev$ 
at $m_{1/2}=220\gev$ , and at $m_{1/2}=300\gev$ for the 
channel $ \tilde{\chi}^+_2  W^-$. Similarly, for $\mgrav=750 \gev$ at $m_{1/2}=260\gev$,
the  channel $\tilde{\chi}^+_2  H^-$ closes, and at $m_{1/2}=340\gev$ the 
channels  $\tilde{g} g  $ and $\tilde{\chi_1}^+  H^-$ close. Just to complete the list for this gravitino mass, 
at  $m_{1/2}=400\gev$  the Higgs boson channels $\tilde{\chi}^0_1 A,H $ close and at 
$m_{1/2}=520\gev$  the chargino  channel $\tilde{\chi}^+_2 W^- $ closes.

\section{Constraints on Metastable Particles}

\label{sect:constraints}

\subsection{Generic Constraints on Abundances of Metastable Particles}

Before discussing constraints on specific supersymmetric models
with a metastable gravitino, we first
discuss constraints on the possible abundance $\zeta_X$ of a generic metastable
particle $X$, as a function of its possible lifetime $\tau_X$.
We discuss exclusively the constraints due to the effects of the electromagnetic and
hadronic showers
produced in $X$ decays, postponing to a later paper discussion of the extra constraints
that are imposed on the abundances of charged metastable particles by their
catalytic effects on light-element abundances due to the formation of bound states.
Thus, the constraints presented in this Section apply exclusively to {\it neutral}
metastable particles including, but not limited to, the gravitino.
Not wishing to commit to any specific model, we give results for
two typical values of $B_h$, which are applicable also outside the context of
specific supersymmetric models. Recall, however, that, as noted above, in specific supersymmetric
models one may calculate $B_h$ and it is {\it not}, in general, constant.

The hadronic decays of metastable particles $X$ affect BBN in different ways, depending
on the stage of BBN in which
the non-thermal decay particles interact with the
background thermal nuclei.
This effectively divides the decay effects according to
the decaying particle's lifetime $\tau_X$.

$\bullet$
Decays much before weak freeze-out ($\tau_X \ll 1 \ {\rm sec}$)
produce showers that are thermalized before BBN commences.
These decays thus have no impact on light-element
abundances, and BBN offers no strong constraints on such
short-lived decays.

$\bullet$
Decays during weak freeze-out ($1 \ {\rm sec} < \tau_X < 100 \ {\rm sec}$) introduce
new interactions that may interconvert the neutrons
and protons, e.g., via $n \pi^+ \rightarrow p \pi^0$.
These interactions prolong the $n \leftrightarrow p$
equilibrium, and hence delay the freeze-out of
the $n/p$ ratio.  In this regime the effect on the light elements is
somewhat similar to the addition of relativistic species,
with the dominant effect being that on \he4. However, because
the upper limit on \he4 is weak \cite{OSk2004}, this effect does not 
induce a constraint on our particle properties.

$\bullet$
Finally, decays following weak freeze-out ($\tau_X \gg 100 \ {\rm sec}$)
generate electromagnetic and hadronic showers that induce new (photo)nuclear
interactions, which in turn may modify the light-element abundances
established previously by BBN.

\begin{figure}[ht!]
\begin{center}
\epsfig{file=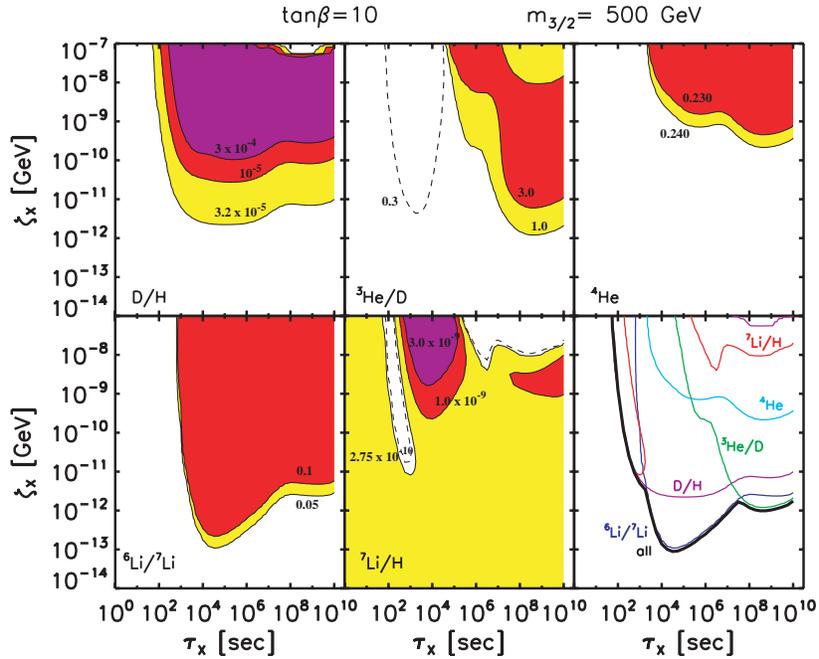,width=0.65\textwidth}
\end{center}
\caption{
\it 
Plots of abundance 
versus lifetime for metastable particles
$X$ with lifetimes $\tau_X$ between 1 and $10^{10}$~sec,
assuming the decay spectra calculated for
$(m_{1/2},m_{3/2},\tan \beta) = (300~{\rm GeV}, 500~{\rm GeV}, 10)$,
in which case $B_p \approx 0.2$
and the electromagnetic branching is $B_{\rm EM} m_{3/2} = 115$ {\rm GeV}.
The $X$ abundance before decay is given by $\zeta_X = m_X n_X/n_\gamma$ (eq.~\ref{eq:zeta}).
The white regions in each panel are those allowed
at face value by the ranges of the light-element abundances
reviewed in Section~\ref{sect:abs},
whilst the yellow, red and magenta regions correspond to
progressively larger deviations from the central values of the abundances.
\label{fig:zeta-tau-std}}
\end{figure}

\begin{figure}[!htbp]
\begin{center}
\epsfig{file=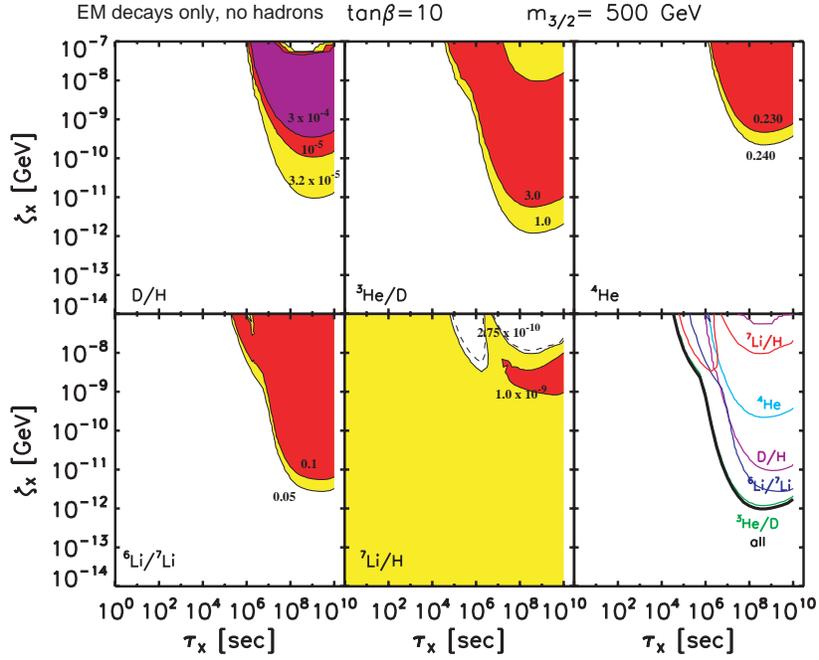,width=0.65\textwidth}
\end{center}
\caption{
\it 
As in Fig.~\ref{fig:zeta-tau-std}, but with 
electromagnetic decay products only.  All hadronic showers 
and resulting interactions with light elements are ignored.
\label{fig:zeta-tau-EMonly}
}
\end{figure}

\begin{figure}[!htbp]
\begin{center}
\epsfig{file=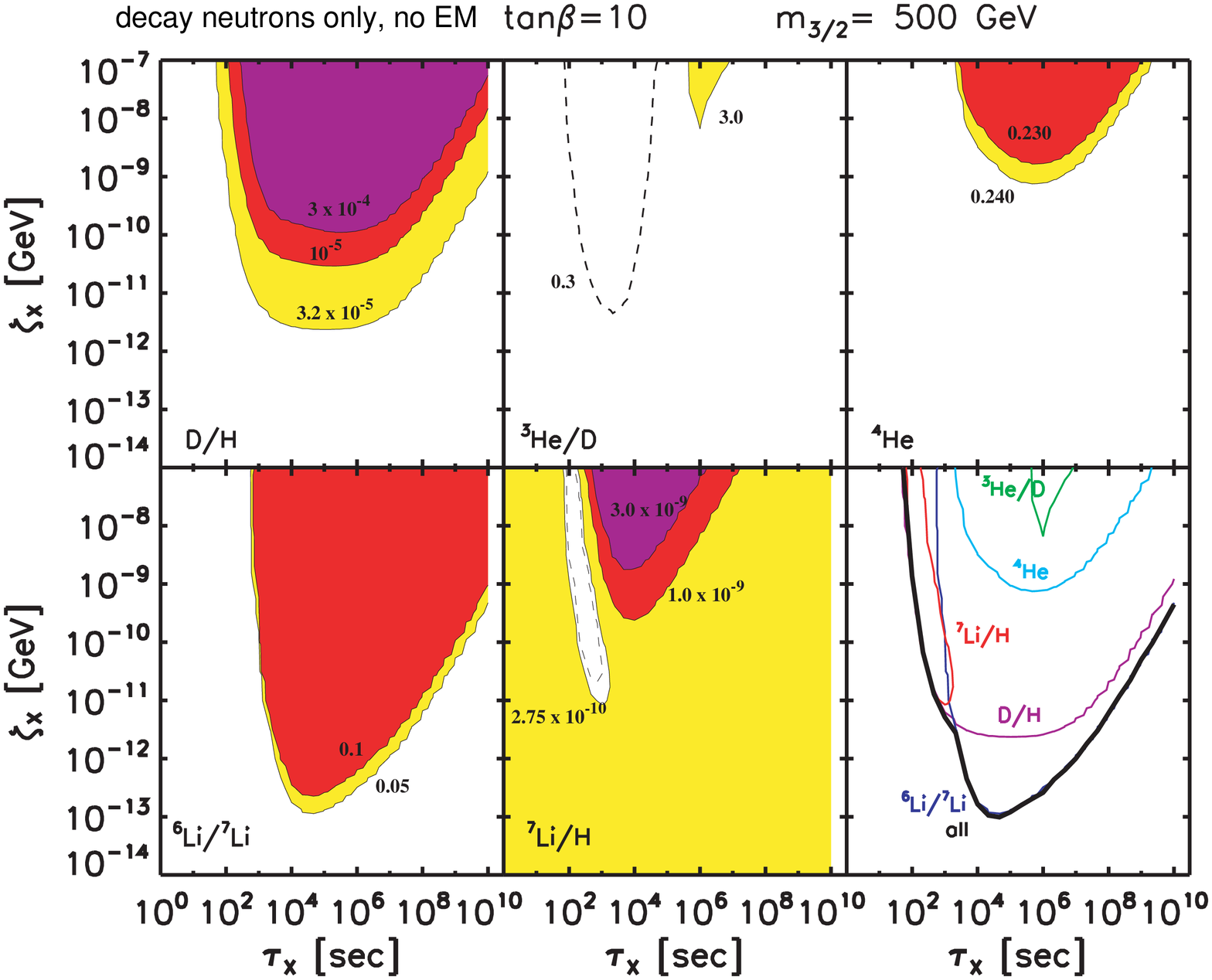,width=0.65\textwidth}
\end{center}
\caption{
\it 
As in Fig.~\ref{fig:zeta-tau-std}, but 
only with decay neutrons.  Decay protons and electromagnetic particles are ignored.
\label{fig:zeta-tau-nonly}
}
\end{figure}

\begin{figure}[!htbp]
\begin{center}
\epsfig{file=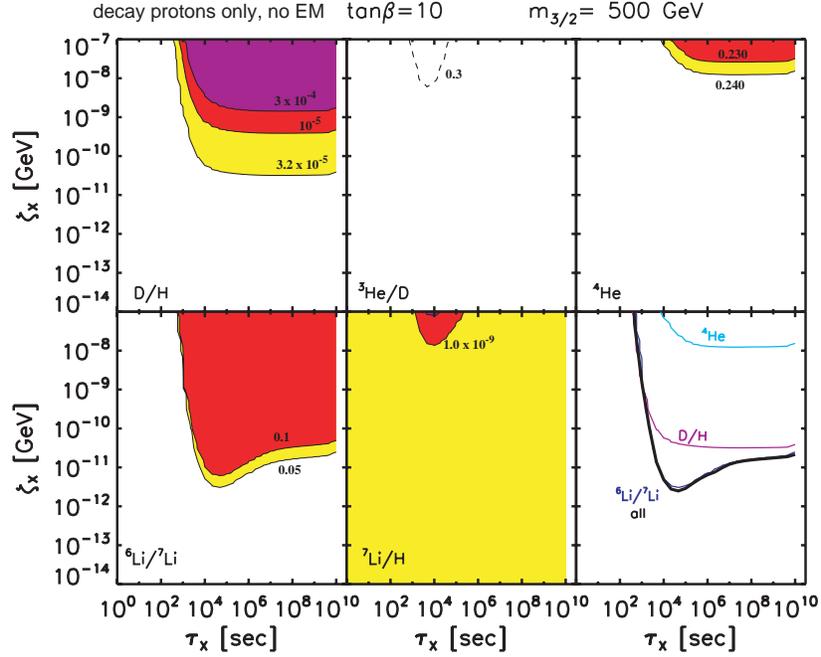,width=0.65\textwidth}
\end{center}
\caption{
\it 
\label{fig:zeta-tau-ponly}
As in Fig.~\ref{fig:zeta-tau-std}, but 
only with decay protons.  Decay neutrons and electromagnetic particles are ignored.
}
\end{figure}

These processes
lead to the constraints seen in 
Figs.~\ref{fig:zeta-tau-std}--\ref{fig:zeta-tau-ponly}.
These plot abundance contours as a function of
pre-decay $X$ abundance $\zeta_X$ as in eq.~(\ref{eq:zeta})
and lifetime $\tau_X$, all for decay spectra 
corresponding to 
$(m_{1/2},m_{3/2},\tan \beta) = (300 \, {\rm GeV}, 500 {\rm GeV},10)$.
In these and subsequent figures, the white regions in each panel are those allowed
at face value by the light-element abundances
reviewed in Section~\ref{sect:abs}.   Specifically, these are:
D/H $< 3.2 \times 10^{-5}$, \he3/D $< 1.0$ (also shown are dashed lines for
\he3/D $< 0.3$),  Y$_{\he4} > 0.240$, 
\li6/\li7 $< 0.05$, and \li7/H $< 2.75 \times 10^{-10}$.
The latter constraint comes from \li7 in globular clusters, and
we also demarcate by a dashed line the region within the white area
where \li7/H $< 1.91 \times 10^{-10}$ as inferred from field stars.
The regions shaded 
yellow, red, and magenta regions correspond to
progressively larger deviations from the central values of the abundances
as noted on the figures.
Fig.~\ref{fig:zeta-tau-std} shows the constraints when 
the effects of both
hadronic and electromagnetic decays are included.
To give a sense of how the various decay modes contribute,
Fig.~\ref{fig:zeta-tau-EMonly} shows constraints when
hadronic effects are omitted,
and only electromagnetic decays are included.
Figs.~\ref{fig:zeta-tau-nonly} (Fig.~\ref{fig:zeta-tau-ponly})
both omit effects of electromagnetic decays, 
and include only hadronic decays to neutrons (protons).

%%%%% FIGURE %%%%%
%%%%%%%%%%

%%%%% FIGURE %%%%%
%%%%%%%%%%

%%%%% FIGURE %%%%%
%%%%%%%%%%

%%%%% FIGURE %%%%%
%%%%%%%%%%

The basic features of these plots are similar to those found
in previous work, but for completeness we summarize them here.
In all plots, we see that at low $\zeta_X$, all constraints
are satisfied except for \li7.  This is reasonable, as
in the limit of $\zeta_X \rightarrow 0$ we recover
the standard BBN abundances, which agree well with observations
except for the \li7 problem which persists.
Also, we see a similar behavior in
all plots for small $\tau_X \la 10^2$ sec.
Here, the decays occur after weak freezeout but
before light element formation occurs, and
so for the most part the abundances are unaffected; the main
exception is perturbations in \he4 due to pion interactions
delaying $n \leftrightarrow p$ freezeout.

For several elements the basic trends at $\tau_X \ga 10^2$ sec are relatively
simple.  
We begin with \he4, which is the only species for which
decays always lead to reduced abundances.
This is physically reasonable, since \he4 is the most abundant
complex species, and the non-thermal reactions represent
sinks but not sources.  That is, photoerosion and spallation
destroy \he4, but $X$ decays cannot produce it; hence \he4
drops as $\zeta_X$ increases.
On the other hand,
we see that decays increase the D/H abundance,
which is readily understood physically. Destruction
of \he4 produces deuteron fragments, some of which
are thermalized before interacting and thus survive.  
Because \he4 is so abundant, even if only a small fraction
of \he4 is destroyed, the resulting deuteron production
can be significant.
For \li6/\li7, the basic trend is also towards increasing production
with increasing $\zeta_X$.  Here, not-yet-thermalized
mass-3 fragments can interact with ambient \he4 to
produce \li6 via $\he3(\alpha,p)\li6$ and $t(\alpha,n)\li6$.
Again, even if only a small fraction of $A=3$ nuclei interact this
way, this can represent a substantial \li6 abundance compared to
observational limits.  Moreover, while other secondary processes
also contribute to \li7, the much larger effect is for \li6, so 
that the \li6/\li7 ratio increases with $\zeta_X$.

By comparing the full constraints (Fig.~\ref{fig:zeta-tau-std})
with those in the electromagnetic-only case of Fig.~\ref{fig:zeta-tau-EMonly},
we see illustrated the well-known result that the 
electromagnetic decays dominate at $\tau_X \ga 10^6$ sec,
but are ineffective at smaller times \cite{cefos}.
Again for D/H, \li7/H, and \he3/D, we see that the hadronic
effects from neutrons and protons (Figs.~\ref{fig:zeta-tau-nonly}
and \ref{fig:zeta-tau-ponly}, respectively) are broadly similar,
though the neutron constraints are generally more restrictive
out to about $\tau_X \sim 10^6$ sec.  As we will see 
in detail below,
this is the timescale for neutron decay to outpace other neutron
interactions.

The cases of \li7 and \he3/D are more complicated.
Turning to \li7 in Fig.~\ref{fig:zeta-tau-std}, 
we see that at intermediate lifetimes ($\sim 10^4$ sec), decays lead to
higher \li7/H; this is due to secondary production from
unthermalized mass-3 spallation products,
$\he3(\alpha,\gamma)\be7$ and $t(\alpha,\gamma)\li7$.
However, at lifetimes around $\tau_X \sim 10^2-10^3$ sec,
there  a narrow ``valley'' emerges in which \li7/H is
{\em reduced}, indeed enough to come into agreement
with observational limits.
By comparing Figs.~\ref{fig:zeta-tau-nonly}
and \ref{fig:zeta-tau-ponly}, we see that this effect 
is entirely due to injected neutrons, and is absent 
when only proton decays are included.
This suggests that neutrons are destroying \be7, and
that is indeed the case.  
As noted already by Jedamizk and in subsequent work \cite{jed,kkm2},
thermalized neutrons can destroy \be7 via
$\be7(n,p)\li7$ and followed by $\li7(p,\alpha)\he4$.
We include this effect, but also, as we discuss below (Appendix~\ref{app:prop}), 
we allow for further \be7 destruction
in which the same reactions are initiated by injected neutrons not yet thermalized.
This addition slightly enhances the low-\li7/H ``valley.''
Notice that the left side of the ``valley" coincides almost exactly with the 
constraint from D/H, so that we find D/H $> 3.2 \times 10^{-5}$
in essentially the entire region where \li7/H $< 2.75 \times 10^{-10}$.
However, along the left side of the ``valley" in Fig.~\ref{fig:zeta-tau-std}, there may be a marginal
solution to the \li7 problem, to which we return later in the context of
the CMSSM.

Finally, turning to \he3/D at low lifetimes, we also see 
a reduction in the ratio for large $\zeta_X$,
with a particularly large effect in the neutron-only case.
As with \li7 destruction, here the free neutrons
preferentially capture on $t$ and \he3.  This leads to a small \he3/D ratio
as $\zeta_X$ increases. Specifically, we see that \he3/D $< 0.3$ in the
narrow strip in Figs.~\ref{fig:zeta-tau-std} and \ref{fig:zeta-tau-nonly} where the \li7 may (almost) be solved.

For a more quantitative understanding we now
apply the analytical model developed above.

\subsection{Non-Thermal Abundance Perturbations:
Analytical Model}

For the reaction
$hb \ra \ell$,
the rate of decrease of the
abundance of the ``target'' background
species $b$ is equal and opposite to the rate
of production of
background species $\ell$,
namely
\beq
-\pd_t Y_b = + \pd_t Y_\ell =
Y_b \Gamma_{hb \ra \ell} ,
\eeq
where the reaction rate per target $b$
is given in the thin- and thick-target limits
in (\ref{eq:rxnlite}).
The net change in these abundances due to this reaction is
the time integral of the rate:
\beq
-\Delta Y_b = + \Delta Y_\ell =
\int Y_b \Gamma_{hb \ra \ell} \ dt
 \approx Y_b \Gamma_{hb \ra \ell} \tau_X .
\eeq
In the last expression
we take the ``exposure time'' to be the $X$ lifetime
$\tau_X = 1/\Gamma_X$; this is the equivalent of the
instantaneous-decay approximation used in the EM case.
Since the non-thermal spectra, and thus non-thermal reaction rates,
scale as $N_h \sim \Gamma_{hb\ra \ell} \sim \Gamma_X$, the
lifetime dependence drops out in the abundance changes.

For the thin-target case, we have
\beqar
\label{eq:thin}
\left. \Delta Y_\ell^{\rm bg}  \right|_{hb \ra \ell}^{\rm thin}
  & = & Y_X
  \frac{Y_b^{\rm bg}}{Y_k^{\rm bg}} \int_{\nrg_{\rm th}} Q_h(\nrg) \
     \frac{\sigma_{hb \ra \ell}(\nrg)}{\sigma_{hk \ra \rm inel}} d\nrg \\
  & = & Y_X \frac{Y_b^{\rm bg}}{Y_k^{\rm bg}} Q_h(>\nrg_{\rm th})
     \left\langle 
         \frac{\sigma_{hb \ra \ell}(\nrg)}{\sigma_{hk \ra \rm inel}}
     \right\rangle ,
\eeqar
where the spectrum-weighted average is
\beq
\avg{\sigma_b/\sigma_k}
 = \frac{\int_{\nrg_{\rm th}} Q_h(\nrg) \sigma_b(\nrg)/\sigma_k(\nrg) \, d\nrg}
        {\int_{\nrg_{\rm th}} Q_h(\nrg) \, d\nrg} .
\eeq
We see that the abundance change is determined by several parameters.
The scaling with $Y_X$ is intuitively clear--the perturbation should
be proportional to the number of non-thermal decays per baryon.
Another important parameter is the number $Q_h(>\nrg_{\rm th}) \approx B_h$ of
non-thermal $h$ particles per decay above threshold, which
is to an excellent approximation the total number of $h$ particles
per $X$ decay, also an intuitive result.
Finally, the cross-section factor  is an analogue of a branching ratio
for the reaction in question relative to all inelastic reactions.

For the thick-target case, we have
\beqar
\label{eq:thick}
\left. \Delta Y_\ell  \right|_{hb \ra \ell}^{\rm thick}
  & = & Y_X Y_b^{\rm bg}  \int_{\nrg_{\rm th}} Q_h(> \nrg) \
      \frac{\sigma_{hb \ra \ell}/m_p}{d\nrg_h/dR} d\nrg \\
  & = & Y_X Y_b^{\rm bg}  Q_h(> \nrg_{\rm th}) \
      \avg{R} \avg{\sigma_{hb \ra \ell}/m_p}  \\
  & \approx & 6 \times 10^{-4} \ Y_X Y_b^{\rm bg}  \
      \pfrac{B_h}{10^{-2}} \
      \pfrac{\avg{R}}{1 \ \rm g \, cm^{-2}} \
      \pfrac{\avg{\sigma}}{30 \ {\rm mbarn}} ,
\eeqar
where we again define weighted averages of the cross section $\avg{\sigma}$
and stopping range $\avg{R}$ (with units $[\rm g/cm^2]$) as
\beqar
\avg{\sigma}
  & = & \frac{\int_{\nrg_{\rm th}} Q_h(> \nrg) \ \sigma_{hb \ra \ell}  \
                (d\nrg_h/dR)^{-1} d\nrg }
        {\int_{\nrg_{\rm th}} Q_h(> \nrg) \ (d\nrg_h/dR)^{-1} d\nrg } ,\\
\avg{R}
  & = & \frac{\int_{\nrg_{\rm th}} Q_h(> \nrg) \ (d\nrg_h/dR)^{-1} d\nrg }
        {Q_h(> \nrg_{\rm th})} .
\eeqar
Recall that $\langle R \rangle \propto Y_{e^\pm}^{-1}$ and the fiducial value we 
choose is appropriate for $Y^{-1}_{e^\pm} \sim 1$.
We see that the abundance perturbations
for hadronic decays scale with--and thus constrain--the
{\em number} of pre-decay $X$ particles (gravitinos, in the present
study) per baryon (or equivalently per photon or per unit entropy).
However, to show hadronic as well as electromagnetic results
on the same plot, it is useful to introduce
$\zeta_X = m_X Y_X \eta$ as in eq.~\pref{eq:zeta},
so that $Y_X = \zeta_X/\eta m_X$.
We now apply the observed abundance
constraints and express them in terms of $\zeta_X$.

Consider some light-element
abundance constraint $Y_\ell^{\rm lim}$.
The perturbation saturates this constraint when
$\Delta Y_\ell = \delta Y_\ell^{\rm obs} 
\equiv  Y_\ell^{\rm lim}-Y_{\ell}^{\rm std}$,
where $Y_{\ell}^{\rm std}$ is the standard BBN result (see e.g. \cite{cfo5}).
This abundance will be reached for
some value of $Y_X$ and thus $\zeta$.
Namely, we have
\beqar
\label{eq:thinpert}
\zeta_{X,{\rm lim}}^{\rm thin}
  & \simeq & m_X \eta \delta Y_\ell^{\rm obs} \frac{Y_k^{\rm bg}}{Y_b^{\rm bg}}
     B_h^{-1}
     \left\langle 
          \frac{\sigma_{hb \ra \ell}}{\sigma_{hk \ra \rm inel}}
     \right\rangle^{-1}  \\
  & = &  10^{-11} \ {\rm GeV}
   \ \pfrac{\delta Y_\ell^{\rm obs}}{0.4 \times 10^{-5}}
   \ \pfrac{m_X}{500 \ \rm GeV}
   \ \pfrac{0.2}{B_h}
   \ \pfrac{Y_k^{\rm bg}}{Y_b^{\rm bg}}
   \ \pfrac{0.5}{\avg{\sigma_{hb \ra \ell}/\sigma_{hk \ra \rm inel}}} ,
\nonumber
\eeqar
where the fiducial values are appropriate for constraints based
on D/H, produced mostly via \he4 spallation.
Here we take the total inelastic interactions
$hk \rightarrow {\rm inel}$
to be dominated by  interactions with background \he4, so
that $Y_k^{\rm bg}/Y_b^{\rm bg} = 1$;
this is the case for nucleons up to $\nrg \sim 10$ GeV,
and beyond this the smaller $NN$ inelastic cross sections
change the result by no more than about a factor of 2.

We compare our analytical estimates to numerical results for
the production of deuterium obtained using our codes,
as shown in 
Figs.~\ref{fig:zeta-tau-std}--\ref{fig:zeta-tau-ponly}.
We see that for short lifetimes,
our prediction from eq.~\pref{eq:thinpert}
agrees with the D/H abundance constraints for
$Y_{\rm D} = 3.2 \times 10^{-5}$
to within a factor of 3.
In particular, our thin-target estimate is a good
description of the neutron-only results (Fig.~\ref{fig:zeta-tau-nonly}).
This was to be expected as neutron propagation is, for these
$X$ lifetimes, well within the thin-target regime.

For the thick-target case, putting $\Delta Y_\ell = \delta Y_\ell^{\rm obs}$
gives
\beqar
\zeta_{X,{\rm lim}}^{\rm thick}
  & \simeq & m_X \eta 
     \frac{\delta Y_\ell^{\rm obs}}{Y_b^{\rm bg}} B_h^{-1} \
      \frac{m_p/\avg{\sigma_{hb \ra \ell}}}{\avg{R}} \\
  & = &   10^{-9} \ {\rm GeV}
   \  \\
   & \times &
   \ \pfrac{\delta Y_\ell^{\rm obs}}{0.4 \times 10^{-5}}
   \ \pfrac{m_X}{500 \ \rm GeV}
   \ \pfrac{0.2}{B_h}
   \ \pfrac{0.1}{Y_b^{\rm bg}}
   \ \pfrac{100 \ \rm mb}{\langle \sigma_{hb \ra \ell} \rangle}
   \ \pfrac{1 \ \rm g \, cm^{-2}}{\langle R \rangle} , \nonumber
  \eeqar
where again we use fiducial values that are appropriate for constraints based
on D/H, produced mostly via \he4 spallation and $Y_{e^\pm}=1$.
We see that the thick-target limits are about a factor $\sim 100$
weaker than the thin-target limits.
This is reasonable, as the thick-target limit
is less efficient in light-element production.
Again comparing Figs.~\ref{fig:zeta-tau-std}-\ref{fig:zeta-tau-ponly},
we see that the proton-only results (Fig.~\ref{fig:zeta-tau-ponly})
lie between the thin- and thick-target estimates.
This is also as expected: for the $X$ lifetimes
of interest, the proton propagation is partially
within the thick-target regime.

Given our success in understanding of D/H, 
it is worth extending the analysis to \li6, which
should be due to secondary production of non-thermalized
mass-3 nuclides via $\he3(\alpha,p)\li6$
and $t(\alpha,n)\li6$.
If this is the case, then 
the \li6 abundance should be given by the product
of the non-thermal mass-3 abundance $\Delta Y_3$
and the probability that the mass-3 nuclide interacts
before it stops.
But for the thin-target case, the probability is small,
and is just given by the product of the mass-3 interaction rate
$\Gamma_{3\alpha} = Y_\alpha^{\rm bg}  n_{\rm B} \sigma_{3\alpha} v$ 
times
its stopping time $\tau_{3,\rm stop} 
= (\int b^{-1} \; d\nrg) \approx (m_p n_{\rm B} v/R_3)^{-1}$,
with $R_3$ the mass stopping grammage.
Thus we have
\beq
\Delta Y(\li6) 
  = 
  \Delta Y_3 \ Y_\alpha^{\rm bg} R_3 \frac{\sigma({{}^3A \alpha \rightarrow N\li6})}{m_p} ,
\eeq
with $Y_\alpha^{\rm bg} \sim 0.1$.
Using the above thin-target expression for $\Delta Y_d$ (eq.~\ref{eq:thin}),
and taking $(\he3/{\rm D})_{\rm spall} \sim 1$ for
spallation production,
this becomes
\beqar
\Delta Y(\li6) 
 & \approx &
\frac{\zeta_X}{\eta m_X}  B_h
     \left\langle 
         \frac{\sigma_{hb \ra \ell}}{\sigma_{hk \ra \rm inel}}
     \right\rangle
\ Y_\alpha^{\rm bg}
    \ R_3 \frac{\sigma({{}^3A \alpha \rightarrow N \li6})}{m_p} .
\eeqar
Assuming the \li7 perturbation is small, so that 
$Y(\li7) \approx Y(\li7)_{\rm BBN,std} \sim 3 \times 10^{-10}$,
we solve for $\zeta_X$ in terms of the observed limit
on $\li6/\li7 = Y(\li6)/Y(\li7)$:
\beqar
\label{eq:67oom}
\zeta_{X,{\rm lim}}^{\rm thin}(\li6)
  & \simeq &  \eta m_X  (\li6/\li7)_{\rm obs} 
  \frac{Y(\li7)}{Y_\alpha^{\rm bg}}   B_h^{-1}
     \left\langle 
         \frac{\sigma_{hb \ra \ell}}{\sigma_{hk \ra \rm inel}}
     \right\rangle^{-1} 
    \ \frac{m_p}{R_3 \sigma({{}^3A \alpha \rightarrow N \li6})}  \\
  & \sim &  2.5 \times 10^{-14} \ {\rm GeV}
   \ \pfrac{\left.\li6/\li7\right|_{\rm obs}}{0.05}
   \ \pfrac{1 \ \rm g/cm^2}{R_3}
   \ \pfrac{30 \, \rm mb}{\sigma({}^3A \alpha \ra N \li6)} ,
\eeqar
where the fiducial values not shown are the same as in eq.~(\ref{eq:thinpert}).
This result within a factor of 3 the strongest \li6/\li7
constraints around $\tau_X \sim 10^4$ sec,
where the neutrons have time to interact before they decay.

The agreement between our order-of-magnitude estimates
and full numerical results serves as a strong
check on our analysis and gives us confidence in
both approaches.

%%%%%  the following command is a workaround for the dreaded 
%%%%%  ``too many unprocessed floats'' error message
%\clearpage

\subsection{Constraints on the Abundance of a Massive Gravitino}

We now apply the above analysis to the specific
class of supersymmetric scenarios with a
metastable massive gravitino and a neutralino LSP $\chi$.

We recall that neutralino LSP scenarios are characterized by narrow strips
along which the relic $\chi$ density lies within the range $0.0975 < \Omega_{\rm CDM} h^2 < 0.1223$
that is favored by WMAP and other
observations~\cite{wmap}.
Since this range is
relatively narrow (namely only a few~$\%$), within any
specific supersymmetric scenario it determines some combination of the model
parameters also to within ${\cal O}$(few)~$\%$. A more detailed discussion
is given in~\cite{bench}, where some explicit parameterizations of such WMAP strips
are given. In the examples given there, the value of $m_0$ is tightly determined in
CMSSM scenarios as a function of $m_{1/2}$ for fixed values of $\tan \beta$ and $A_0$.
Many properties of these supersymmetric models change little as one varies $m_0$
across such a narrow WMAP strip. Specifically, the branching ratios for massive gravitino decay
and hence $B_h$ vary little across a strip, and one may usefully represent the
cosmological constraints on such CMSSM scenarios as functions of $m_{1/2}$
alone, using a representative value of $B_h$ that is calculated as a function of $m_{1/2}$.

\begin{figure}[ht!]
\begin{center}
\epsfig{file=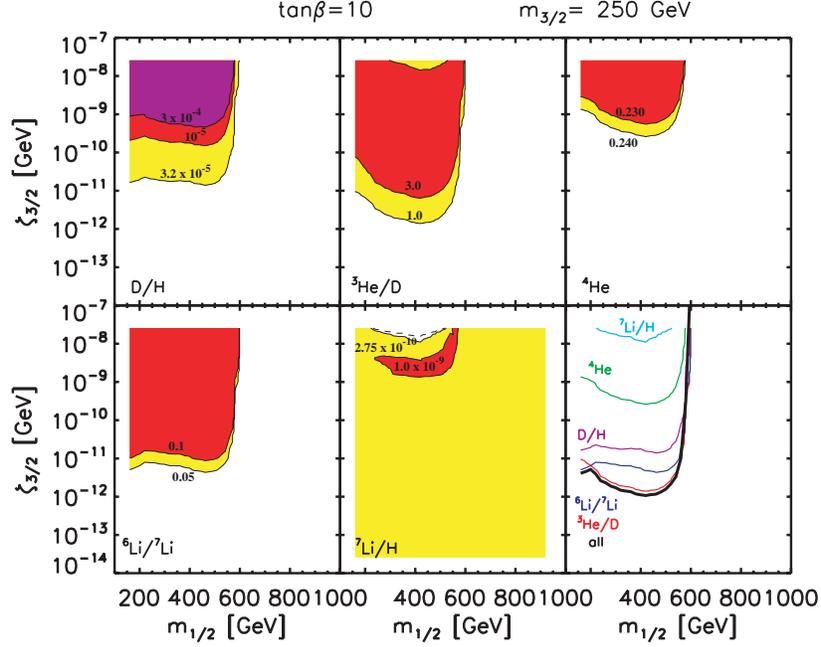,width=0.65\textwidth}
\caption{\it
The effects of the decays of a gravitino with a mass $m_{3/2} = 250$~GeV
on the different light-element abundances (D/H, \he3/D, \he4, \li7/H and
\li6/\li7) as a function of $m_{1/2}$ along the WMAP coannihilation strip for a CMSSM scenario
with $\tan \beta = 10, A_0 = 0$. As in Figs.~\ref{fig:zeta-tau-std},
the white regions in each panel are those allowed
at face value by the light-element abundances
reviewed in Section~\ref{sect:abs}, and the yellow, red, and magenta regions correspond to
progressively larger deviations from the central values of the abundances.
}
\label{fig:elements250tb10}
\end{center}
\vspace{-1em}
\end{figure}

One representative example is shown in Fig.~\ref{fig:elements250tb10}: the 
first five panels, with shadings,
display the effects of the decays of a gravitino with a mass $m_{3/2} = 250$~GeV
on the different light-element abundances (D/H, \he3/D, \he4, 
\li6/\li7 and \li7/H) as functions of $m_{1/2}$ along the WMAP strip 
in the coannihilation region for a CMSSM scenario
with $\tan \beta = 10, A_0 = 0$. 
As in Fig.~\ref{fig:zeta-tau-std} - \ref{fig:zeta-tau-ponly}, 
the white regions in each panel are those allowed
at face value by the ranges 
of the light-element abundances
reviewed in Section~\ref{sect:abs}, and the yellow, red, and magenta regions correspond to
progressively larger deviations from the central values of the abundances. 
The final panel (bottom right), shows the strongest constraints
from each abundance, as labeled, and from these one infers the
strongest overall constraint, shown by the thick black curve. 

%%%%%%%%%%%%%%%%%%%%%%%%% F I G U R E %%%%%%%%%%%%%%%%%%%%%%%%%%%%%%%%%%%%%%%%%
%%%%%%%%%%%%%%%%%%%%%%%%% F I G U R E %%%%%%%%%%%%%%%%%%%%%%%%%%%%%%%%%%%%%%%%%

We first 
note that all constraints weaken abruptly as 
$m_{1/2} \rightarrow 600$ GeV.
This
corresponds to the limiting case when $m_\chi \sim 0.42 m_{1/2} \to m_{3/2} = 250$~GeV.
In this limit, the energies in the EM and HD showers vanish, and the effects of
gravitino decay disappear. 
Note that $m_{1/2} = 250$ GeV is the smallest
value we consider; for larger values this effect is less important. 

Consider now the D/H constraints in the region $m_{1/2} \la 600$ GeV
in the top left panel of Fig.~\ref{fig:elements250tb10}.
We see that the lowest contour, at $\rm D/H = 3.2 \times 10^{-5}$,
lies at about $\zeta_{3/2} \simeq 10^{-11} \gev$ between $m_{\rm 1/2} = 180$ GeV and just under
600 GeV~\footnote{We recall that the gravitino becomes the LSP
for $m_{1/2} \ga 600$ GeV, in which case a different analysis is necessary.}.
To understand this behavior, we recall from 
Fig.~\ref{fig:lifetimes} that for $m_{1/2} = 180-580$ GeV,
the gravitino lifetime grows from about $10^7$ sec to $10^{10}$ sec,
and in Fig.~\ref{fig:branching} we see that the
nucleon branching ratios $B_p \approx B_n \sim 6 \times 10^{-2}$
for $m_{1/2} \la 400$ GeV.  
Combining these with the generic lifetime dependence in
Fig.~\ref{fig:zeta-tau-std}, we see that for $B_p \approx B_n \sim 2 \times 10^{-1}$,
the lowest D/H contour for $\tau \ga 10^7$ sec is roughly constant
at $\zeta_{3/2} \sim 3 \times 10^{-12} \gev$.
In our case, these constraints weaken due to the lower branching ratios
by a factor $\sim 3$, yielding the value $\zeta_{3/2} \sim 10^{-11} \gev$ seen 
in Fig.~\ref{fig:elements250tb10}.
The higher D/H contours 
in Fig.~\ref{fig:elements250tb10} can be understood in a similar manner.

As another example, consider the \li7/H panel
in the bottom middle panel of Fig.~\ref{fig:elements250tb10}.
There we see, for $m_{1/2} \la 600$ GeV, an ``island'' of the lowest
constraint where $\zeta_{3/2} \sim 10^{-10}-10^{-8} \gev$.
As $\zeta_{3/2}$ drops, \li7/H then grows, peaking to
highest level where $\zeta_{3/2} \sim 3 \times 10^{-9} \gev$.
Then as $\zeta_{3/2}$ continues to decrease,
\li7/H continues drops down to its standard BBN level.
From the generic plot Fig.~\ref{fig:zeta-tau-std}, we see that for $\tau \ga 10^{7}$ sec
and with $B_p \sim 0.2$, the \li7/H constraint forms an island
around $\zeta_{3/2} \sim 10^{-9} \gev$, with a width that grows with $\tau$.
Scaling down to the lower branching, we again can understand
the behavior in Fig.~\ref{fig:elements250tb10}.

The fact that this qualitative analysis works so well indicates that
the generic results are a good approximation to the full ones,
modulo changes in branching ratios.  However, the generic results are for
a particular decay spectrum whereas, as we have seen, there
are significant variations in the decay spectrum.
Thus we infer that our results are not strongly dependent on the detailed 
shapes of the decay spectra beyond the sensitivity to 
the branching ratios.
This also makes sense in terms of our analytic approximations, which
suggest that the decay spectra enter principally via their integral properties
and particularly their branching ratios.

This understanding of the non-thermal particle effects leading
to our constraints gives confidence in our results, 
and thus to their implications for supersymmetry.
Namely, again in the case of Fig.~\ref{fig:elements250tb10}
where $\tan \beta = 10$ and $m_{3/2} = 250$ GeV,
the fact that 
the observed abundances generally agree with the standard BBN
calculations implies that the allowed (white) regions are generally at low gravitino
abundance. The exception is the bottom middle panel, where the observational
discrepancy with the standard BBN calculation of the
\li7 abundance is reflected in the fact that the yellow region extends
from a vanishing gravitino abundance up to quite large values, and the ``preferred" white region
appears only when $\zeta_{3/2} > 10^{-9}$~GeV
and $m_{1/2}$ is not too large~\footnote{We also note
the existence of a more disfavored (red) region in the bottom middle panel
showing  the \li7/H ratio, appearing when $\zeta_{3/2} > 10^{-9}$~GeV
and $m_{1/2} \sim  500$~GeV.}. 
It is clear
that this ``preferred" region for \li7/H is incompatible with the allowed regions for
the other light-element abundances shown in the other panels.
Thus, there is no value of $m_{1/2}$ in this
particular CMSSM scenario that solves the \li7/H problem. As already mentioned,
we disregard the
\li7/H problem in the compilation of constraints shown in the bottom right
panel of Fig.~\ref{fig:elements250tb10}, and in the following discussion. 

The weakening of the constraints as $m_{1/2} \to 600$~GeV
implies that a large abundance is
allowed for $m_{3/2} = 250$~GeV if $m_{1/2} \sim 600$~GeV,
apart from the issue of the \li7 abundance. We recall that if $m_{1/2} > 600$~GeV with fixed
$m_{3/2} = 250$~GeV, the gravitino becomes the LSP and the lightest
neutralino becomes the NLSP. In this case, the constraint due to the
cosmological relic density should be applied to the gravitino, the WMAP
strip is no longer relevant, and a different analysis would be required.

In the final, summary panel of Fig.~\ref{fig:elements250tb10} (bottom right),
we see explicitly that the \li7 constraint (light blue) is incompatible with the
other constraints due to \he4 (green), D/H (magenta), \he3/D (red) and
\li6/\li7 (dark blue).
The weakest of these constraints is that due to \he4, the two strongest
constraints at smaller and larger $m_{1/2}$, respectively, are those due to \li6/\li7
and \he3/D, and the combined constraint is shown in black. We see
that it hovers in the range $\zeta_{3/2} = m_{3/2} n_{3/2}/n_\gamma \sim 3 \times 10^{-12}$ 
to $10^{-12}$~GeV,
corresponding to $n_{3/2}/n_\gamma \sim 10^{-14}$, except in the limit as
$m_{1/2} \to 600$~GeV.

\begin{figure}[th!]
\begin{center}
\epsfig{file=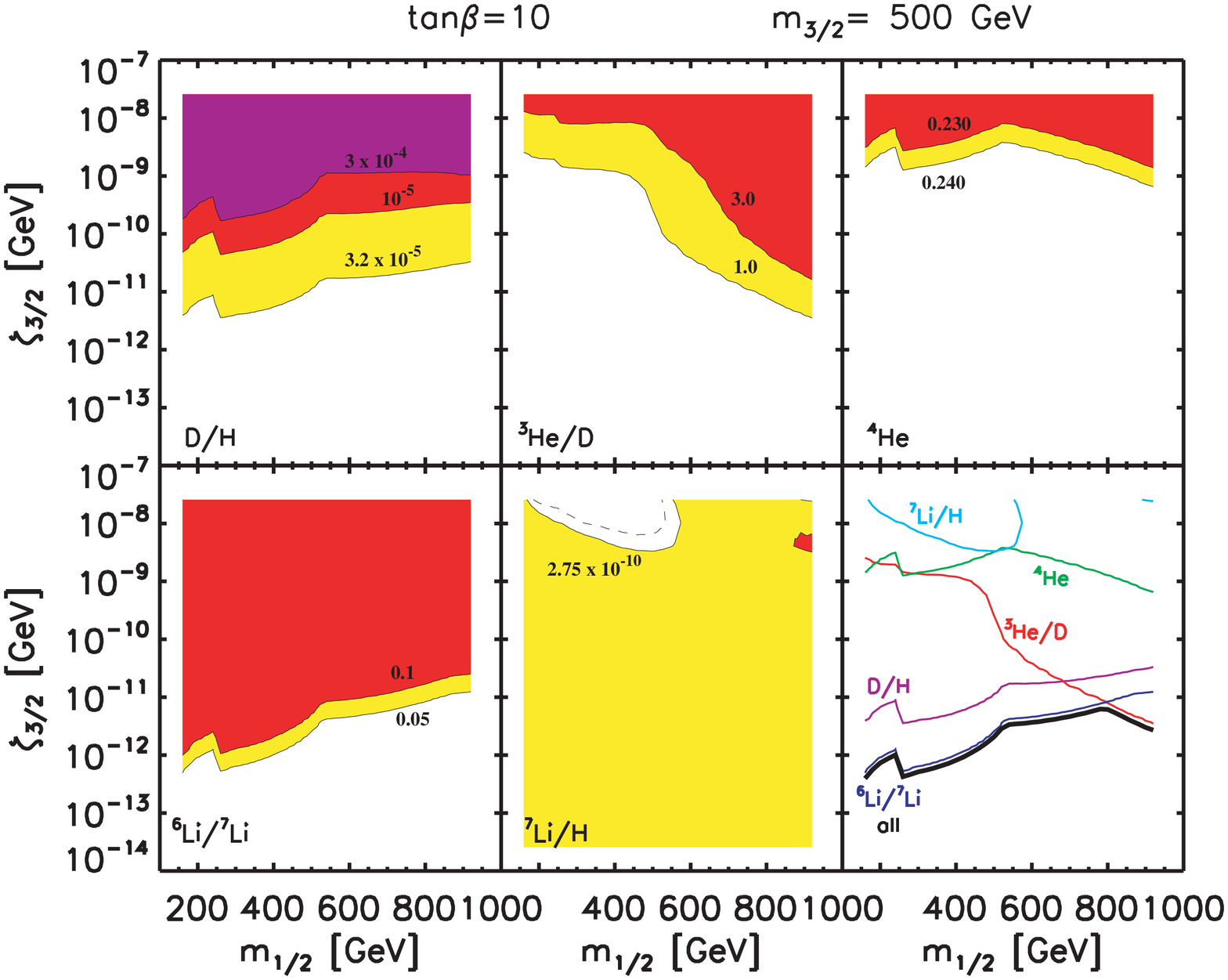,width=0.65\textwidth}
\caption{\it
As for Fig.~\protect\ref{fig:elements250tb10}, with $m_{3/2} = 500$~GeV
but unchanged values for the CMSSM parameters.
}
\label{fig:elements500tb10}
\end{center}
\vspace{-1em}
\end{figure}

\begin{figure}[tbh!]
\begin{center}
\epsfig{file=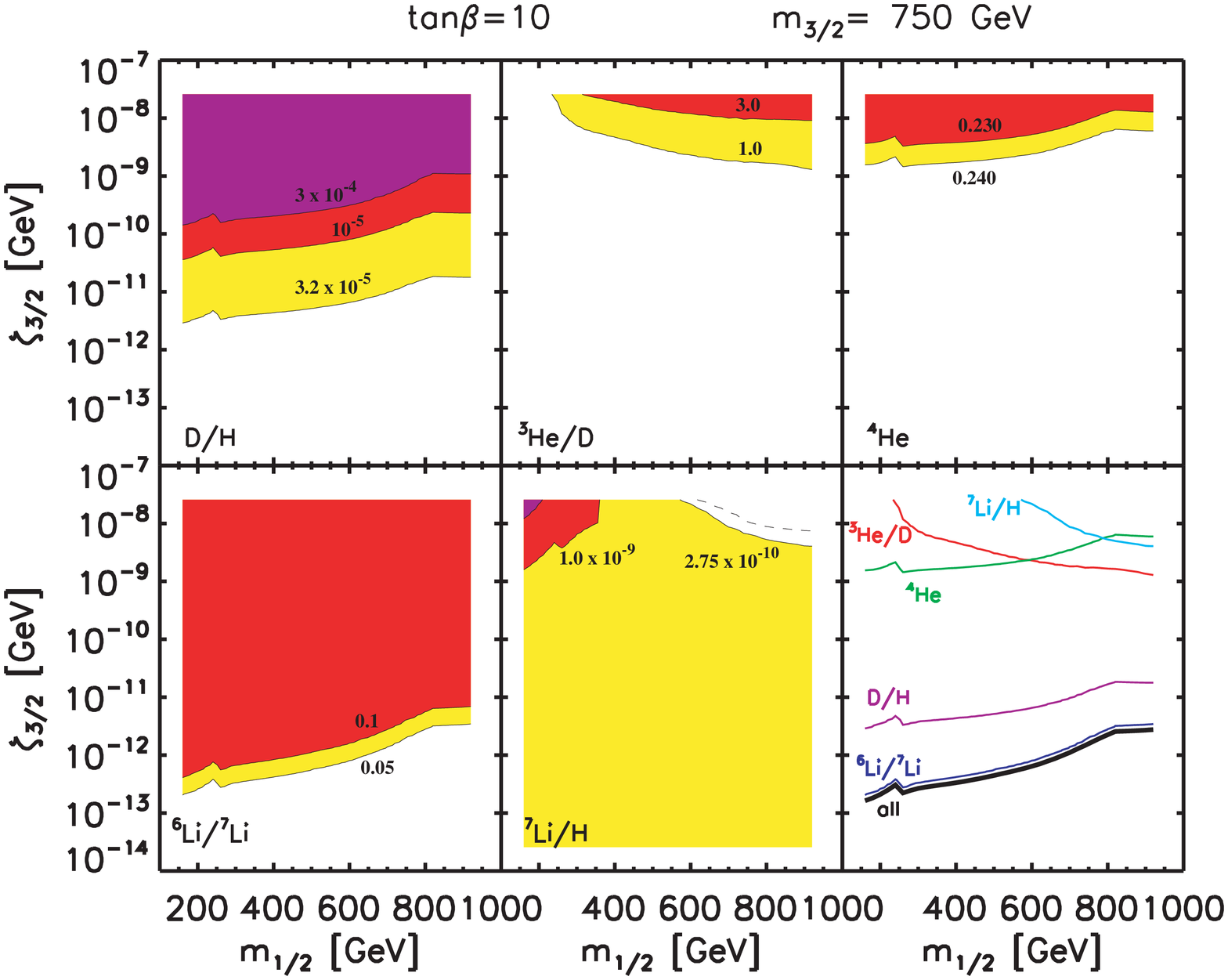,width=0.65\textwidth}
\caption{\it
As for Fig.~\protect\ref{fig:elements250tb10}, with $m_{3/2} = 750$~GeV
but unchanged values for the CMSSM parameters.
}
\label{fig:elements750tb10}
\end{center}
\vspace{-1em}
\end{figure}

\begin{figure}[tbh!]
\begin{center}
\epsfig{file=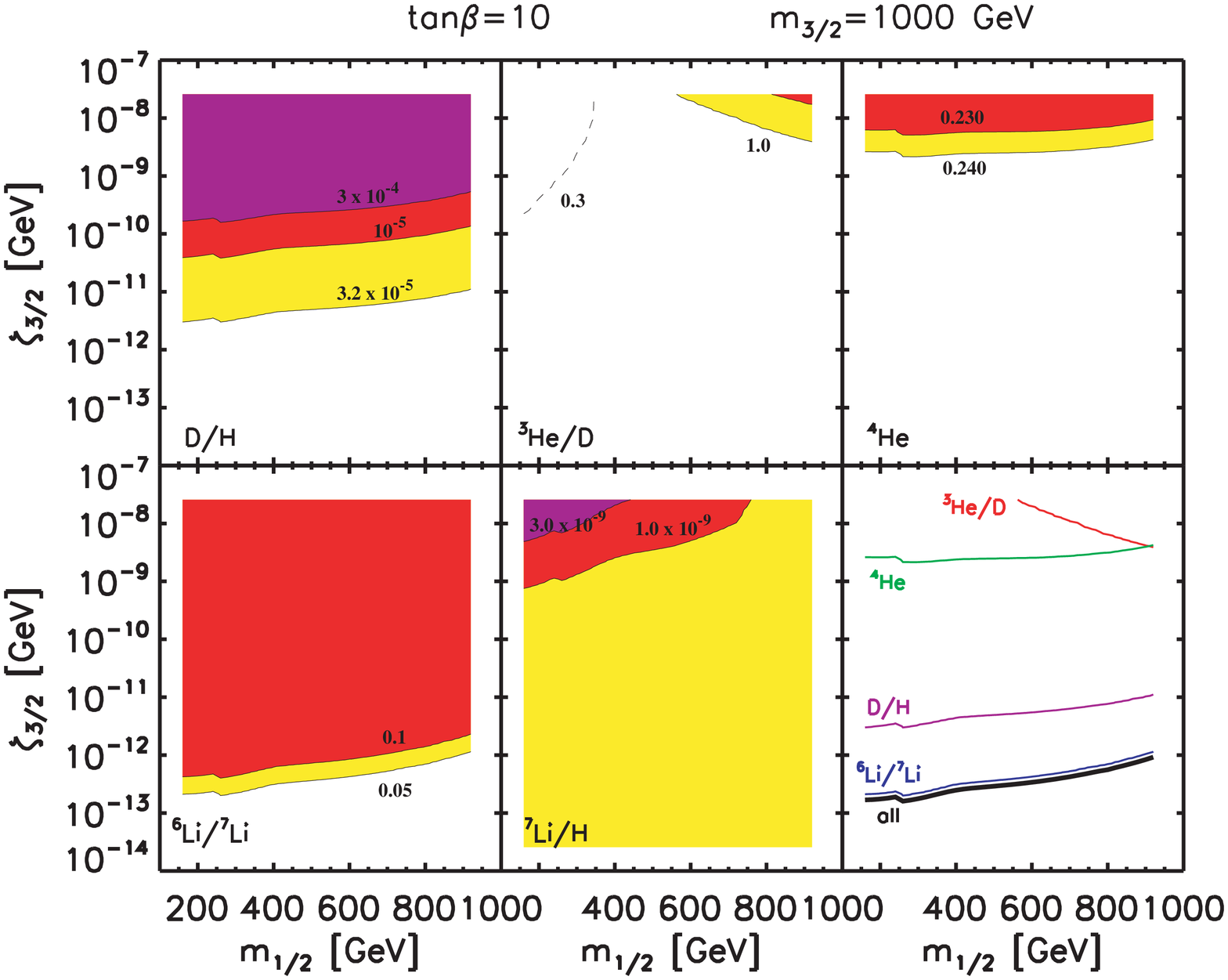,width=0.65\textwidth}
\caption{\it
As for Fig.~\protect\ref{fig:elements250tb10}, with $m_{3/2} = 1000$~GeV
but unchanged values for the CMSSM parameters.
}
\label{fig:elements1000tb10}
\end{center}
\vspace{-1em}
\end{figure}

\begin{figure}[tbh!]
\begin{center}
\epsfig{file=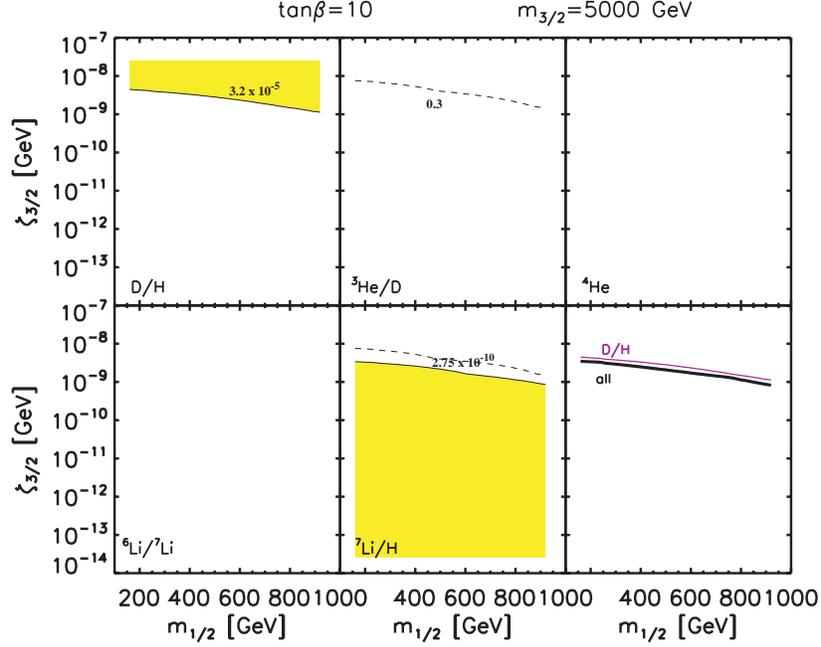,width=0.65\textwidth}
\caption{\it
As for Fig.~\protect\ref{fig:elements250tb10}, with $m_{3/2} = 5000$~GeV
but unchanged values for the CMSSM parameters.
}
\label{fig:elements5000tb10}
\end{center}
\vspace{-1em}
\end{figure}

Figs.~\ref{fig:elements500tb10}, \ref{fig:elements750tb10}, 
\ref{fig:elements1000tb10}, and
 \ref{fig:elements5000tb10} show the correspondingly coloured regions of
 varying discomfort for the different light-element abundances for the larger values
 of $m_{3/2} = 500$~GeV, $750$~GeV, $1000$~GeV
 and $5000$~GeV, respectively.
 The constraints are shown for the full length of the WMAP strip up to $m_{1/2}
 \sim 900$~GeV, along which $m_\chi < m_{3/2}$ for all the displayed values of $m_{3/2}$.
For $m_{3/2} \le 1000$ GeV,
we see that the D/H constraint (top left panel in each plot) and the \he4 constraint
 (top right in each plot) are relatively stable,
reflecting the rough constancy of the D/H results for
long lifetimes (see Fig.~\ref{fig:zeta-tau-std}),
and the rough constancies of the branching ratios
(see Fig.~\ref{fig:branching}).
For $m_{3/2} = 5000$~GeV, the short lifetimes
(see Fig.~\ref{fig:lifetimes})
severely weaken the constraints for all elements
(cf.~Fig.~\ref{fig:zeta-tau-std}), with D/H surviving as the strongest for all $m_{1/2}$.
We discuss later the case of large $m_{3/2}$.

%%%%%%%%%%%%%%%%%%%%%%%%% F I G U R E %%%%%%%%%%%%%%%%%%%%%%%%%%%%%%%%%%%%%%%%%
%%%%%%%%%%%%%%%%%%%%%%%%% F I G U R E %%%%%%%%%%%%%%%%%%%%%%%%%%%%%%%%%%%%%%%%%

%%%%%%%%%%%%%%%%%%%%%%%%% F I G U R E %%%%%%%%%%%%%%%%%%%%%%%%%%%%%%%%%%%%%%%%%
%%%%%%%%%%%%%%%%%%%%%%%%% F I G U R E %%%%%%%%%%%%%%%%%%%%%%%%%%%%%%%%%%%%%%%%%

%%%%%%%%%%%%%%%%%%%%%%%%% F I G U R E %%%%%%%%%%%%%%%%%%%%%%%%%%%%%%%%%%%%%%%%%
%%%%%%%%%%%%%%%%%%%%%%%%% F I G U R E %%%%%%%%%%%%%%%%%%%%%%%%%%%%%%%%%%%%%%%%%

%%%%%%%%%%%%%%%%%%%%%%%%% F I G U R E %%%%%%%%%%%%%%%%%%%%%%%%%%%%%%%%%%%%%%%%%
%%%%%%%%%%%%%%%%%%%%%%%%% F I G U R E %%%%%%%%%%%%%%%%%%%%%%%%%%%%%%%%%%%%%%%%%

In the case of \li7/H
 (bottom middle panels), the ``preferred" white region changes position as $m_{3/2}$
 increases, moving to larger $m_{1/2}$ for $m_{3/2} = 500, 750$~GeV, but reverting
 to low $m_{1/2}$ at high $m_{3/2}$. 
The $m_{1/2}$ mass ranges are those which have lifetimes
near $\sim 3\times 10^{6} \ \rm sec$
(see Fig.~\ref{fig:zeta-tau-std}), whereas the \li7 constraint is
strongest for long lifetimes.  However, there is no overlap
 between the white regions in this and the other panels, implying that the \li7
 problem cannot be solved for any value of $m_{3/2}$ for the particular values of
 $\tan \beta$ and $A_0$ chosen here. As stated previously, we do not include
 the \li7 constraint in our compilation.
As \li7 is problematic in standard BBN, the inclusion of this constraint
would give the false impression that nearly every supersymmetric model
with a decaying gravitino is excluded.  We are therefore implicitly assuming
that there is another solution for the \li7 problem,
e.g., due to observational or astrophysical uncertainties as discussed in
Section~\ref{sect:abs}.  In contrast, the constraints from the
other light elements perturb the previous concordance of BBN with respect to those elements.

Finally, we note that 
for $m_{3/2} = 500$ to 1000 GeV, the \li6/\li7 constraint
is significantly stronger than for $m_{3/2} = 250$ GeV,
becoming progressively stricter.
This reflects the stronger limits arising when the gravitino
lifetime is  shorter (see Fig.~\ref{fig:zeta-tau-std}).
In all cases the \li6/\li7 ratio  provides the most
restrictive limit on $\zeta_{3/2}$, and
strengthens by a factor $\sim 10$
as $m_{1/2}$ increases from 250 to 1000~GeV.
Specifically, we find $\zeta_{3/2} \la 10^{-12}-10^{-11} \gev$
for $m_{3/2} = 500$ GeV,
$\zeta_{3/2} \la 3 \times 10^{-13}-3\times 10^{-12} \gev$
for $m_{3/2} = 750$ GeV, and
$\zeta_{3/2} \la 10^{-13}-10^{-12} \gev$
for $m_{3/2} = 1000$ GeV.

In the case of Fig.~\ref{fig:elements5000tb10}, we see that the most
significant upper limit on the gravitino abundance is that from D/H.
Comparing this with the {\it lower} limit  on the gravitino abundance coming from \li7/H,
we see that they are marginally compatible over essentially the full range of $m_{1/2}$
displayed. However, this conclusion is crucially dependent on the precise
implementations of the D/H and \li7/H constraints: for example, if the upper limit
on the \li7/H abundance is strengthened to $1.91 \times 10^{-10}$, as suggested by
field stars and indicated by the dashed line in the middle lower panel of 
Fig.~\ref{fig:elements5000tb10}, compatibility becomes more difficult.

\begin{figure}[th!]
\begin{center}
\epsfig{file=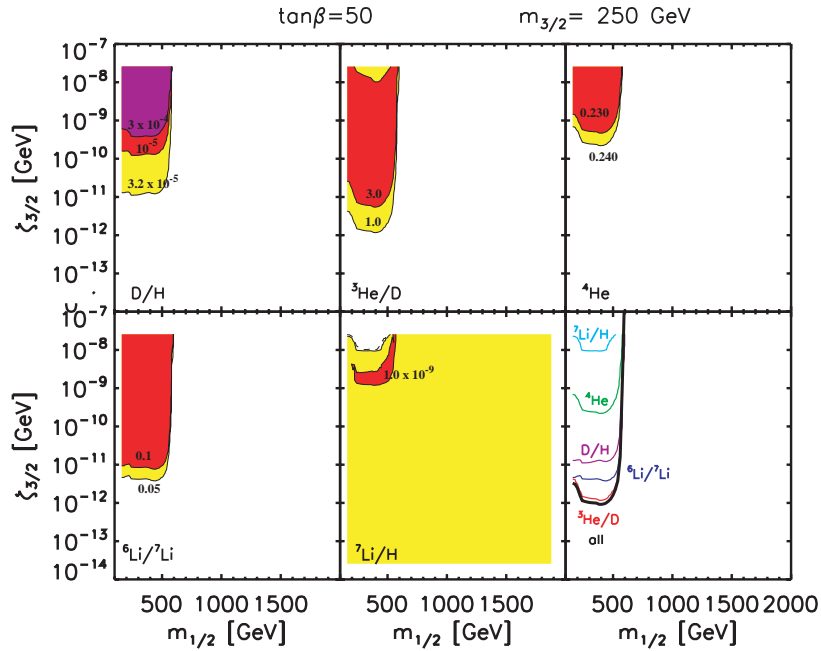,width=0.65\textwidth}
\end{center}
\caption{\it
As for Fig.~\protect\ref{fig:elements250tb10}, with $m_{3/2} = 250$~GeV
but $\tan \beta = 50$ and
unchanged values for the CMSSM parameters.
}
\label{fig:elements250tb50}
\vspace{-1em}
\end{figure}

\begin{figure}[tbh!]
\begin{center}
\epsfig{file=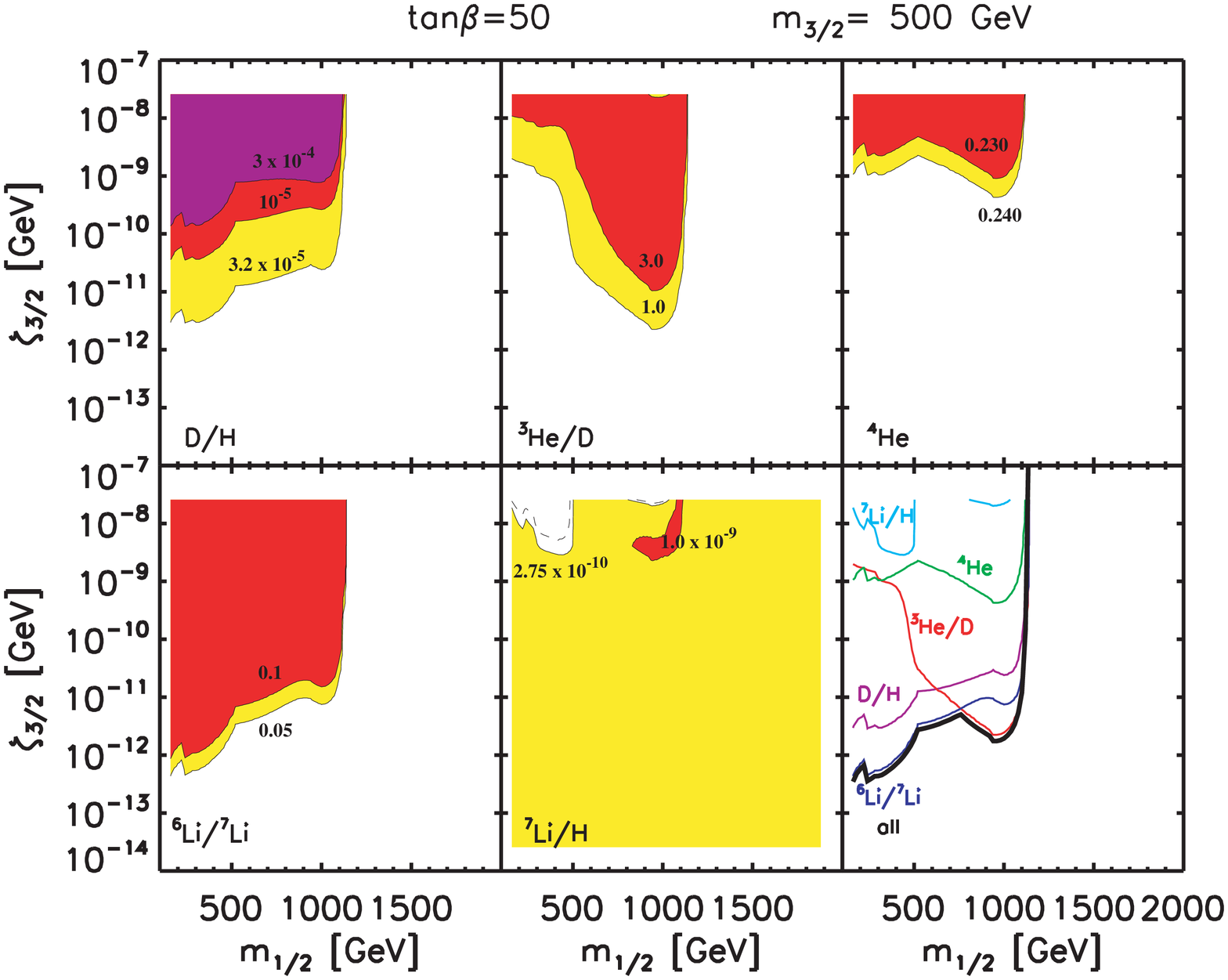,width=0.65\textwidth}
\end{center}
\caption{\it
As for Fig.~\protect\ref{fig:elements250tb10}, with 
$m_{3/2} = 500$~GeV, but $\tan \beta = 50$ and
unchanged values for the CMSSM parameters.
}
\label{fig:elements500tb50}
\vspace{-1em}
\end{figure}

\begin{figure}[tbh!]
\begin{center}
\epsfig{file=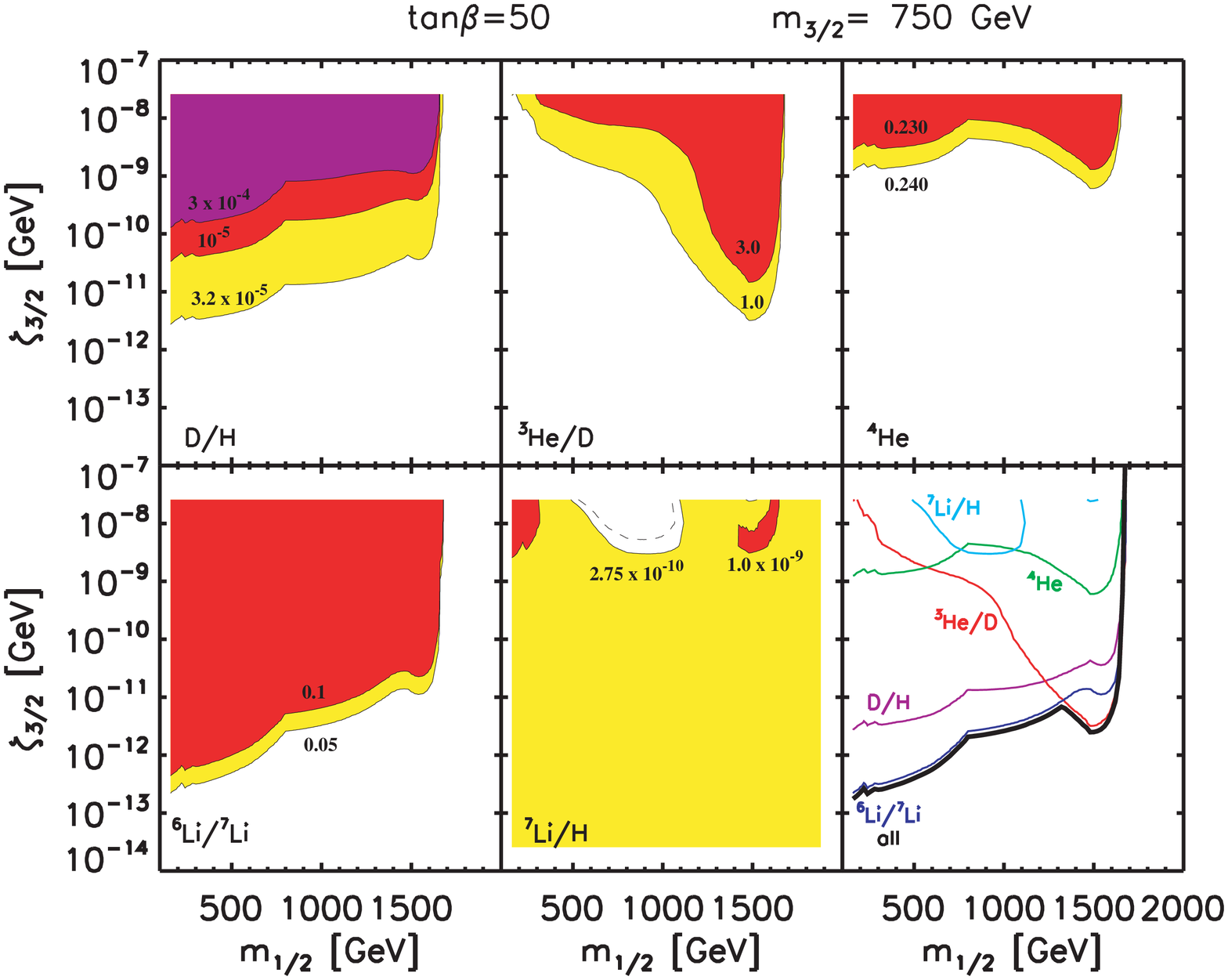,width=0.65\textwidth}
\end{center}
\caption{\it
As for Fig.~\protect\ref{fig:elements250tb10}, with $m_{3/2} = 750$~GeV, 
but $\tan \beta = 50$ and
unchanged values for the CMSSM parameters.
}
\label{fig:elements750tb50}
\vspace{-1em}
\end{figure}

\begin{figure}[tbh!]
\begin{center}
\epsfig{file=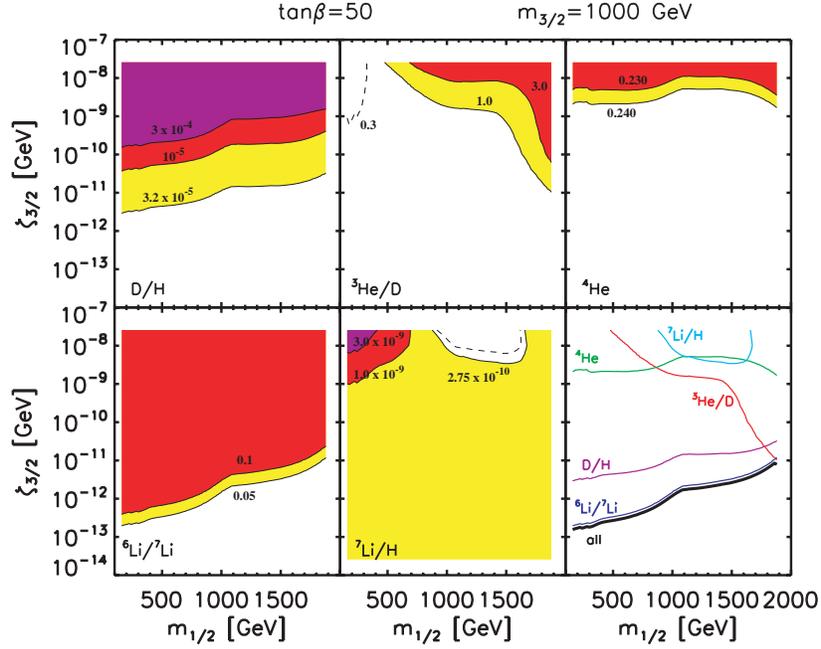,width=0.65\textwidth}
\end{center}
\caption{\it
As for Fig.~\protect\ref{fig:elements250tb10}, with (top left) 
$m_{3/2} = 1000$~GeV, but $\tan \beta = 50$ and
unchanged values for the CMSSM parameters.
}
\label{fig:elements1000tb50}
\vspace{-1em}
\end{figure}

\begin{figure}[tbh!]
\begin{center}
\epsfig{file=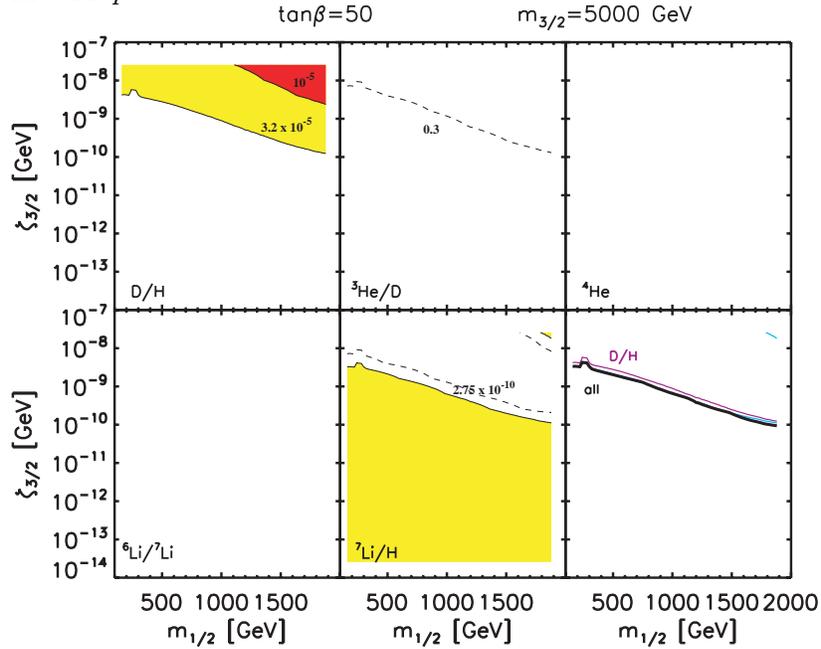,width=0.65\textwidth}
\end{center}
\caption{\it
As for Fig.~\protect\ref{fig:elements250tb10}, with 
$m_{3/2} = 5000$~GeV, but $\tan \beta = 50$ and
unchanged values for the CMSSM parameters.
}
\label{fig:elements5000tb50}
\vspace{-1em}
\end{figure}

Figs.~\ref{fig:elements250tb50}--\ref{fig:elements5000tb50}
are the same as Figs.~\ref{fig:elements250tb10}--\ref{fig:elements5000tb10},
but for $\tan \beta = 50$.
Similar trends emerge as for the $\tan \beta = 10$ results, 
with some differences of detail due to
the more rapid rise in lifetimes with $m_{1/2}$
(Fig.~\ref{fig:lifetimes}).
The range in $m_{1/2}$ is extended to almost 1.9 TeV
and includes the rapid annihilation funnel in addition to 
the coannihilation region.
We again see that no regions allow for a solution to the
\li7 problem while simultaneously satisfying the
other light-element constraints.
The constraints on $\zeta_{3/2}$ are 
again dominated by the \li6/\li7 ratio, and are weak for $m_{3/2} = 5000$ GeV.
As in the $\tan \beta = 10$ case, the \li7/H constraint is incompatible
with the others for $m_{3/2} \le 1000$~GeV, but may be marginally
compatible for $m_{3/2} = 5000$~GeV.

%%%%%%%%%%%%%%%%%%%%%%%%% F I G U R E %%%%%%%%%%%%%%%%%%%%%%%%%%%%%%%%%%%%%%%%%

%%%%%%%%%%%%%%%%%%%%%%%%% F I G U R E %%%%%%%%%%%%%%%%%%%%%%%%%%%%%%%%%%%%%%%%%

%%%%%%%%%%%%%%%%%%%%%%%%% F I G U R E %%%%%%%%%%%%%%%%%%%%%%%%%%%%%%%%%%%%%%%%%

%%%%%%%%%%%%%%%%%%%%%%%%% F I G U R E %%%%%%%%%%%%%%%%%%%%%%%%%%%%%%%%%%%%%%%%%

%%%%%%%%%%%%%%%%%%%%%%%%% F I G U R E %%%%%%%%%%%%%%%%%%%%%%%%%%%%%%%%%%%%%%%%%

%%%%%%%%%%%%%%%%%%%%%%%%% F I G U R E %%%%%%%%%%%%%%%%%%%%%%%%%%%%%%%%%%%%%%%%%

\begin{figure}[tbh!]
\begin{center}
\epsfig{file=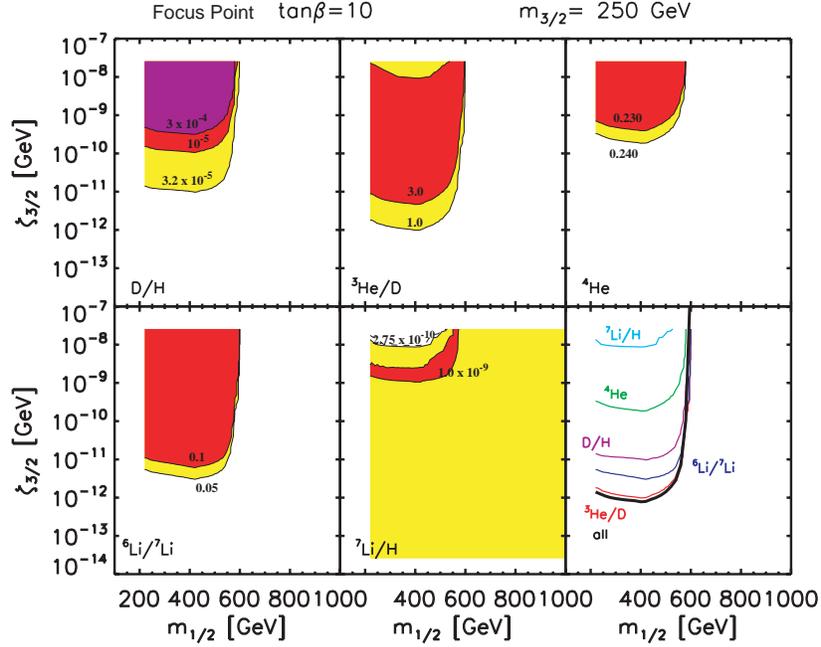,width=0.65\textwidth}
\end{center}
\caption{\it
As for Fig.~\protect\ref{fig:elements250tb10}, with $m_{3/2} = 250$~GeV,
but $\tan \beta = 10$ and
CMSSM parameters appropriate for the WMAP strip in the focus-point region.
}
\label{fig:elements250tb10fp}
\vspace{-1em}
\end{figure}

\begin{figure}[tbh!]
\begin{center}
\epsfig{file=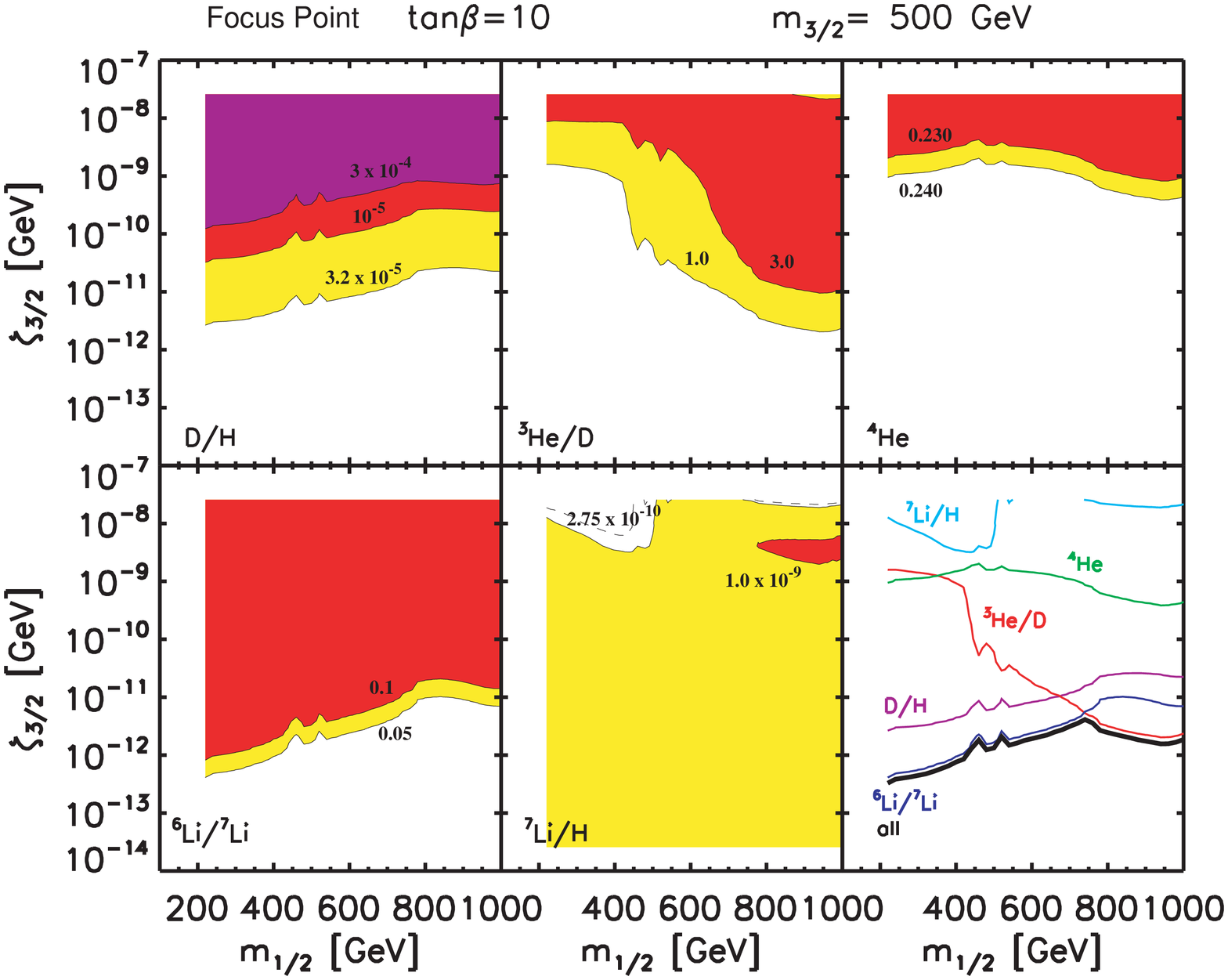,width=0.65\textwidth}
\end{center}
\caption{\it
As for Fig.~\protect\ref{fig:elements250tb10}, with $m_{3/2} = 500$~GeV, 
but $\tan \beta = 10$ and
CMSSM parameters appropriate for the WMAP strip in the focus-point region.
}
\label{fig:elements500tb10fp}
\vspace{-1em}
\end{figure}

\begin{figure}[tbh!]
\begin{center}
\epsfig{file=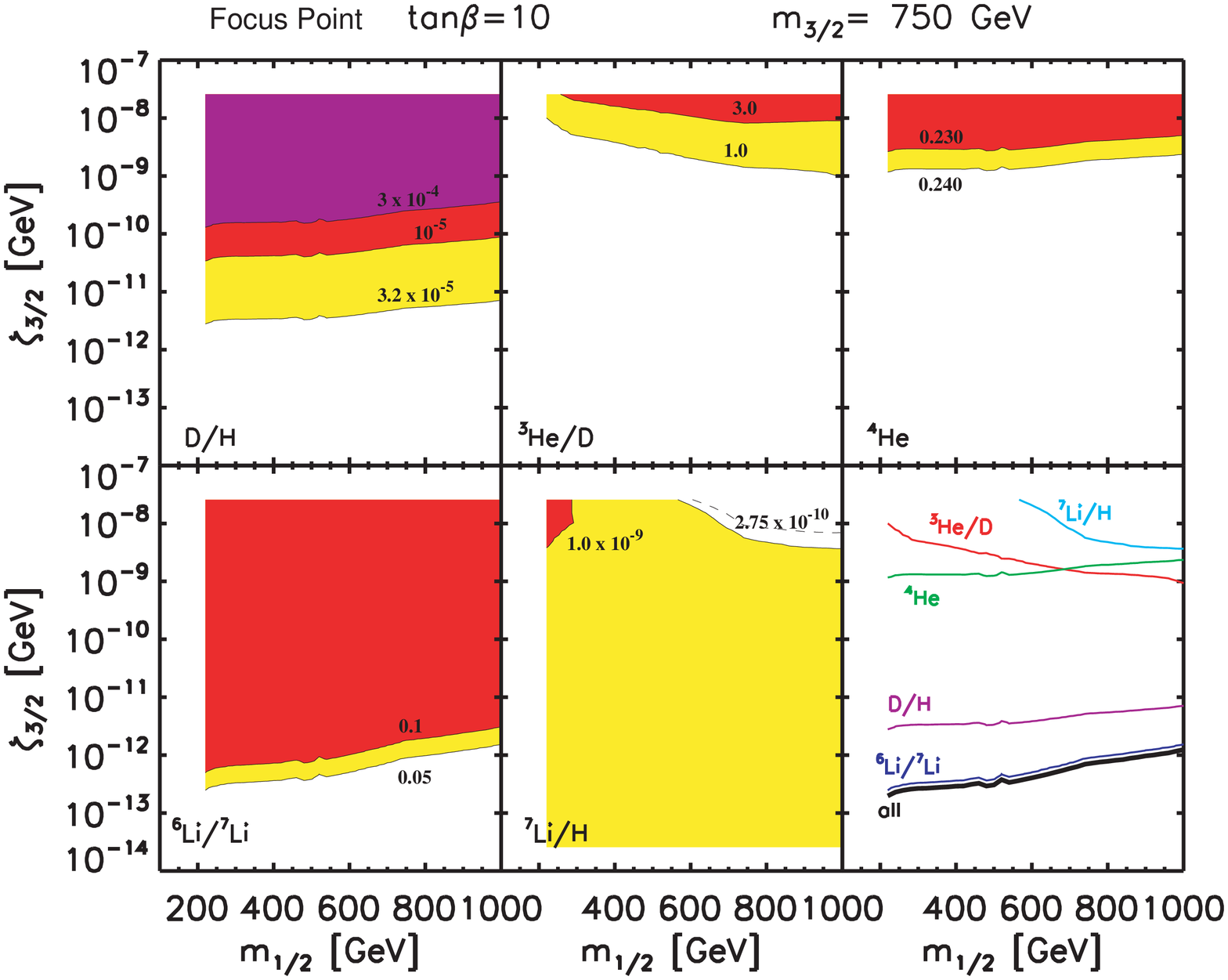,width=0.65\textwidth}
\end{center}
\caption{\it
As for Fig.~\protect\ref{fig:elements250tb10}, with $m_{3/2} = 750$~GeV, 
but $\tan \beta = 10$ and
CMSSM parameters appropriate for the WMAP strip in the focus-point region.
}
\label{fig:elements750tb10fp}
\vspace{-1em}
\end{figure}

\begin{figure}[tbh!]
\begin{center}
\epsfig{file=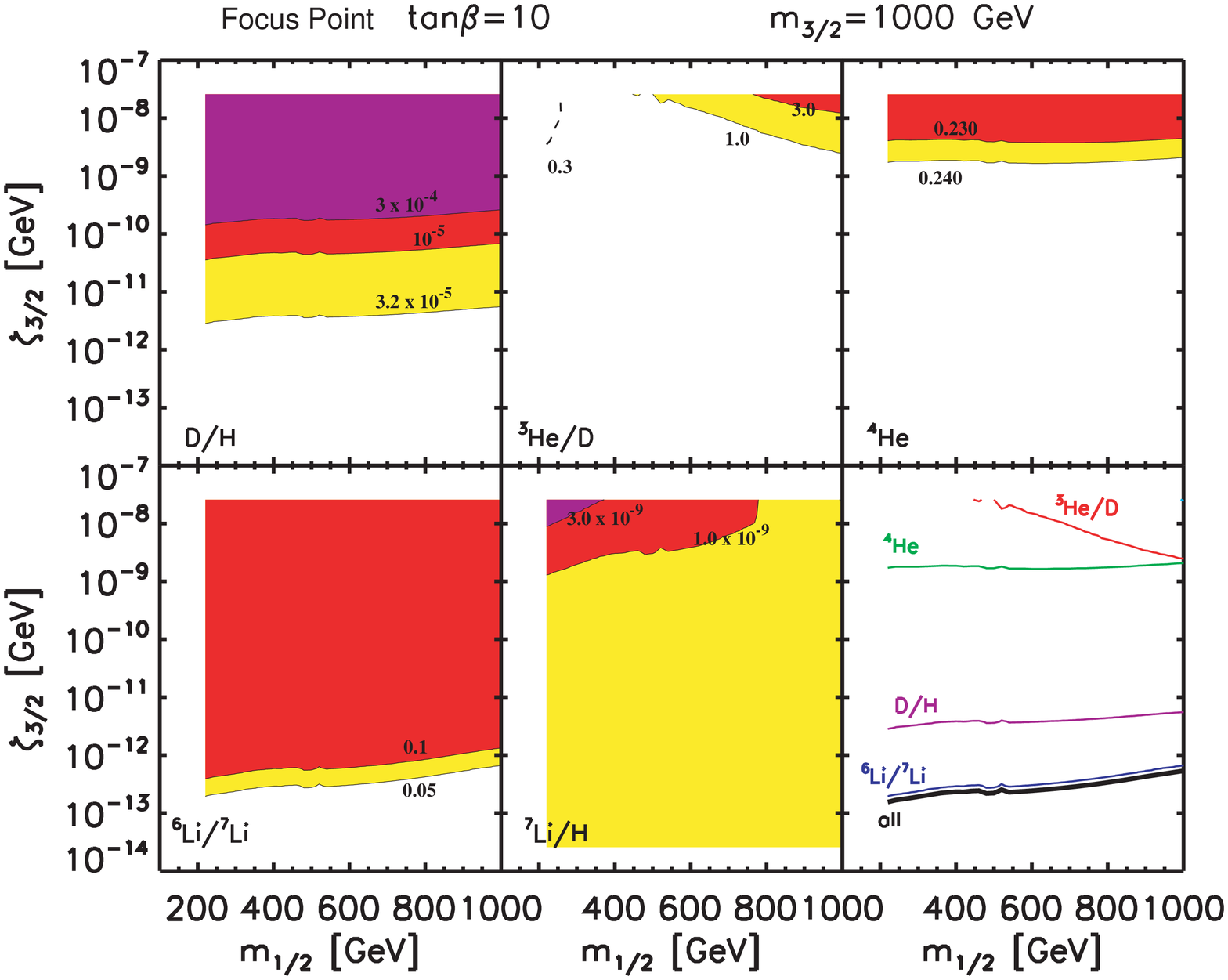,width=0.65\textwidth}
\end{center}
\caption{\it
As for Fig.~\protect\ref{fig:elements250tb10}, with 
$m_{3/2} = 1000$~GeV, but $\tan \beta = 10$ and
CMSSM parameters appropriate for the WMAP strip in the focus-point region.
}
\label{fig:elements1000tb10fp}
\vspace{-1em}
\end{figure}

\begin{figure}[tbh!]
\begin{center}
\epsfig{file=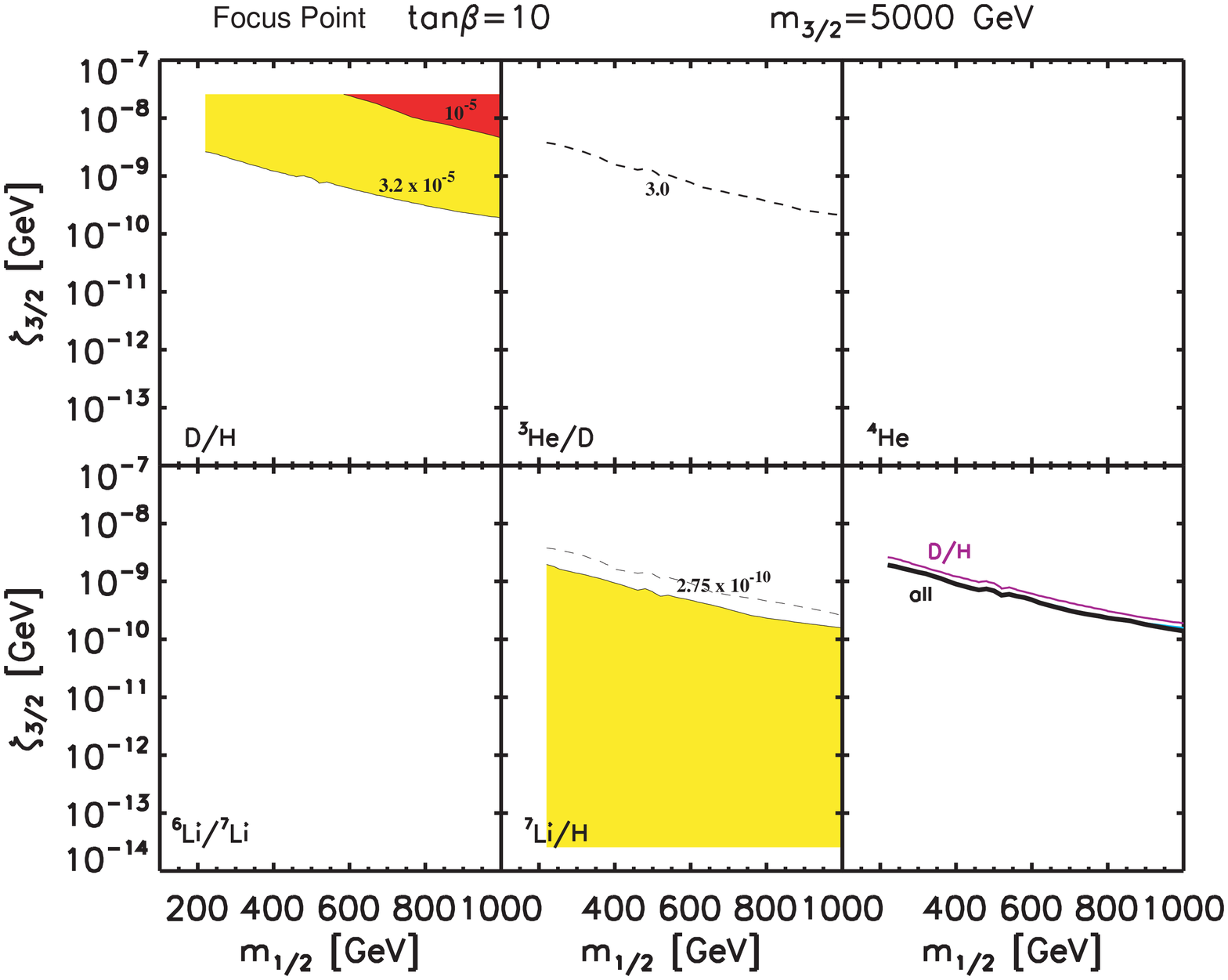,width=0.65\textwidth}
\end{center}
\caption{\it
As for Fig.~\protect\ref{fig:elements250tb10}, with 
$m_{3/2} = 5000$~GeV, but $\tan \beta = 10$ and
CMSSM parameters appropriate for the WMAP strip in the focus-point region.
}
\label{fig:elements5000tb10fp}
%\vspace{-1em}
\end{figure}

\begin{figure}[tbhp!]
\begin{center}
\epsfig{file=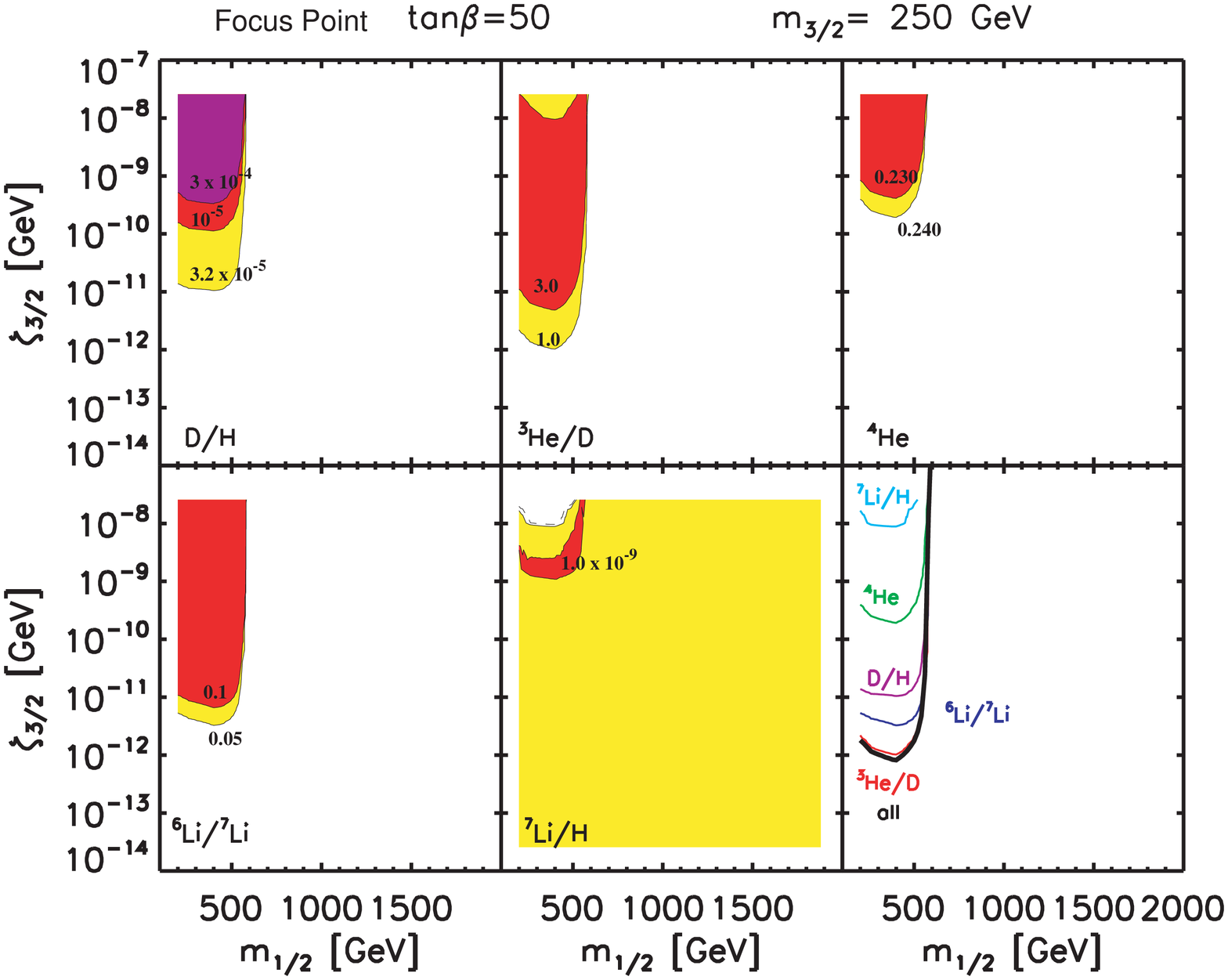,width=0.65\textwidth}
\end{center}
\caption{\it
As for Fig.~\protect\ref{fig:elements250tb10}, with 
$m_{3/2} = 250$~GeV,
but $\tan \beta = 50$ and
CMSSM parameters appropriate for the WMAP strip in the focus-point region.
}
\label{fig:elements250tb50fp}
\vspace{-1em}
\end{figure}

\begin{figure}[tbhp!]
\begin{center}
\epsfig{file=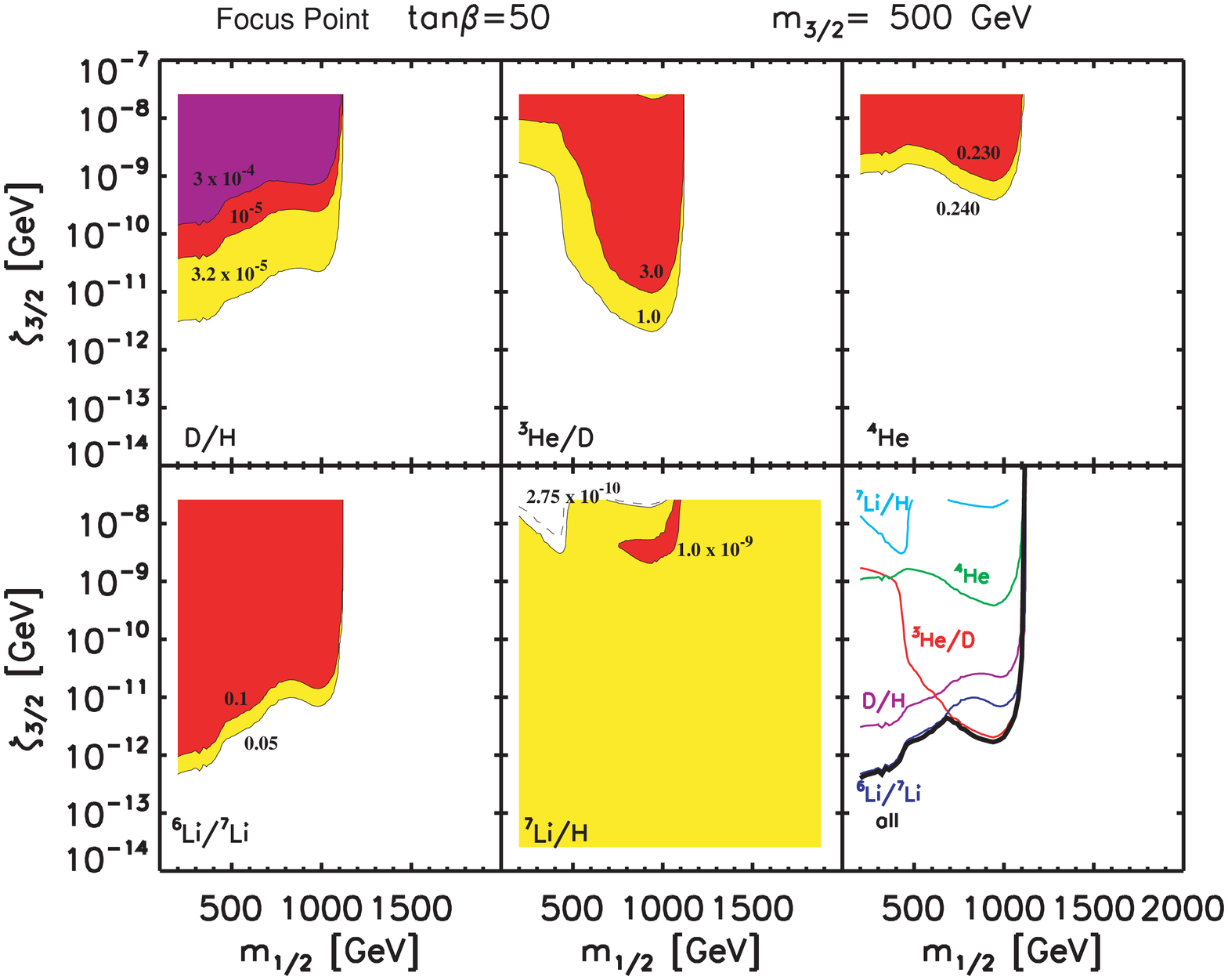,width=0.65\textwidth}
\end{center}
\caption{\it
As for Fig.~\protect\ref{fig:elements250tb10}, with 
$m_{3/2} = 500$~GeV, 
but $\tan \beta = 50$ and
CMSSM parameters appropriate for the WMAP strip in the focus-point region.
}
\label{fig:elements500tb50fp}
\vspace{-1em}
\end{figure}

\begin{figure}[tbhp!]
\begin{center}
\epsfig{file=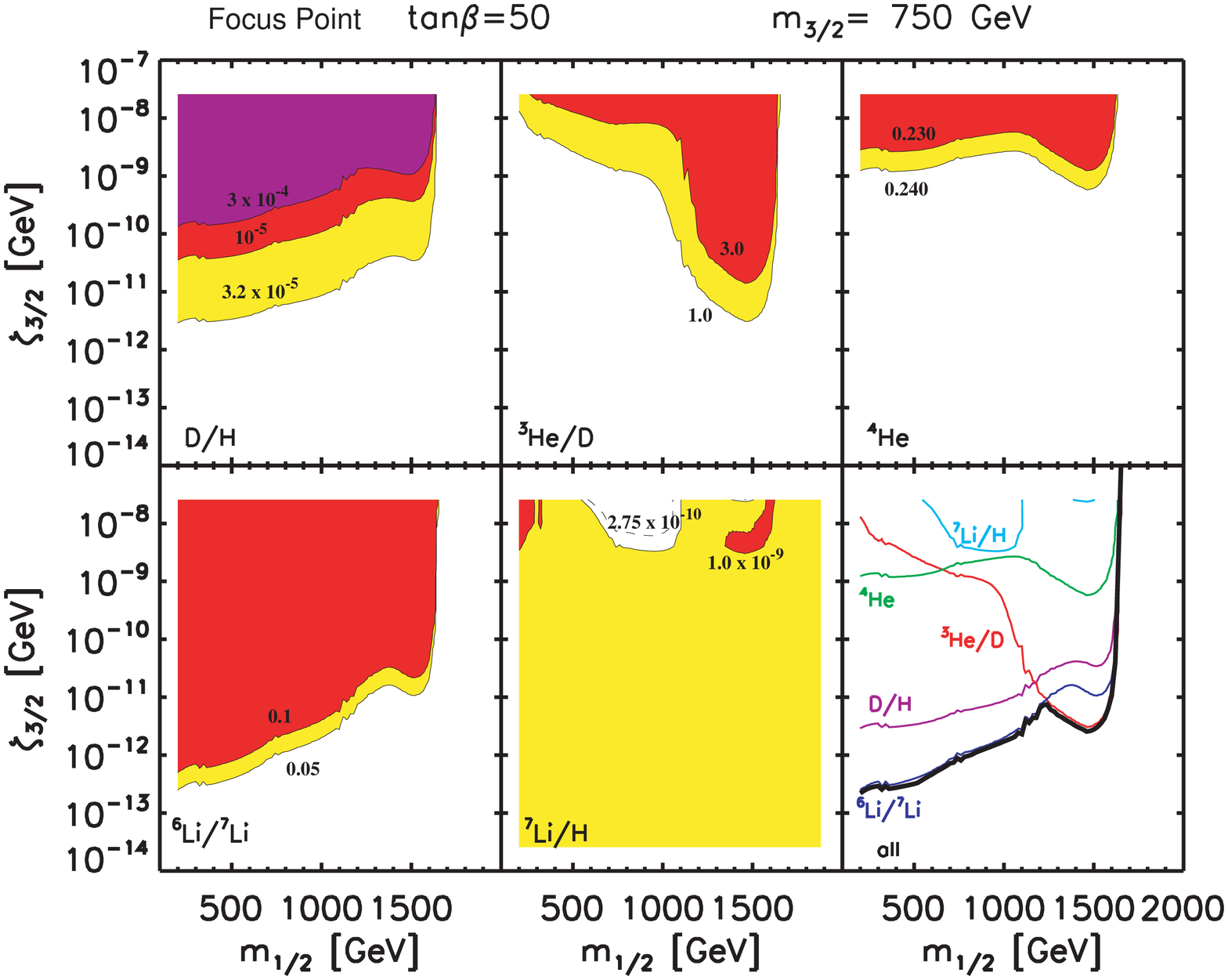,width=0.65\textwidth}
\end{center}
\caption{\it
As for Fig.~\protect\ref{fig:elements250tb10}, with 
$m_{3/2} = 750$~GeV,
but $\tan \beta = 50$ and
CMSSM parameters appropriate for the WMAP strip in the focus-point region.
}
\label{fig:elements750tb50fp}
\vspace{-1em}
\end{figure}

\begin{figure}[tbhp!]
\begin{center}
\epsfig{file=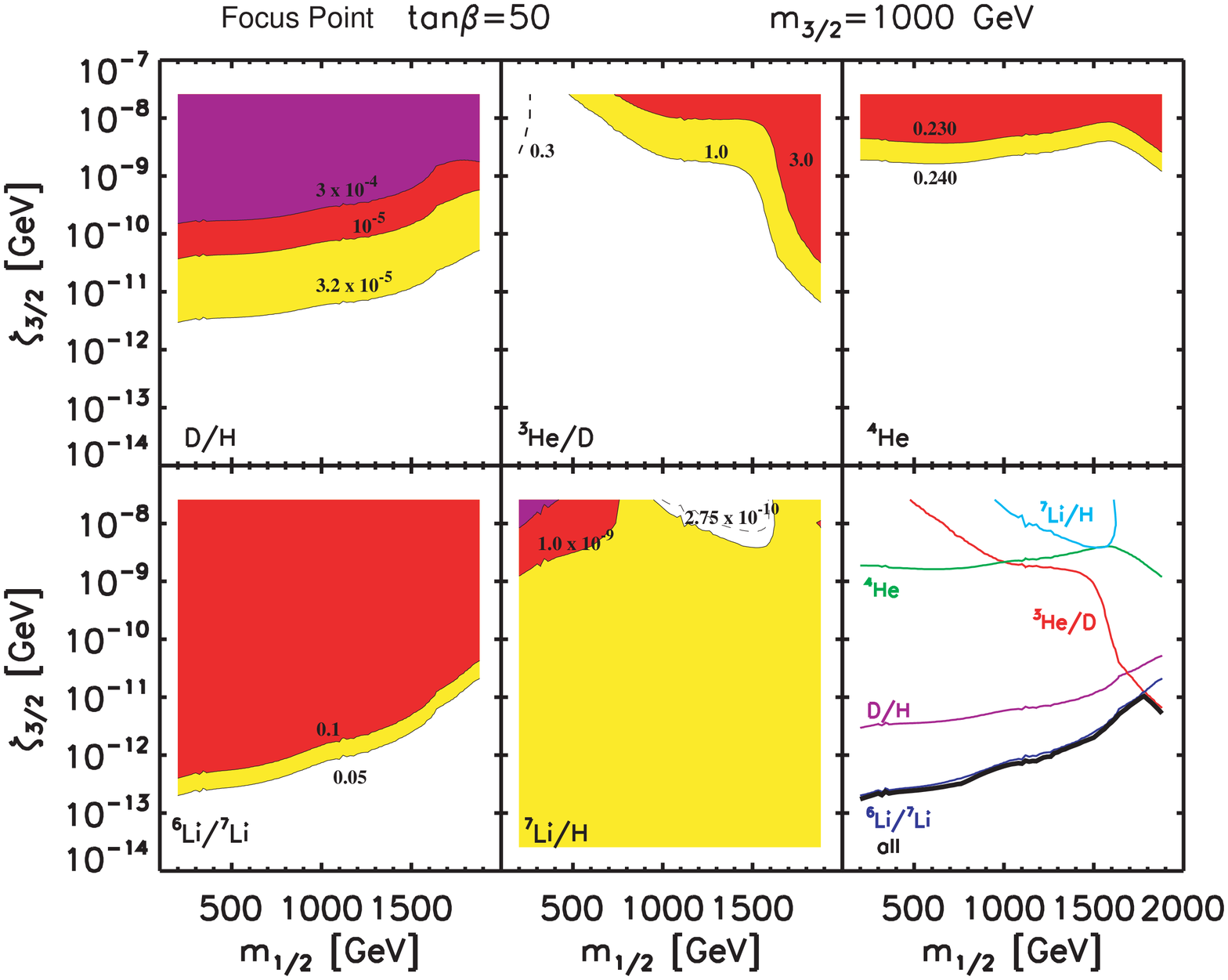,width=0.65\textwidth}
\end{center}
\caption{\it
As for Fig.~\protect\ref{fig:elements250tb10}, with 
$m_{3/2} = 1000$~GeV, 
but $\tan \beta = 50$ and
CMSSM parameters appropriate for the WMAP strip in the focus-point region.
}
\label{fig:elements1000tb50fp}
\vspace{-1em}
\end{figure}

\begin{figure}[tbhp!]
\begin{center}
\epsfig{file=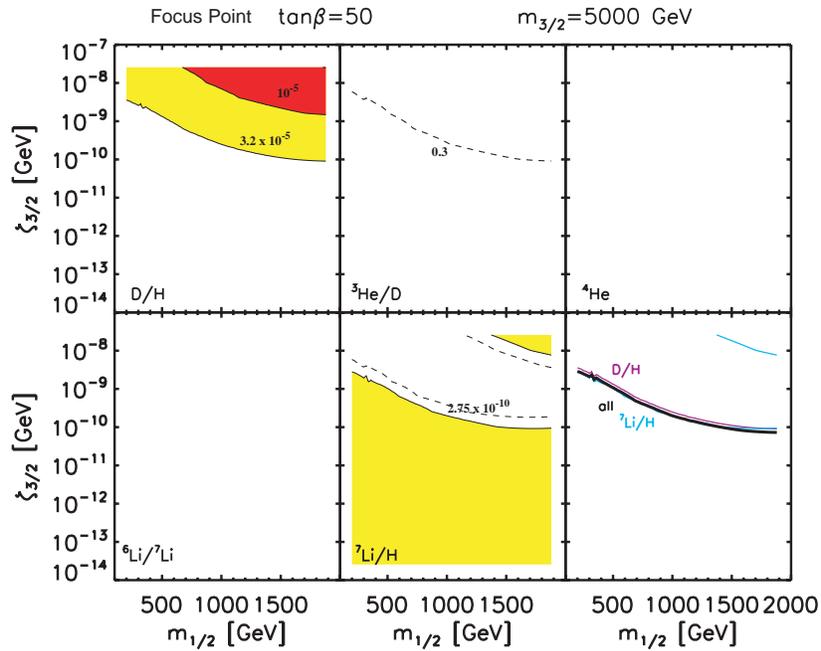,width=0.65\textwidth}
\end{center}
\caption{\it
As for Fig.~\protect\ref{fig:elements250tb10}, with 
$m_{3/2} = 5000$~GeV, 
but $\tan \beta = 50$ and
CMSSM parameters appropriate for the WMAP strip in the focus-point region.
}
\label{fig:elements5000tb50fp}
\vspace{-1em}
\end{figure}

Figs.~\ref{fig:elements250tb10fp}--\ref{fig:elements5000tb10fp} present a similar analysis to that in
Figs.~\ref{fig:elements250tb10}--\ref{fig:elements5000tb10},
but for the focus-point region of the CMSSM with $\tan \beta = 10$.
The results are also largely similar to those for the previous cases.
The same is true for the focus-point region of the CMSSM with $\tan \beta = 50$,
shown in Fig.~\ref{fig:elements250tb50fp}--\ref{fig:elements5000tb50fp}.
As in the previous cases, the \li7/H constraint is incompatible with all the
other constraints, except possibly for $m_{3/2} = 5000$~GeV.

\clearpage

%%%%%%%%%%%%%%%%%%%%%%%%% F I G U R E %%%%%%%%%%%%%%%%%%%%%%%%%%%%%%%%%%%%%%%%%

%%%%%%%%%%%%%%%%%%%%%%%%% F I G U R E %%%%%%%%%%%%%%%%%%%%%%%%%%%%%%%%%%%%%%%%%

%%%%%%%%%%%%%%%%%%%%%%%%% F I G U R E %%%%%%%%%%%%%%%%%%%%%%%%%%%%%%%%%%%%%%%%%

%%%%%%%%%%%%%%%%%%%%%%%%% F I G U R E %%%%%%%%%%%%%%%%%%%%%%%%%%%%%%%%%%%%%%%%%

%%%%%%%%%%%%%%%%%%%%%%%%% F I G U R E %%%%%%%%%%%%%%%%%%%%%%%%%%%%%%%%%%%%%%%%%

%\clearpage

%%%%%%%%%%%%%%%%%%%%%%%%% F I G U R E %%%%%%%%%%%%%%%%%%%%%%%%%%%%%%%%%%%%%%%%%

%%%%%%%%%%%%%%%%%%%%%%%%% F I G U R E %%%%%%%%%%%%%%%%%%%%%%%%%%%%%%%%%%%%%%%%%

%%%%%%%%%%%%%%%%%%%%%%%%% F I G U R E %%%%%%%%%%%%%%%%%%%%%%%%%%%%%%%%%%%%%%%%%

%%%%%%%%%%%%%%%%%%%%%%%%% F I G U R E %%%%%%%%%%%%%%%%%%%%%%%%%%%%%%%%%%%%%%%%%

%%%%%%%%%%%%%%%%%%%%%%%%% F I G U R E %%%%%%%%%%%%%%%%%%%%%%%%%%%%%%%%%%%%%%%%%

\subsection{Varying $m_{3/2}$ and the \li7 Problem}

\begin{figure}[tbhp!]
\begin{center}
\epsfig{file=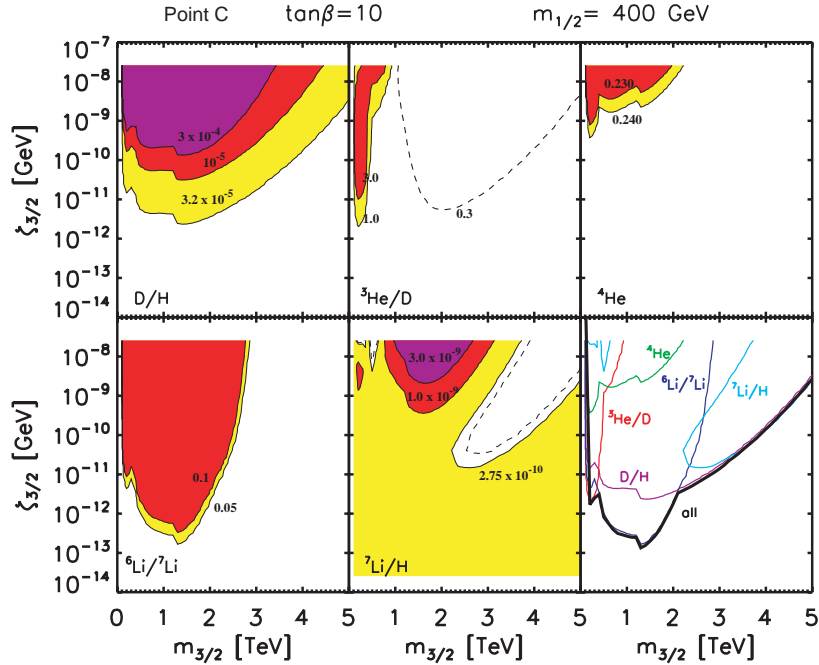,width=0.65\textwidth}
\end{center}
\caption{\it
The effects of the decays of a gravitino with variable mass $m_{3/2}$
on the different light-element abundances for a specific point (benchmark C) with
$m_{1/2} = 400$~GeV on the WMAP coannihilation strip for a CMSSM scenario
with $\tan \beta = 10, A_0 = 0$. As in previous figures,
the white regions in each panel are those allowed
at face value by the light-element abundances
reviewed in Section~\ref{sect:abs}, and the yellow, red, and magenta regions correspond to
progressively larger deviations from the central values of the abundances.
We see marginal compatibility between the \li7 constraint (light blue)
and the other constraints for $m_{3/2} \ga 3$~TeV.
}
\label{fig:elementsCm32}
\vspace{-1em}
\end{figure}

The previous plots displayed the gravitino constraints along
WMAP strips in the $(m_{1/2}, m_0)$ planes for a few discrete choices of
$m_{3/2}$. We now display some results as continuous functions
of $m_{3/2}$ for a few WMAP-compatible points with certain discrete values of $m_{1/2}$. 
Fig.~\ref{fig:elementsCm32} shows our first choice, $m_{1/2} = 400$~GeV and $\tan \beta = 10$,
in which case the WMAP point is essentially benchmark point C defined in~\cite{bench}. As expected from the previous
figures, we see that the other light-element constraints are incompatible with
the \li7/H constraint for low $m_{3/2}$. Disregarding the \li7 problem, we find, e.g., an upper limit
on $\zeta_{3/2} \sim 10^{-13}$ when $m_{3/2} \sim 1.4$~TeV.
However, marginal compatibility with the cosmological \li7 abundance is
approached for $m_{3/2} \ga 3$~TeV with $\zeta_{3/2} > 10^{-11} \gev$, as was to be expected from
Fig.~\ref{fig:elements5000tb10}. This is a realization within the CMSSM of
the marginal compatibility between the \li7 abundance and the D/H ratio
that was noted earlier in the context of Fig.~\ref{fig:zeta-tau-std}.

%%%%%%%%%%%%%%%%%%%%%%%%% F I G U R E %%%%%%%%%%%%%%%%%%%%%%%%%%%%%%%%%%%%%%%%%

\begin{figure}[tbhp!]
\begin{center}
\epsfig{file=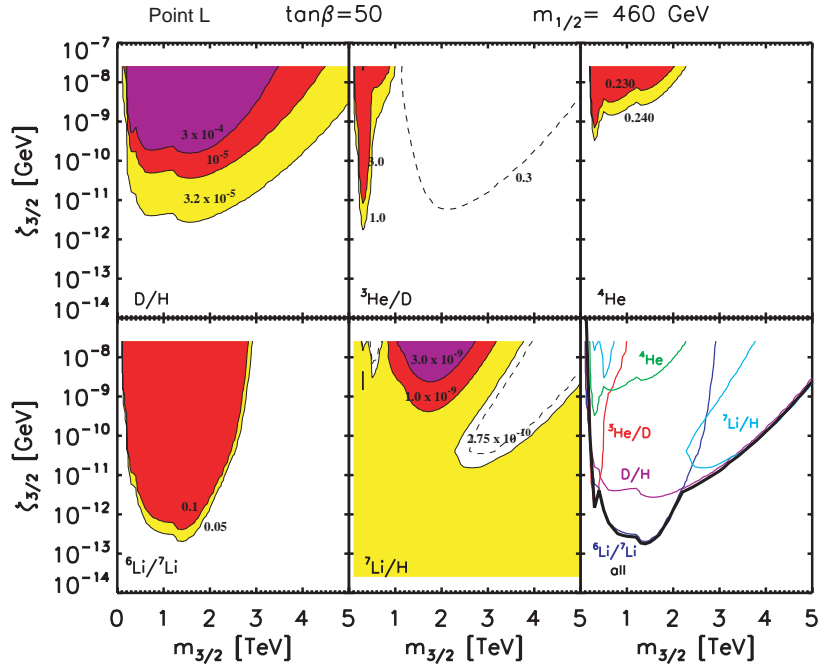,width=0.65\textwidth}
\end{center}
\caption{\it
As for Fig.~\protect\ref{fig:elementsCm32}, for CMSSM benchmark point L with 
$m_{1/2} = 460$~GeV and $\tan \beta = 50$, on the WMAP strip in the coannihilation region.
}
\label{fig:elementsLm32}
\vspace{-1em}
\end{figure}

\begin{figure}[tbhp!]
\begin{center}
\epsfig{file=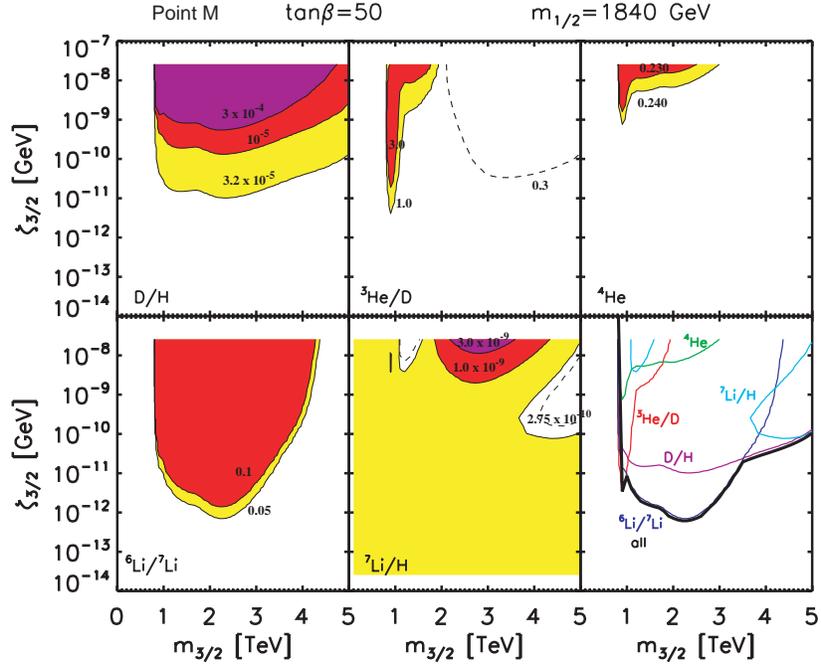,width=0.65\textwidth}
\end{center}
\caption{\it
As for Fig.~\protect\ref{fig:elementsCm32}, for CMSSM benchmark point M with 
$m_{1/2} = 1840$~GeV and $\tan \beta = 50$, on the WMAP strip in the rapid-annihilation
funnel region.
}
\label{fig:elementsMm32}
\vspace{-1em}
\end{figure}

\begin{figure}[tbhp!]
\begin{center}
\epsfig{file=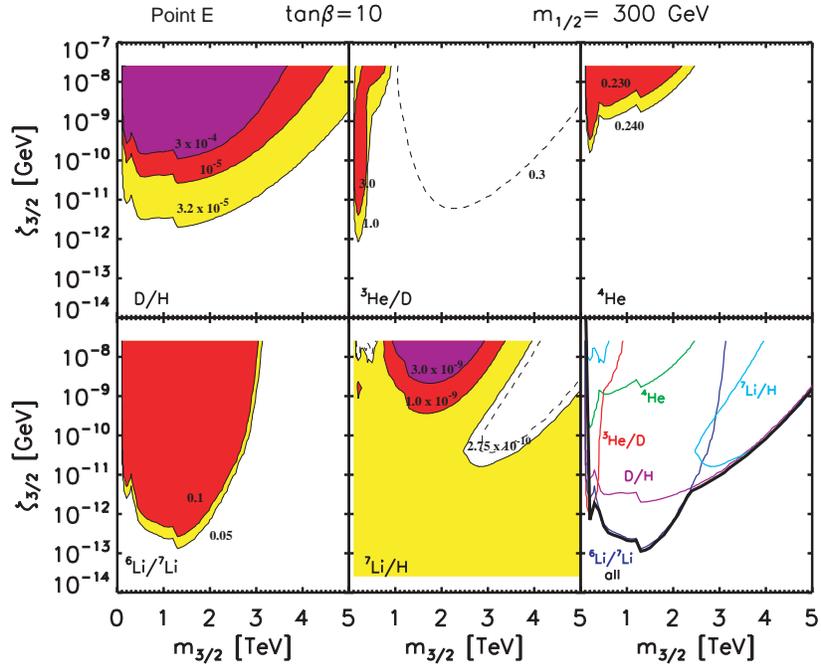,width=0.65\textwidth}
\end{center}
\caption{\it
As for Fig.~\protect\ref{fig:elementsCm32}, for CMSSM benchmark point E with 
$m_{1/2} = 300$~GeV and $\tan \beta = 10$, on the WMAP strip in the focus-point region.
}
\label{fig:elementsEm32}
\vspace{-1em}
\end{figure}

Figs~\ref{fig:elementsLm32}, \ref{fig:elementsMm32} and \ref{fig:elementsEm32}
display similar features for other choices of WMAP-compatible benchmark points~\cite{bench}.
In the case of Fig.~\ref{fig:elementsLm32}, we choose benchmark point L with
$m_{1/2} = 460$~GeV and $\tan \beta = 50$, which is also in a coannihilation strip.
We see very similar features to Fig.~\ref{fig:elementsCm32}, e.g., an upper limit
on $\zeta_{3/2} \sim 2 \times 10^{-13} \gev$ when $m_{3/2} \sim 1.6$~TeV if the \li7 constraint
is disregarded, and a marginal solution of the \li7 problem for $m_{3/2} \ga 3$~TeV
with $\zeta_{3/2} > 10^{-11} \gev$.
Fig.~\ref{fig:elementsMm32} is based on benchmark point M with $m_{1/2} = 1840$~GeV and 
$\tan \beta = 50$, which is in a rapid-annihilation funnel. The features seen in the previous
figures shift to higher $m_{3/2}$, e.g., we see an upper limit
on $\zeta_{3/2} \sim 10^{-12} \gev$ when $m_{3/2} \sim 2.4$~TeV if the \li7 constraint
is disregarded, and a marginal solution of the \li7 problem for $m_{3/2} \ga 4$~TeV
with $\zeta_{3/2} > 5 \times 10^{-11} \gev$.
Finally, Fig.~\ref{fig:elementsEm32} is based on benchmark point E with $m_{1/2} = 300$~GeV and 
$\tan \beta = 10$, which is in a focus-point region.
We see in general results that are very similar to those for benchmark point C shown in
Fig.~\ref{fig:elementsCm32}.

%%%%%%%%%%%%%%%%%%%%%%%%% F I G U R E %%%%%%%%%%%%%%%%%%%%%%%%%%%%%%%%%%%%%%%%%

%%%%%%%%%%%%%%%%%%%%%%%%% F I G U R E %%%%%%%%%%%%%%%%%%%%%%%%%%%%%%%%%%%%%%%%%

%%%%%%%%%%%%%%%%%%%%%%%%% F I G U R E %%%%%%%%%%%%%%%%%%%%%%%%%%%%%%%%%%%%%%%%%

The stability of the marginal ``solution" of the \li7 problem reflects the fact that
the gravitino lifetime $\sim 10^2 - 10^3$~s and the number of nucleons per decay are relatively stable
for heavy gravitino masses as $m_{1/2}$ and $\tan \beta$ are varied: see
Figs~\ref{fig:lifetimes} and \ref{fig:branching}. The underlying physics
of this marginal ``solution",
as discussed in~\cite{jed,kkm2}, is that for this range of gravitino lifetimes,
thermalized neutrons can destroy \be7 via $\be7(n,p)\li7$, following which
\li7 is destroyed by the $\li7(p,\alpha)\he4$ reaction. Here, as discussed in
Appendix~\ref{app:prop}, we also allow for supplementary \be7 destruction
by non-thermalized neutrons.

\section{Conclusions}

\label{sect:conclude}

The first objective of this paper has been to document a new suite of
codes for the development of non-thermal particle showers in the
early universe. These codes treat both electromagnetic and hadronic
components of showers, and can be applied to the decays of unstable
particles within a wide range of possible lifetimes, that may decay either
during or subsequent to standard BBN. We have also shown
how this suite of codes may be used to analyze the possible effects of
generic particles with typical decay modes.

As a specific application of this suite of codes,
we have re-examined the cosmological constraints on unstable
gravitinos in $R$-conserving CMSSM scenarios with a neutralino LSP
arising from their effects on the light-element abundances.
The first step in this re-examination was the recalculation of some
two-body gravitino decay modes and the calculation of some important
three-body decay modes. The next step was the simulation of the
electromagnetic and hadronic products of these decays using {\tt PYTHIA}.
We have then used our new suite of codes to simulate the interactions
of these decay products with the cosmological plasma, including both
energy losses and interactions with the nuclei of light elements that affect
their abundances. Generally, we find that details of the decay spectra are
less important than the overall fractions of baryons produced in the gravitino
decays. These fractions are model-dependent, and vary considerably with
the gravitino mass and the CMSSM parameters. In addition to the baryon
fractions, the constraints on the gravitino abundance inferred from the
observed light-element abundances depend sensitively on the gravitino
lifetime, and are generally weaker in scenarios with a shorter-lived gravitino.

For $m_{3/2} < 3$~TeV,
none of the CMSSM scenarios we study solves the cosmological \li7
problem. Accordingly, one must either reject these CMSSM scenarios,
or re-evaluate the observational data on \li7, or postulate some other
mechanism for bringing the \li7 abundance into line with standard BBN
predictions, or find some way to modify these predictions.
Setting the \li7 problem aside, we find that the strongest constraints are 
usually those imposed by the \li6/\li7 ratio,
with weaker limits coming from the D/H, \he3/D and \he4 constraints.
We have evaluated these constraints along WMAP strips in the $(m_{1/2}, m_0)$
plane where the relic neutralino density falls within the expected range for cold
dark matter, for $m_{3/2} = 250, 500, 750, 1000$ and 5000~GeV,
$A_0 = 0$ and two representative choices $\tan \beta = 10, 50$. There
are two strips for each choice of $\tan \beta$: one in the coannihilation
and funnel region, and one in the focus-point region.

We find that the upper limits on $\zeta_{3/2} = m_{3/2} n_{3/2}/n_\gamma$ are
quite similar along these WMAP strips, that they weaken as
$m_{1/2}$ increases for fixed $m_{3/2}$ (particularly for $\tan \beta = 50$), 
that they strengthen by about an order of magnitude as $m_{3/2}$ increases 
from 250~GeV towards 1000~GeV, and that the constraints are significantly
weaker for $m_{3/2} = 5000$~GeV. The constraints also weaken considerably as
the neutralino mass approaches $m_{3/2}$ from below, as the phase space
for gravitino decay disappears. For larger values of $m_{1/2}$, the
neutralino is no longer the LSP, and the WMAP strips are inapplicable.

Extending these studies to larger $m_{3/2}$, we find that the
D/H and \li7/H constraints may become marginally compatible for $m_{3/2} \ga 3$~TeV and a very
narrow range of gravitino abundance that increases with $m_{3/2}$.
In this region, both thermal and non-thermal neutrons destroy \be7 via the $\be7(n,p)\li7$ reaction, 
which is followed by \li7 destroyed via the $\li7(p,\alpha)\he4$ reaction~\cite{jed,kkm2}
for a finely-tuned range of lifetimes around $\tau_{X} \sim 10^3$ sec.
We use here the recently updated $\he3(\alpha,\gamma)\be7$
rate \cite{cd} which makes the \li7 problem worse,
since it requires \be7 destruction by a larger factor.
We find that the competing upward perturbations
to D/H and \he3/D drive these species away from
their observed levels in the regimes where the \li7 problem is solved.
This highlights the continued importance of the nuclear reaction
data, particularly at energies beyond the $\la 1$ MeV range
important for standard, thermal BBN reactions.
Additional data extending from $\sim 3$ MeV to higher energies $\sim 30$ MeV
are urgently needed,
particularly for reactions such as $\he3(\alpha,\gamma)\be7$,
and for $\be7+n \rightarrow {\rm anything}$.
Further studies of both the light-element abundances as well as the relevant nuclear
reactions is required in order to verify whether this possible ``solution" to
the \li7 problem is viable and, if so, how finely one must tune the gravitino
abundance within the CMSSM.

The extension of this analysis to the case of  a gravitino LSP depends on
the nature of the NLSP. If the NLSP
is neutral, much of the suite of codes developed here would be directly
applicable. On the other hand, if the NLSP is charged, e.g., the lighter
stau or some other slepton, it may form electromagnetic bound states
before decaying, and catalyze nuclear interactions that are not included
in the codes developed here. We will return to this issue in a future
publication.

\renewcommand{\theequation}{A-\arabic{equation}}
  % redefine the command that creates the equation no.
  \setcounter{equation}{0}  % reset counter
\renewcommand{\thesection}{A \arabic{equation}}
  % redefine the command that creates the equation no.
%  \setcounter{section}{+1}  % reset counter

\section*{Acknowledgments}
\noindent
We would like to thank M. Pospelov and M. Voloshin for helpful discussions.
The work of R.H.C.~was supported  by the U.S. National Science Foundation
Grants No.~PHY-01-10253 (NSCL) and Nos.~PHY-02-016783 
and PHY-08-22648 (JINA).
B.D.F.~is grateful to acknowledge the hospitality of CERN TH group,
and of the NASA Goddard Space Flight Center, 
where some of this work was completed.
The work of K.A.O. and F.L. is supported in part by DOE grant DE-FG02-94ER-40823 at the 
University of Minnesota.
The  work of V.C.S. was supported by Marie Curie International Reintegration grant
ÒSUSYDM-PHENÓ, MIRG-CT-2007-203189 and Marie Curie Excellence grant MEXT-CT-2004-014297.

\

\appendix

\noindent{\bf{\Large Appendices}}

\section{Propagation of Non-Thermalized Particles in the Cosmic Plasma}

\label{app:prop}

We wish to find the non-thermal spectra
$N_h$ which are solutions of the propagation equation
(\ref{eq:cascade}).
As discussed in Section \ref{sect:showers},
the equation is self-regulating, with $N_h$
adjusting itself so that sources come into equilibrium
with sinks, and hence that $\pd_t N_h = 0$.
In this case, the equation we wish to solve is
\beq
\label{eq:prop}
J_{h}(\epsilon) - \Gamma_{h}(\epsilon)N_{h}(\epsilon)-\pd_\nrg\left[b_{h}(\epsilon) N_{h}(\epsilon)\right]
 = 0 .
\eeq
The sink term has two contributions, due to elastic and inelastic scattering:
\beq
\label{eq:sink}
\Gamma_{h}(\epsilon) = \sum_{b} n_{b}
[\sigma v]_{hb \rightarrow h}(\nrg) + \sum_{h^\prime \ne h,b} n_{b}
[\sigma v]_{hb \rightarrow h^\prime}(\nrg) .
\eeq
The source term has three contributions, due to direct injection, elastic
down-scattering and inelastic down-scattering, respectively, given by:
\beqar
\label{eq:source}
 J_{h}(\epsilon) & = &
  J_{X,h}(\epsilon) + 
  J_{{\rm elastic},h}(\nrg) + J_{{\rm inelastic},h}(\nrg) \nonumber  \\
 & = & \Gamma_X Q_{h}(\epsilon) + \sum_{b} n_{b} \int_\nrg^\infty
N_{h} \frac{d[\sigma v]_{hb \rightarrow
h}(\nrg^\prime,\nrg)}{d\nrg} d\nrg^\prime \nonumber  \\
 && \quad \quad   \;     \,    +  \sum_{h^\prime \ne h,b} n_{b}
\int_\nrg^\infty N_{h^\prime} \frac{d[\sigma v]_{h^\prime b
\rightarrow h}(\nrg^\prime,\nrg)}{d\nrg} d\nrg^\prime ,
\eeqar
where $Q_h(\nrg)$ gives the number of hadron of type $h$
per $X$ decay as discussed in Section \ref{sect:nuke}, so
that $J_{X,h} = \Gamma_X Q_h$ gives the $h$ injection
rate per $X$ particle.
The second term in eq.~(\ref{eq:source})
gives the rate at which $h$ particles at energy $\nrg$ 
are produced by elastic scatterings of other $h$ particles,
integrated over initial energies $\nrg^\prime$.  The third 
term in eq.~(\ref{eq:source}) gives the
rate at which $h$ particles at energy $\nrg$ are produced
via inelastic processes $h^\prime b \rightarrow h$,
integrated over initial energies $\nrg^\prime$.
Note that the
elastic and inelastic sources in eq.~\pref{eq:source}
are both integrals over proton and neutron spectra $N_p$
and $N_n$.
Thus, in the cascade equation eq.~\pref{eq:cascade},
there is one factor of a hadronic spectrum $N_h$
in every term except the decay injection term.
Thus, on dividing the cascade equation by $\Gamma_X$,
one infers an equation for $\tilde{N}_h = N_h/\Gamma_X$
which has the same energy losses and scattering
sources and sinks, but whose decay injection term is $Q_h(\nrg)$,
which is
independent of $\Gamma_X$.
In other words, we find that
the physical spectrum
depends in the following way on the decay properties:
$N_h(\nrg) = \Gamma_X \tilde{N}_h(\nrg)$,
i.e., the scaling separates $N_h(\nrg)$ into two factors: the
$X$ decay rate, and
the branching and spectral shapes of the $h$ daughters.

We now look at each of the source terms in detail.

\subsection{Elastic- and Inelastic-Scattering Terms}

Hadronic processes are described in \pref{eq:sink} and
\pref{eq:source} by the differential cross section $d\sigma_{h_i b
\rightarrow h_j}(\nrg_i,\nrg_j)/d\nrg_j$ for hadron $h_i$ with energy
$\nrg_i$ to produce $h_j$ with energy $\nrg_j$.  The corresponding
total cross section for the process is
\begin{equation}
\sigma_{h_i b \rightarrow
h_j}(\nrg_i) = \int d\nrg_j d\sigma_{h_i b \rightarrow
h_j}(\nrg_i,\nrg_j)/d\nrg_j.
\end{equation}
We find it useful to split
the collisional loss terms into the rates for inelastic
collisions with background nuclei:
$\lambda_{h}^{\rm inel} = \sum_b n_{b} \sigma_{hb}^{\rm inel}(\nrg)$
and elastic collisions: $\lambda_{h}^{\rm el} =
\sum_b n_{b} \sigma_{hb \rightarrow hb}(\nrg)$, where $h$ loses energy.
Similarly, we have divided the source terms into a double sum over all
inelastic processes $j_{\rm shower} k_{\rm bg} \rightarrow i$ and a
sum over elastic scatterings.

\begin{table}
\caption{\it Nuclear reactions used in this analysis.
\label{tab:reactions}}
\begin{center}
\begin{tabular}{cc}
\multicolumn{2}{c}{Reactions used in non-thermal propagation} \\
\hline\hline
Reaction & Reference \\
\hline
$NN \rightarrow NN$ (elastic) &Kawasaki et al.~\cite{kkm2} \\
$NN \rightarrow NN $ (total) &  Kawasaki et al.~\cite{kkm2}\\
$p\he4 \rightarrow p\he4$ (elastic) & Meyer~\cite{meyer} \\
$n\he4 \rightarrow p\he4$ (elastic) & Meyer~\cite{meyer}\\
$np \rightarrow$ (inelastic) & Meyer~\cite{meyer}; Ando, Cyburt, Hong, and Hyun~\cite{ando} \\
$p\he4 \rightarrow $ (inelastic) & Meyer~\cite{meyer} \\
$p\he4 \rightarrow d\he3$ & Meyer~\cite{meyer}\\
$p\he4 \rightarrow np\he3$ & Meyer~\cite{meyer}\\
$p\he4 \rightarrow ddp$ & Meyer~\cite{meyer}\\
$p\he4 \rightarrow dnpp$ & Meyer~\cite{meyer}\\
$p\he4 \rightarrow nnppp$ & Meyer~\cite{meyer}\\
$p\he4 \rightarrow N\he4 \pi$ & Meyer~\cite{meyer}\\
$n\he4 \rightarrow $ (inelastic) & Meyer~\cite{meyer}\\
$n\he4 \rightarrow npt$ & Meyer~\cite{meyer}\\
$n\he4 \rightarrow ddn$ & Meyer~\cite{meyer}\\
$n\he4 \rightarrow dnnp$ & Meyer~\cite{meyer}\\
$n\he4 \rightarrow nnppp$ & Meyer~\cite{meyer}\\
$n\he4 \rightarrow N\he4 \pi$ & Meyer~\cite{meyer}\\
\hline\hline
\end{tabular}

\bigskip

\begin{tabular}{cc}
\multicolumn{2}{c}{Additional reactions with background nuclides} \\
\hline\hline
Reaction & Reference \\
\hline
$pn \rightarrow d \gamma$ & Ando, Cyburt, Hong, and Hyun~\cite{ando} \\
$pd \rightarrow$ (inelastic) & Kawasaki et al.~\cite{kkm2}\\  %% ok
$pd \rightarrow \he3 \gamma$ & Cyburt et al.~\cite{cefo} \\
$pt \rightarrow n \he3$ & Cyburt~\cite{cyburt} \\ %%% XXXXX upgrade needed XXX
$pt \rightarrow$ (inelastic) & Kawasaki et al.~\cite{kkm2}\\
$p\he3 \rightarrow$ (inelastic) & Kawasaki et al.~\cite{kkm2}\\ %% ok
$p \li6 \rightarrow \be7 \gamma$ & Cyburt et al.~\cite{cefo} \\ %% ok
$p \li7 \rightarrow \he4 \he4 \gamma$ & Cyburt~\cite{cyburt} \\ 
$p \be7 \rightarrow \iso{B}{8} \gamma$ & this work \\
$d\he4 \rightarrow \li6 \gamma$ & Mohr~\cite{mohr}\\ %%  Peter Mohr J,PR/C,50,1543,1994
$t\he4 \rightarrow \li6 n$ & Cyburt et al.~\cite{cefo}\\
$t\he4 \rightarrow \li7 \gamma$ & Cyburt~\cite{cyburt} \\
$\he3\he4 \rightarrow \li6 p$ & Cyburt et al.~\cite{cefo} \\
$\he3\he4 \rightarrow \be7 \gamma$ & Cyburt and Davids~\cite{cd} \\
$nd \rightarrow$ (inelastic) & Kawasaki et al.~\cite{kkm2}\\ %% ok
$nd \rightarrow t \gamma$ & Cyburt et al.~\cite{cefo}\\ %% XX update needed XXXX
$nt \rightarrow$ (inelastic) & Kawasaki et al.~\cite{kkm2}\\ %% ok
$n \he3 \rightarrow$ (inelastic) & Kawasaki et al.~\cite{kkm2}\\ %% ok
$n \li6 \rightarrow \li7 \gamma$ & Cyburt et al.~\cite{cefo} \\
\hline\hline
\end{tabular}
\end{center}
\end{table}

The kinematics of the final state particles
are treated as in \cite{kkm2}.

\subsection{Energy Losses}

A decay particle loses energy not only loses discontinuously (in time and in
energy space) through two-body
scatterings, but also continuously (in time and energy space)
through collective electromagnetic interactions with the background plasma.
These are the analogs of the usual ionization energy losses
of particles in laboratory matter, but modified to reflect the
fully-ionized state of the
thermal plasma in the early universe.
The collective interactions amount to a sum of soft, long-range two-body
interactions with the surrounding particles, which occur out to some screening
radius.  As a result, the energy-loss rates are linearly proportional
to the density of the background species in question, and to the
electromagnetic coupling for the interaction in question.

Thus, the energy-loss rate $b_h(\nrg) = -(d\nrg_h/dt)_{\rm background}$
of hadron species $h$ sums over the non-thermal particle's
electromagnetic interactions with the background plasma:
\beq
\label{eq:Eloss}
b_h(\nrg) = b_{h,e}(\nrg)
 + b_{h,\gamma}(\nrg)
 + b_{h,{\rm pair}}(\nrg) ,
\eeq
where the terms account for interactions with
background electrons, positrons and
photons, as well as pair-production.
Before going into the details of the energy dependences of
these terms, some general effects are
important to note.

The dominant electromagnetic losses for both
protons and neutrons are due to
interaction with the background
electrons and positrons, which are the most
mobile and abundant charged species.
Consequently, the key loss rates are proportional to the total
number density $n_{e,\rm tot} = n_{e^-}+n_{e^+}$
of background electrons and positrons, itself set by the plasma temperature
and the excess $n_{e,\rm net} = n_{e^-}-n_{e^+}
= \avg{Z}_{\rm B} n_{B} \approx (Y_{\rm proton}+2Y_{{}^{4}{\rm He}})
  n_{\rm B}$ of
electrons over positrons required by charge balance
with background protons and nuclei.
As the universe undergoes pair annihilation,
the electron/positron abundance and thus the electromagnetic losses
change drastically, and pass through three regimes \cite{fh}.
At $T \ga m_e$, pairs are relativistic
and have numbers comparable to photons:
$n_{e,\rm tot} \approx 3 n_\gamma/2 \gg n_{e,\rm net}$.
At $m_e \ga T \ga m_e/25$,
electrons are nonrelativistic but pairs remain much more abundant than baryons,
though less than photons,
with $n_{e,\rm tot} \approx \sqrt{n_{e,\rm net}^2+(2n_{e^+})^2}$ and
$n_{e^+} = 2(m_eT/2\pi)^{3/2} e^{-m_e/T}$.
At $T \la m_e/25$, pairs are nonrelativistic, the positron density is vanishingly small,
and the electron density is set by charge balance:
$n_{e,\rm tot} = n_{e,\rm net}$.

Non-thermal ions suffer all of the
losses represented by the terms of
eq.~\pref{eq:Eloss}, while neutrons see only the term
that accounts for interactions via their magnetic moment.
A consequence of this distinction is that charged decay hadrons
have a qualitatively different energy-loss behavior from that of
neutrons; we thus discuss the two cases separately.

For non-thermal ions, the dominant energy losses
are due to Coulomb interactions with the background
electrons and positrons.
Such losses of fast ions
in plasmas have
a large
literature of dedicated study
\cite{tsytovich,gould,inokuti}
in addition to calculations made in support of
early-universe applications \cite{Reno,Jedamzik06}.
Following the general approach of
\cite{gould},
we write Coulomb losses for an ion with
energy $\gamma M$ and charge $Z$ in the form
\beq
b_{\rm h,e} = - \pfrac{d\nrg_h}{dt}_{\rm Coulomb}
 \equiv \frac{ Z^2 e^2 \omp^2}{v} B(\gamma) ,
\eeq
where $\omp$ is the
plasma frequency
for non-relativistic electrons:
\beq
\omp^2 = \left. \frac{4\pi e^2 n_e}{m_e} \right|_{\rm cgs}
 = \frac{4\pi \alpha \hbar c n_e}{m_e c^2} .
\eeq
For a relativistic electron plasma, where $T_e > m_e$,
then \cite{tsytovich,gould}
\beq
\omega_{\rm p,rel}^2 \rightarrow \frac{4\pi e^2 n_e}{3 kT_e}
  = \frac{m_e}{3T_e} \left(\omp^{\rm nonrel}\right)^2 \ .
\eeq
The dimensionless function $B(v)$ or $B(\gamma)$
encodes the effects of different physical regimes.
We are interested in various limits, specifically
ion speed $v \sim 1$ versus $v \ll 1$,
and relativistic $T_e \gg m_e$ versus non-relativistic $T_e \ll m_e$
plasmas.
The $B$ function generally features a Coulomb logarithm,
which is roughly $\Lambda \sim \ln (q_{\rm max}/q_{\rm min})$,
where $q$ is the momentum transfer from the fast particle
to a plasma electron.
For example, in the non-relativistic case,
$q_{\rm max} \sim 2 m_e \gamma v$, and $q_{\rm min} \sim \hbar \omega_p/v$,
so that $\Lambda \sim \ln (2 m_e \gamma v^2/\hbar \omp)$.
Then, for $T_e \ll m_e$, i.e., $n_e \ll T_e^3$,
we put
\beq
B(\gamma) =   \ln \pfrac{2\gamma m_e v^2}{\hbar \omp}
  \left[ {\rm erf}\pfrac{v}{v_T}
  - \frac{2}{\sqrt{\pi}}
  \frac{v}{v_T} \left(1+\frac{m}{M}\right) e^{-(v/v_T)^2} \right] ,
\eeq
where $v_T^2 = 2T_e/m_e$.
The term in square brackets $\to 1$ for relativistic nuclides,
but gives the appropriate reduction in losses for non-relativistic
particles.

At high energies $\ga 1$ GeV, two other loss mechanisms are important, and
our treatment here closely follows that of \cite{kkm2}.
Photo-pair production, e.g., $p \gamma \rightarrow p e^+ e^-$,
becomes important for
$\nrg_p \ga 3 \ {\rm GeV} \ (T/0.1 \ {\rm MeV})$
\cite{blumenthal}.
We include this using the procedure outlined in
\cite{chod}.
Energy losses due to photopion production,
e.g., $p \gamma \rightarrow p \pi^0$,
become important at
energies $\nrg_p \ga  500 \ {\rm GeV} \ (T/0.1 \ {\rm MeV})$.
We include these following \cite{bg}.

For neutrons, Coulomb scattering is of course absent,
but magnetic-moment interactions with the ambient
radiation background do act to slow the non-thermal neutrons
\cite{Reno,gould93}.
We adopt the loss rate of \cite{turner}:
\beq
\label{eq:dEdt_n}
b_n = - \frac{d\nrg_n}{dt}
  = \frac{2\pi e^4 g_n^2 n_e}{m_e} \pfrac{m_e}{m_n}^2 v^2 (1-v^2) \gamma^2 ,
\eeq
where $m_n$ is the neutron mass
and $g_n = -1.91$ is the magnetic moment of the neutron
in units of the nuclear magneton.
This expression is similar to the result of
\cite{Jedamzik06},
except that the latter has a numerical prefactor
$3/2$ compared to that in eq.~\pref{eq:dEdt_n} and omits the factor $(1+v^2)$.
These differences are not dramatic, and
for mildly relativistic speeds these factors compensate
to give very similar loss rates.

\subsection{Solution of the Propagation Equation}

All the non-thermal species
$h$ evolve according to cascade equations
of the form \pref{eq:cascade}, and together these constitute a coupled set of
partial differential equations (PDEs) in energy and time.
Because the source term includes the elastic term with
$N_{h}$ inside the integral, these equations
are of integro-differential form.

Fortunately, major simplifications occur due to the
hierarchy of timescales in the problem.
The equation is self-regulating, and the timescale
to reach quasi-steady equilibrium is much faster
than the decay time or the expansion timescale. Hence
one can treat the spectrum as quasi-steady, so that the
coupled PDEs in energy and time become coupled ordinary differential
equations (ODEs) at each time step,
a great simplification. Moreover, redshift and expansion effects can safely be ignored.

Despite these considerable improvements, the equation is still of
integro-differential form, and the
integration is therefore not immediate.
We follow the usual procedure to solve such
integral equations, namely iteration.
We begin with a zeroth-order approximate solution,
in which we neglect the elastic and inelastic source terms.
In this approximation, eq.~\pref{eq:cascade}
becomes
\beq
\label{eq:leakybox}
\pd_t N_{h}(\epsilon) =J_{X,h}(\epsilon) - \Gamma_h(\epsilon)
N_{h}(\epsilon) - \pd_\nrg(b_{h}(\epsilon) N_{h}(\epsilon)) .
\eeq
For secondary nuclear species not present in the original decays,
(i.e., all but nucleons), the solution of eq.~\pref{eq:leakybox} can
be written in terms of the following quadrature
\beq
N_h(\nrg,t) = \frac{1}{b(\nrg)} \int_\nrg^\infty
   d\nrg^\prime \ J_{X,h}(\nrg^\prime,t) \ e^{-\tau(\nrg^\prime,t)} ,
\eeq
where the exponential ``optical depth" factor
\beq
\tau(\nrg^\prime,t)
 = \int_\nrg^{\nrg^\prime} d \nrg^{\prime \prime} \
    \frac{\Gamma(\nrg^{\prime \prime},t)}{b(\nrg^{\prime\prime})}
\eeq
is a measure of the average number of
inelastic interactions over the time taken to
lose energy electromagnetically from $\nrg^\prime$ to $\nrg$.
The secondary and tertiary cascades can then be treated iteratively,
correcting the initial omission of these terms:
\beqar
N_{h}^{(0)}(\nrg,t) & = & \frac{1}{b_h(\nrg,t)} \int_{\nrg}^{\infty}
 d\nrg^\prime \ J_{X,h}(\nrg^\prime,t) \ e^{-\tau(\nrg^\prime,t)} ,\\
N_{h}^{(1)}(\nrg,t) & = & \frac{1}{b_h(\nrg,t)} \int_{\nrg}^{\infty}
 d\nrg^\prime \ J_h^{(0)}(\nrg^\prime,t) \ e^{-\tau(\nrg^\prime,t)} ,\\
& \vdots & \\
N_{h}^{(i)}(\nrg,t) & = & \frac{1}{b_h(\nrg,t)} \int_{\nrg}^{\infty}
 d\nrg^\prime \ J_h^{(i-1)}(\nrg^\prime,t) \ e^{-\tau(\nrg^\prime,t)} ,
\eeqar
where $J_h^{(i)}$ is the source term from eq.~(\ref{eq:source}),
replacing $N_h$ with $N_h^{(i)}$.

Once these distributions converge,
one can insert them into equation~(\ref{eq:dYdthad}) and solve for the
hadro-dissociation rate.  This iterative procedure generalizes the procedure
adopted in some previous studies~\cite{kkm2}.
Rather than including the exponent in the
integral, they treat the exponential $R$ term as a $\delta$ function,
evaluated at $\nrg_*$, defined as the energy
at which the optical depth is unity: $\tau(\nrg_*,\nrg,t)=1$.

\begin{figure}[!htbp]
\epsfig{file=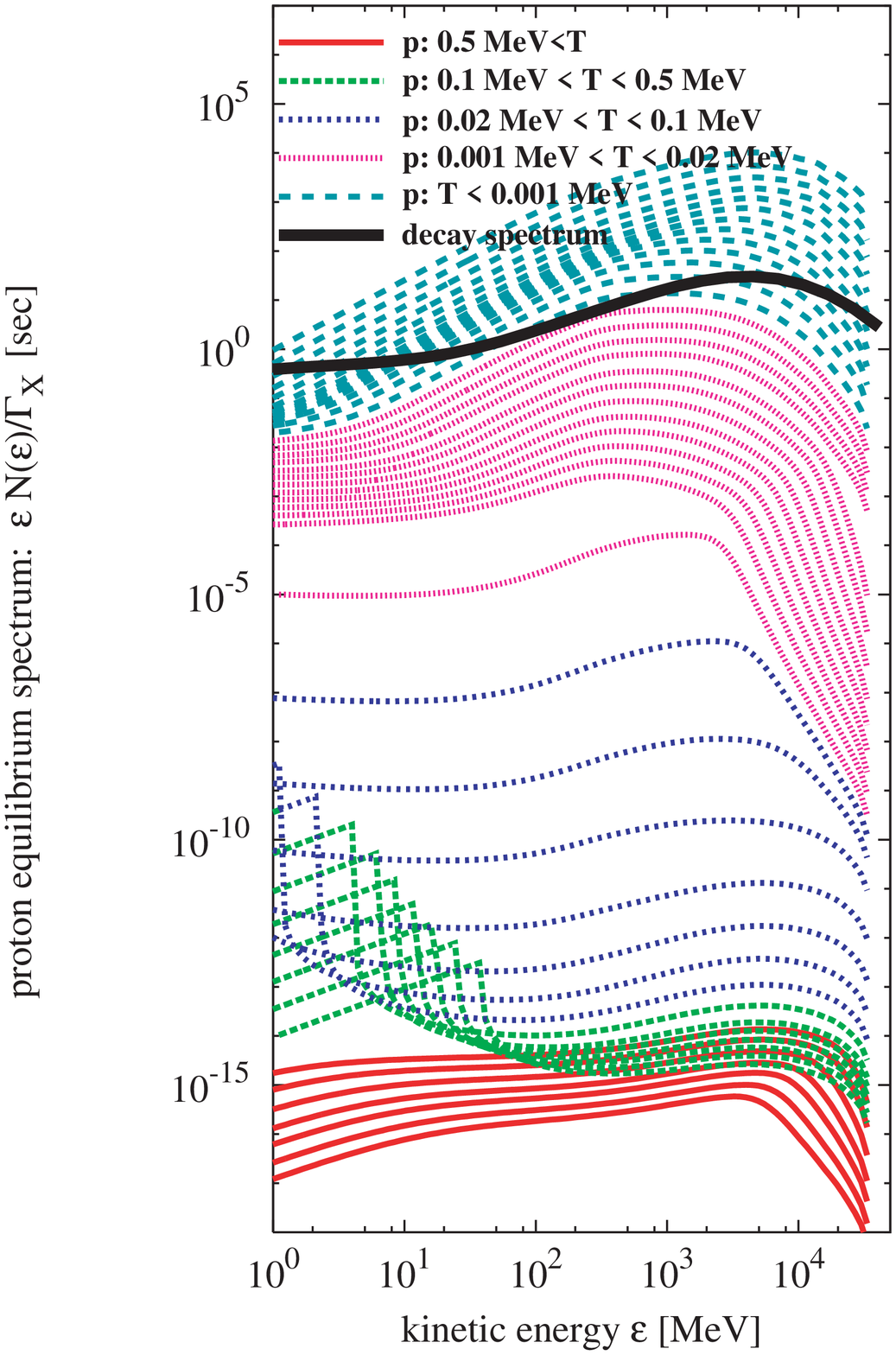,width=0.45\textwidth}
\epsfig{file=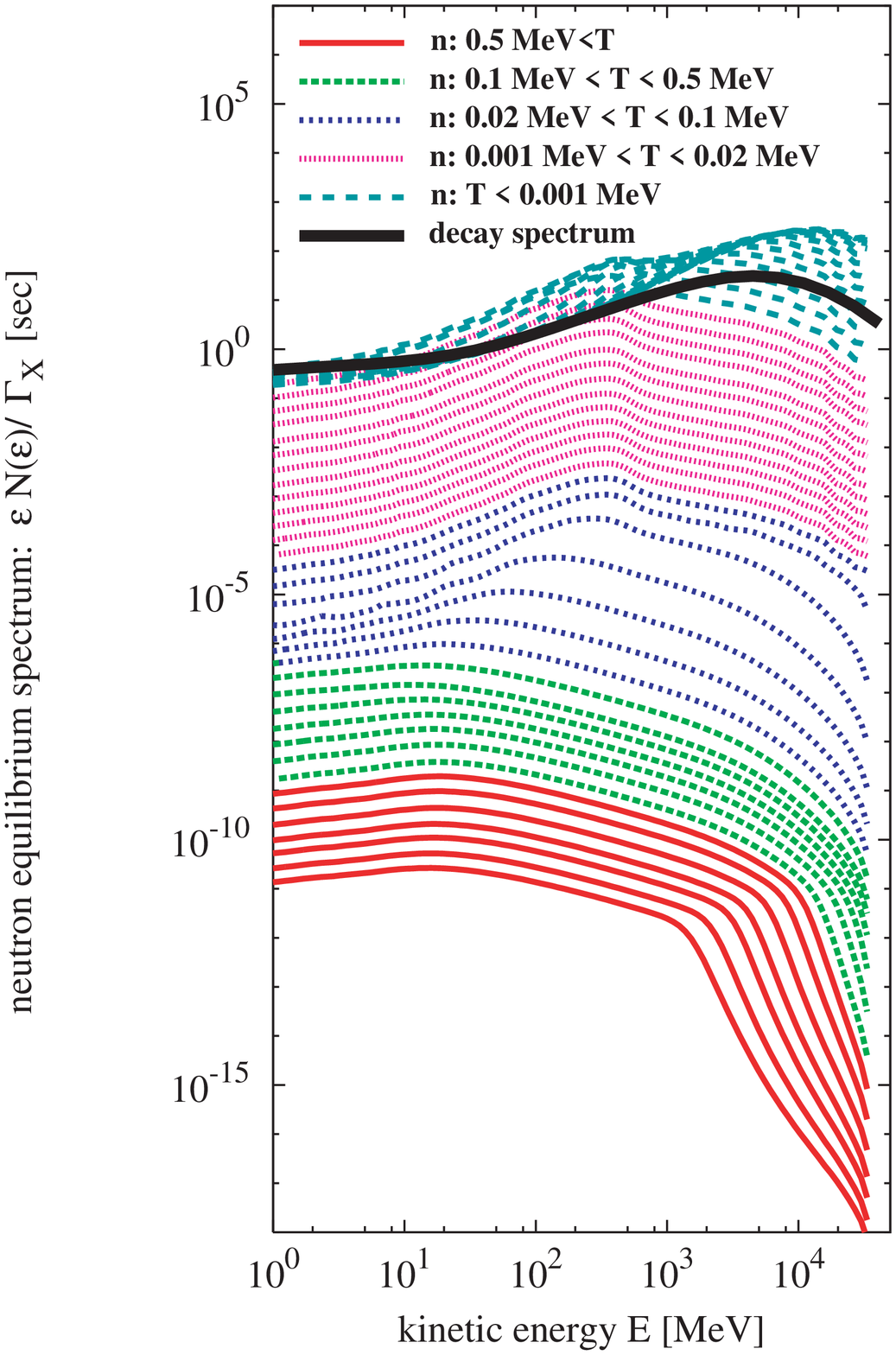,width=0.45\textwidth}
\caption{
\it Spectra of non-thermal hadrons after propagation in
the cosmic plasma; we plot $\nrg N_h(\nrg)$, which
indicates the particle number per logarithmic energy interval.
Results are shown for a large range of cosmic epochs
(i.e., temperatures, densities, background abundances).
Propagated spectra scale with the decay rate 
$\Gamma_X$, so we plot
$N_h/\Gamma_X$.
The decay injection spectra
are for
$(m_{1/2},m_{3/2},\tan \beta) = (120 {\rm GeV}, 500 {\rm GeV}, 50)$.
\label{fig:non-thermal}
}
\end{figure}

We plot the resulting non-thermal
neutron and proton
spectra in Fig.~\ref{fig:non-thermal}.
The thick solid black curve gives (in arbitrary units)
the shape of the source spectra $\nrg Q_p(\nrg)$ and $\nrg Q_n(\nrg)$.
The colored curves give the corresponding propagated spectra
for a wide range of cosmic epochs, labeled by the
temperature.
The highest temperature, $T = 2.5$ MeV, corresponds
to the lowest (red) curve, while the
lowest temperature, $T = 6 \times 10^{-5}$ MeV, corresponds to the
highest (bluegreen) curve in each plot.
Note that we plot $\nrg N_h(\nrg)/\Gamma_X$:
(1)
we use the product $\nrg N_h(\nrg)$
so that one can best assess the particle number per logarithmic interval; and
(2) as noted above in Section \ref{sect:analytic},
the propagated spectra are proportional to the decay rate 
$\Gamma_X$, and so by dividing by this factor we
remove this dependence.

Most striking in Fig.~\ref{fig:non-thermal}
is the enormous dynamic range in the spectra
amplitudes as the cosmic environment evolves.
Indeed, the highest
and lowest curves in the plots are separated
by more than 20 orders of magnitude!
This can be understood as a combination of two
factors.
First, 
the propagated spectra scale inversely with
the cosmic baryon density, either directly via 
$N_{h,{\rm thin}} \propto 1/n_B$
(eq.~\ref{eq:specthin}), 
or indirectly via the density dependence of energy
losses, 
$N_{h,{\rm thick}} \propto 1/b \propto 1/Y_{e^\pm}n_B$
(eq.~\ref{eq:specthick}).
This inverse density dependence leads to the steady rise in the
spectral amplitudes at late times.
Since the temperature range shown 
in Fig.~\ref{fig:non-thermal}
varies over a factor
$T_{\rm hi}/T_{\rm low} =  4 \times 10^{4}$,
the baryon density varies by a factor
$(T_{\rm hi}/T_{\rm low})^3 = 6 \times 10^{13}$.
For neutrons at $\nrg \la 100$ MeV,
the spectra are dominated by thin-target interactions
and this factor gives the range over which these spectral
amplitudes vary.

It is important to note that while the density changes lead to enhanced
spectra amplitudes at low temperatures and densities,
the non-thermal interaction rates scale as the product
of the non-thermal
flux times the density of background targets.  
This compensating factor of density means that
the overall interaction rates do not directly vary
with density (though the spectral shapes vary with 
cosmic epoch).

For neutrons at $\nrg \ga 100$ MeV, and for protons at
all energies, there is an additional evolutionary effect
on the spectral amplitudes.  Namely, these particles
are dominated by thick-target propagation, where
the spectral amplitudes are inversely proportional to
energy losses which are themeselves dependent
on the $e^\pm$ density and thus the total lepton per 
baryon ratio $Y_{e^\pm}$:  
$N_{h,{\rm thick}} \propto 1/b \propto 1/Y_{e^\pm}n_B$
(eq.~\ref{eq:specthick}).
Before pair annihilation, pairs are comparable in number to
photons,
$Y_{e^\pm} \sim Y_\gamma \sim 1/\eta \sim 10^9$;
after annihilation is complete, the remaining electrons 
provide charge balance and thus $Y_{e^{\pm}} = Y_{e^-} \sim 1$.
Thus as pairs annihilate and energy losses are greatly reduced,
the propagated spectra rise dramatically.
The pair number changes most when
temperature drops from $T \sim 0.1$ MeV down to
$\sim 0.02$ MeV,
which leads to the large increases
in the neutron spectral amplitudes at $\nrg \ga 100$ MeV
and in the proton amplitudes at all energies.
This additional annihilation factor 
contributes on top of the baryon density effect
and leads to the full variation seen in the spectral amplitudes.
Note that since the thin-target neutrons are responsible
for most of the abundance perturbations,
and since we are interested in times $\ga 10^3$ sec,
this annihilation effect is not very important for 
our results, though it is included in all of our calculations.

Beyond the overall spectral amplitudes, the spectral
shapes versus energy are set by the combination of
the source shape and the energy dependence of
the dominant loss curve.  
For example, in the case of neutrons at $\nrg \la 100$ MeV,
the losses are dominated by inelastic scattering
which is nearly energy-independent, and thus the
nearly flat source shape $\nrg Q_n(\nrg)$ leads to
a nearly flat propagated shape.  On the other hand,
for protons at late times and at $\nrg \la 1$ GeV,
losses are dominated by Coulomb interactions,
with $N_p \propto Q_p(>\nrg)/b_{\rm coul}$.
We have $b_{\rm coul} \propto 1/v \propto 1/\nrg^{1/2}$,
and for this decay spectrum
$Q_p(>\nrg) \simeq \nrg Q_p(\nrg) \sim$ constant.
Thus we have $\nrg N_p(\nrg) \propto \nrg^{3/2}$,
which is exactly the scaling behavior in
Fig.~\ref{fig:non-thermal} for $\la 1$ GeV  protons
at late times and low temperatures.

\subsection{Rates for Non-Thermal Reactions on Background Light Nuclei}

As non-thermal particles propagate in the cosmic plasma
they interact with background nuclides and
change light element abundances.
As shown in eq.~(\ref{eq:non-thermal}),
the rates for these processes 
depend on the non-thermal spectra we calculate,
but also on the cross sections for the processes.
The largest such rates are between the non-thermal
nucleons and the most abundant background species--nucleons
and \he4 nuclei; these rates are important in determining
the non-thermal spectra themselves, and are included
in the propagation calculations.
Reactions with less abundant background nuclei are neglected in
the non-thermal propagation calculations.
All of the non-thermal production and rates on
background nuclei are evaluated as a function of
cosmic time and environment, and  are incorporated
as additional channels in the usual (thermal) nucleosynthesis code.

Table \ref{tab:reactions} lists the nuclear reactions we
use in our analysis, as well as the references for the
cross sections we adopt.  We show the reactions used in
the propagation calculations as well as in the non-thermal rate
evaluations, and reactions only used for non-thermal rates.

Several reactions
have been updated since previous studies or were not present.
We adopt the recent revised fit to 
$\he3(\alpha,\gamma)\be7$ from Cyburt and Davids \cite{cd}.
The cross section is somewhat higher than in previous assessments,
leading to larger non-thermal production of \be7
as well as a boosted thermal production and thus
a higher standard (unperturbed) BBN value \cite{cfo5}.
All else being equal, this change increase worsens the \li7 problem,
and also somewhat reduces the \li6/\li7 ratio 
at where the \li7 
level remains dominated by the primordial production.

We have also added non-thermal reactions not present
in our previous treatments, and as far as we are aware, also
not present in those of other groups.
Namely, we have included new channels for non-thermal
$A=7$ destruction:
$\be7(n,p)\li7$, $\be7(n,\alpha)\he4$, and 
$\li7(p,\alpha)\he4$.
Of these, by far the largest cross section is for
$\be7(n,p)\li7$, which as a thermal rate is important in
standard BBN. This reaction leads to $A=7$ reduction
when followed by $\li7(p,\alpha)\he4$ which has
a reduced Coulomb barrier relative to $\be7+p$.
As noted above (Section \ref{sect:constraints})
and in refs.~\cite{jed,kkm2},
$A=7$ destruction via $\be7(n,p)\li7$
with {\em thermalized} decay
neutrons is responsible for the ``valley'' of concordant \li7/H
at lifetimes $10^2-10^3$ sec as seen in Fig.~{\ref{fig:zeta-tau-std}}.
It thus is reasonable to include the same process due to {\em non-thermal}
neutrons.
In fact, we find that these rates do not
appreciably change the final \li7/H abundance.

\section{Gravitino Decay Amplitudes}
\label{app:gravdec}

In calculating the relevant gravitino decay rates, we use the
interaction Lagrangian given in \cite{Pradlerthesis}:
\beqar
  \label{eq:grav int L}
  \mathcal{L}_\text{int} &=&
  -\frac{i}{\sqrt{2}\mpl} \left[
  \mathcal D_{\mu} \phi^{*(C)} \overline{\psi}_\nu \gamma^{\mu}
  \gamma^{\nu} \chi^{(C)}_{L}
  -\mathcal D_{\mu} \phi^{(C)} \overline{\chi}^{(C)}_{L} \gamma^\nu \gamma^\mu \psi_\nu
  \right] \nonumber \\&&
  - \frac{i}{8\mpl}\overline{\psi}_\mu [\gamma^{\rho}\,,\gamma^{\sigma}]
  \gamma^{\mu} \lambda^{(G)} F_{\rho\sigma}^{(G)},
\eeqar
 where $\psi_\mu$ is gravitino field, $\phi$ and $\chi$ are chiral
matter fields, $F_{\rho\sigma}$ is the field strength of a gauge boson
field, and $\lambda$ is the corresponding gaugino field. 
We denote $\chi_L = P_{L} \chi$ and
$\overline{\chi}_L = \overline{\chi} P_R$, where $P_L$ and $P_R$ are the
usual chiral projectors. The indices (C) and (G) denote chiral and
gauge multiplets, respectively, and $\mpl$ is the reduced Planck mass
defined as $\mpl = 1/\sqrt{8\pi G_N}$, where $G_N$ is the Newton
gravitational constant. The Feynman rules derived from this
Lagrangian are given in Appendix B of~\cite{Pradlerthesis}.

We use the relations
 \beqar
  && P_L {\tilde W}= P_L \left( V^{*}_{11} {\tilde \chi}_1 + V^{*}_{21}
  {\tilde \chi}_2 \right) \, , \,
  P_L {\tilde H} = P_L \left( V^{*}_{12} {\tilde \chi}_1 + V^{*}_{22} {\tilde \chi}_2
  \right), \nonumber  \\ 
&&  P_R {\tilde W} = P_R \left( U_{11} {\tilde \chi}_1 + U_{21} {\tilde \chi}_2
  \right)\, , \,
  P_R {\tilde H} = P_R \left( U_{12} {\tilde \chi}_1 + U_{22} {\tilde \chi}_2
  \right) \, ,
\eeqar
 to transform the interaction-eigenstate charged wino $\tilde W^+$
and charged Higgsino $\tilde H^+$ to the mass-eigenstate charginos
${\tilde \chi}^+_j$, where $U$ and $V$ are the unitary matrices used to
diagonalize the chargino mass matrix. We note that all the
fields in the above relations are positively charged. To get the
transformation relations for negatively charged 
conjugate states of the positively charged fields, one needs to exchange $U$
and $V$ in the above relations, e.g., 
\beq
  P_L {\tilde W^{c}} = P_L \left( U^{*}_{11} {\tilde \chi}^{c}_1 + U^{*}_{21} {\tilde
\chi}^{c}_2 \right). \eeq 
For neutralinos, the transformation from the
interaction eigenstates $\tilde H_{1,2}$, $\tilde W_3$ and $\tilde B$ to
the neutralino mass eigenstates $\tilde{\chi}^0_{i}$ is given by the
relations
 \beqar
&&  P_L {\tilde H}_i = P_L N^{*}_{j,i+2} {\tilde{\chi}}^0_{j}\,,\,
  P_L {\tilde B} = P_L N^{*}_{j1} {\tilde \chi}^0_{j}\,,\,
  P_L {\tilde W}_3 = P_L N^{*}_{j2} {\tilde \chi}^0_{j}\,, \nonumber \\
&&  P_R {\tilde H}_i = P_R N_{j,i+2} {\tilde \chi}^0_{j}\,, \,
  P_R {\tilde B} = P_R N_{j1} {\tilde \chi}^0_{j}\,,\,
  P_R {\tilde W}_3 = P_R N_{j2} {\tilde \chi}^0_{j}\,,
\eeqar
 where the repeated index $j$ is summed over from $1$ to $4$, and
the unitary matrix $N$ diagonalizes the neutralino mass
matrix. To obtain the mass eigenstates of the Higgs bosons, we use the
relations~\cite{GH1986}
 \beqar
  \label{eq: higgs gauge-mass}
   && H^1_2 = H^{+} \cos \beta, \,\,
    H^2_1 = H^{-} \sin \beta, \nonumber \\
   && H^1_1 = v_1 + \frac{1}{\sqrt{2}}\left(H^0_1 \cos \alpha - H^0_2 \sin \alpha + {i} H^0_3 \sin \beta \right),
    \nonumber \\
   && H^2_2 = v_2 + \frac{1}{\sqrt{2}}\left(H^0_1 \sin \alpha + H^0_2 \cos \alpha + {i} H^0_3 \cos \beta \right),
\eeqar 
where the $H^\pm$ are the charged Higgs fields, $H^0_1$ and
$H^0_2$ are the CP-even neutral Higgs fields, and $H^0_3$ is the CP-odd
neutral Higgs field. We denote by $v_1$ and $v_2$ the vacuum expectation values of
the two neutral Higgs fields, $\tbt$ is defined as $\tbt = v_2/v_1$, and $\alpha$ is
the mixing angle which leads to the two CP-even mass eigenstates.
Together with the relation $m^2_W = \frac{1}{2} \, g^2 (v^2_1 +
v^2_2)$, one gets $v_1 = \frac{\sqrt{2}\,m_W}{g}\cos \beta$ and $v_2
= \frac{\sqrt{2}\,m_W}{g}\sin \beta$, where $g$ is the gauge
coupling for $SU(2)_L$ and $m_W$ is the mass of the $W$ boson. For
squarks and sleptons, we do not consider mixing between generations, and
we take into account only left-right mixing for the third
generation. This simplification is sufficient within the CMSSM framework
we consider. The relation between the mass eigenstates
$\stop_{1,2}$ and the interaction eigenstates $\stop_{L,R}$ is 
\beqar
    \left( \begin{array}{c}
            \stop_L \\ \stop_R
        \end{array} \right) =
    \left( \begin{array}{cc}
            U_{\stop \, 1L} & U_{\stop \, 2L} \\
             U_{\stop \, 1R} & U_{\stop \,
             2R}
        \end{array} \right)
    \left( \begin{array}{c}
            \stop_1 \\ \stop_2
        \end{array} \right),
\eeqar
where the transformation matrix is unitary. There are similar interaction/mass-eigenstate 
transformations for $\sbot$ and $\stau$. We
treat the neutrinos as in the Standard Model, i.e., as massless, purely
left-handed neutrinos (and right-handed anti-neutrinos) only. For
details of the above transformations, we direct the reader
to~\cite{GH1986}.

We now list the amplitudes for all the two-body gravitino decay
channels. We denote by $m_b$ and $m_f$ the masses of the final-state bosons
and fermions, respectively. For $\grav \to \tilde{\chi}^0_i \,
\gamma$ and $\grav \to \tilde{g} \, g$, we have:
\beq
    \overline{\left| {\mathcal {M}} \right|^2} =
    \frac{1}{6\, \mgrav^2 \mpl^2} \left| {B} \right|^2
      \left( 3 \,
    \mgrav^2 + m^2_f \right) \left( \mgrav^2 - m^2_f \right)^2,
\eeq
 where $B=N^\prime_{i1}$ for the $\tilde{\chi}^0_i$ final states,
with $N^\prime_{i1} = N_{i1} \cosw + N_{i2} \sinw$, and $\theta_W$
is the weak mixing angle, and $B=1$ for the $\tilde{g}$ final state. For $\grav
\to \tilde{\chi}^\pm_j \, W^\mp$ and $\grav \to \tilde{\chi}^0_i \,
Z$, we have:
\beqar
   \overline{\left| {\mathcal {M}} \right|^2} &=&
   \frac{1}{12 \, \mgrav^2 \mpl^2} \,\bigg\{
    12  \left(C\,D + C^* D^* \right) \mgrav^3 m^2_b m_f \nonumber \\ &&
   + \,\left(\left| {C} \right|^2 + \left| {D}
   \right|^2 \right) \left[ 3\, \mgrav^6 - \mgrav^4 \left(5\, m^2_f
   + m^2_b \right) + \mgrav^2 \left(m^4_f - m^4_b \right) +
   \left(m^2_f - m^2_b \right)^3 \right] \bigg\} \nonumber \\&&
   + \,\frac{G^2}{24\,\mgrav^2 m^2_b \mpl^2} \bigg\{ \left[ \mgrav^4 - 2\, \mgrav^2 \left(m^2_f
   - 5\,m^2_b \right) + \left(m^2_f - m^2_b \right)^2 \right] \nonumber \\&&
   \times \Big[2\, \left(E\, F + E^* F^* \right) \mgrav m_f +
   \left(\left| {E} \right|^2 + \left| {F} \right|^2 \right)
   \left(\mgrav^2 + m^2_f - m^2_b \right) \Big]\bigg\} \nonumber \\ &&
   + \,\frac{G}{6\, \mgrav \mpl^2} \bigg\{ \left(D E^* + E D^* + C F^* +
   F C^* \right)   \nonumber \\ &&
\times  \left[-2\, \mgrav^4 + \mgrav^2 \left(m^2_f +
   m^2_b\right) + \left(m^2_f - m^2_b \right)^2 \right] \nonumber \\&&
   -\,3\left(C\, E + D\,F + C^* E^* + D^* F^* \right)\mgrav m_f
   \left(\mgrav^2 - m^2_f + m^2_b \right) \bigg\},
\eeqar
where $C=V_{j1}$, $D=U_{j1}$, $E=-v_1 U_{j2}$, $F=-v_2
V_{j2}$ and $G=g$ for the $\tilde{\chi}^\pm_j$ final states; and
$C=D=N^\prime_{i2}$, with $N^\prime_{i2}= - N_{i1} \sinw + N_{i2}
\cosw$, $E=F=\frac{1}{\sqrt{2}}\left(-v_1 N_{i3} + v_2
N_{i4}\right)$ and $G= g/\cosw$ for the $\tilde{\chi}^0_i$ final states.
In the above expression, the first and second lines come from the gauge
part of eq.~(\ref{eq:grav int L}) (the second line), the third and fourth
lines come from the matter part of eq.~(\ref{eq:grav int L}) (the
first line), with the gauge bosons coming from the covariant
derivative and the Higgs field $\phi$ taking its vacuum expectation
value (see eq.~(\ref{eq: higgs gauge-mass})), and the fifth and sixth
lines come from the interference of these two parts. For $\grav \to
\tilde{f}\,f$, $\grav \to \tilde{\chi_j}^\pm \,H^\mp$ and $\grav \to
\tilde{\chi}^0_i \, H^0_{1,2,3}$, we have:
\beqar
  \overline{\left| {\mathcal {M}} \right|^2} &=&
  \frac{1}{12\,\mgrav^2 \mpl^2} \Big[ \left( \left| {H} \right|^2 + \left| {K} \right|^2 \right)
  \left(\mgrav^2 + m^2_f - m^2_b \right) - 2\, \left(H\,K + H^* K^*
  \right) \mgrav m_f \Big] \nonumber \\&&
  \times \left[\mgrav^4 - 2\, \mgrav^2 \left(m^2_f + m^2_b \right) +
  \left(m^2_f -m^2_b \right)^2 \right],
  \label{gravitinoM2}
\eeqar 
where $H=$ $K=\frac{1}{\sqrt{2}}\left( N_{i4} \sin \alpha  +
N_{i3} \cos \alpha \right)$ for the $H^0_1$ final state;
$H=K=\frac{1}{\sqrt{2}}( N_{i4} \cos \alpha  - $ $N_{i3} \sin
\alpha)$ for the $H^0_2$ final state;
$H=-K=\frac{1}{\sqrt{2}}\left( N_{i4} \cos \beta  + N_{i3} \sin
\beta \right)$ for the $H^0_3$ final state; $H=U_{j2} \sin \beta$ and
$K=V_{j2} \cos \beta$ for the $H^\pm$ final state; $H=U^*_{\stop \, kR}$
and $K=U_{\stop \, kL}$ for the $\stop_k$ final state (and similarly for
$\sbot_k$ and $\stau_k$ with their corresponding mixing matrices
$U_{\sbot}$ and $U_{\stau}$, respectively); $H=0$ and $K=1$ for all
the $\tilde f_L$ and $\tilde \nu$ final states; and $H=1$ and $K=0$ for
all the $\tilde f_R$ final states.
To get the partial widths, one need to multiply (\ref{gravitinoM2}) by the phase-space 
factor:
\beq
  \frac{N_c}{16\,\pi \mgrav^3}\sqrt{\left[\mgrav^2 - \left(m_f - m_b
  \right)^2 \right] \left[\mgrav^2 - \left(m_f + m_b
  \right)^2 \right]}\,,
\eeq where $N_c$ is the color factor ($3$ for $q\,\tilde q$ channels,
$8$ for the $g\,\tilde g$ channel, and $1$ otherwise).

The amplitude for the three-body decay $\grav \to  \tilde{\chi}^0_i  \, \gamma^*  \to
\tilde{\chi}^0_i \, q  \qbar$ is 
\beqar
  \overline{\left| {\mathcal {M}} \right|^2} &=&
  \frac{g^2 \sin^2 \theta_W Q^2_q}{3\, s^2 \mgrav^2 \mpl^2}
  ~ \Bigg\{ 4\,s\, m_{\tilde{\chi}^0_i} \mgrav^3 \left(N^{\prime \,2}_{i1} + N^{\prime * \,2}_{i1}
  \right) \left(s + 2\,m^2_q \right) \nonumber \\ &&
  + ~ \left| {N^\prime_{i1}} \right|^2 \bigg[ 3\, \mgrav^6 \Big(s +
  2\,m^2_q \Big) + \mgrav^4 \Big( -3\, s \left(s + 2\,t \right)
  + m^2_{\tilde{\chi}^0_i} \left(s - 10\, m^2_q \right) \Big) \nonumber \\ &&
  + ~ \mgrav^2 \Big(6\,s \, m^4_q + 2\, m^2_q \big(-3\,s \left(s + 2\,
  t \right) + 2\, s\, m^2_{\tilde{\chi}^0_i} + m^4_{\tilde{\chi}^0_i} \big) \nonumber \\&&
  + ~ s \big(s^2 + 8\,s t
  + 6\, t^2 - 4\, m^2_{\tilde{\chi}^0_i} \left(s+ 2\, t \right) + 3\, m^4_{\tilde{\chi}^0_i}
  \big) \Big) - \left(s- m^2_{\tilde{\chi}^0_i} \right) \Big(2\, s \, m^4_q - 2\, m^2_q \nonumber \\&&
  \times \big(2\, s t + s \, m^2_{\tilde{\chi}^0_i} - m^4_{\tilde{\chi}^0_i} \big)
  + s\, \big(s^2 + 2\, s t + 2\, t^2 -2\,m^2_{\tilde{\chi}^0_i} \left(s + t\right) + m^4_{\tilde{\chi}^0_i}
  \big) \Big) \bigg] \Bigg\},
\eeqar 
where $s$ and $t$ are the invariant masses of the $q \qbar$ and
$q \tilde{\chi}^0_i$ systems, respectively, $m_{\tilde{\chi}^0_i}$
is the mass of the neutralino $\tilde{\chi}^0_i$, where the LSP corresponds to
$i=1$, and $Q_q = 2/3$ or $-1/3$ for the corresponding quarks. The
differential decay rate is 
\beq
 \label{three-B phase space}
 \frac{d\Gamma}{d s\,d t} = \frac{N_c}{256\,\pi^3 \mgrav^3}~ \overline{\left| {\mathcal {M}}
 \right|^2}.
\eeq 
These three-body and two-body analytic expressions were also
given in~\cite{kmy} and~\cite{0404231}.

For $\grav \to  \tilde{\chi}^0_i  \, W^+ W^-$, there are contributions
from four generically different diagrams: contact, $\gamma/Z$
exchange, $\tilde{\chi}^\pm_j$ exchange and $H^0_1/H^0_2$ exchange
diagrams. The amplitude is too long to be listed here, so we just
write the matrix elements for each of these diagrams (suppressing the
polarization indices):
\beqar
  i \mathcal{M}_{\mathrm {contact}} &=& \frac{i}{4 \mpl} \, g \,
  \overline{u}(q') \left(P_R N_{i2} + P_L N^*_{i2} \right) \,
  \gamma^\mu \, [\gamma_{\rho} \, ,\gamma_{\sigma}] \, \psi_\mu (p) \,
  \epsilon^{\rho \, *} (k) \, \epsilon^{\sigma \,*} (k')\,, \nonumber
\eeqar

\beqar
  i \mathcal{M}_{\grav \to \tilde{\chi}^0_i \gamma^* \to \tilde{\chi}^0_i  \, W^+
  W^-} &=& \frac{i}{4 \mpl} \, \frac{1}{(k+k')^2} \, g^{\alpha \beta} g
  \sinw \, \overline{u}(q') \,
  \left(P_R N^\prime_{i1} + P_L N^{\prime *}_{i1} \right) \,
  \gamma^\mu \, [\slashed{k}+\slashed{k'} \, ,\gamma_\alpha ] \, \psi_\mu (p) \nonumber \\&&
  \times \left[\left(2\,k + k' \right)_\sigma g_{\beta \rho} + \left(k'
  -k \right)_\beta g_{\rho \sigma} - \left(k + 2\,k'\right)_\rho g_{\sigma
  \beta} \right] \epsilon^{\rho \, *}(k) \, \epsilon^{\sigma \,
  *}(k')\,, \nonumber
\eeqar

\beqar
  i \mathcal{M}_{\grav \to \tilde{\chi}^0_i Z^* \to \tilde{\chi}^0_i  \, W^+
  W^-} &=&  \frac{i}{4 \mpl} \, \frac{1}{(k+k')^2 - m^2_Z + i m_Z
  \,\Gamma_Z} \left(g^{\alpha \beta} - \frac{(k+k')^\alpha (k+k')^\beta}{m^2_Z} \right) \,g
  \cosw   \nonumber \\&&
  \times \, \overline{u}(q') \, \bigg\{ \left(P_R N^\prime_{i2} + P_L N^{\prime *}_{i2} \right)
  \gamma^\mu \, [ \slashed{k}+\slashed{k'} \, , \gamma_\alpha ] \nonumber \\&&
  + \, \frac{\sqrt{2} \, g}{\cosw} \Big[ P_R \left(-v_1 N_{i3} + v_2 N_{i4}
  \right) + P_L \left(-v_1 N^*_{i3} + v_2 N^*_{i4}
  \right) \Big]  \gamma^\mu  \gamma_\alpha \bigg\} \, \psi_\mu (p)
  \nonumber \\&&
  \times \left[\left(2\,k + k' \right)_\sigma g_{\beta \rho} + \left(k'
  -k \right)_\beta g_{\sigma \rho} - \left(k + 2\, k'\right)_\rho g_{\beta \sigma}
  \right] \epsilon^{\rho \, *} (k) \, \epsilon^{\sigma \,*} (k')\,, \nonumber
\eeqar

\beqar
  i \mathcal{M}_{\grav \to W^+  \tilde{\chi}^{- \, *}_j \to \tilde{\chi}^0_i  \, W^+
  W^-} &=&  \frac{i}{4 \mpl} \, \frac{1}{(p-k)^2 - m_{\tilde{\chi}^\pm_j}^2 +
  i m_{\tilde{\chi}^\pm_j}  \Gamma_{\tilde{\chi}^\pm_j}} \, g \, \overline{u}(q') \,
   \gamma_\sigma \left[P_R O^R_{ij} + P_L O^L_{ij} \right] \nonumber \\&&
  \cdot \left(\slashed{p}-\slashed{k} + m_{\tilde{\chi}^\pm_j} \right)
  \bigg\{ \left(P_R V_{j1} + P_L U^*_{j1} \right) \gamma^\mu
  [\gamma_\rho \, , \slashed{k} ]  \nonumber \\&& + \, 2 g \left(P_R v_2 V_{j2} + P_L v_1
  U^*_{j2} \right) \gamma^\mu \gamma_\rho \bigg\} \,
  \psi_\mu (p) \, \epsilon^{\rho \, *} (k) \, \epsilon^{\sigma \,*}
  (k')\,, \nonumber
\eeqar

\beqar
  i \mathcal{M}_{\grav \to W^-  \tilde{\chi}^{+ \, *}_j \to \tilde{\chi}^0_i  \, W^+
  W^-} &=&  - \, \frac{i}{4 \mpl} \, \frac{1}{(p-k')^2 - m_{\tilde{\chi}^\pm_j}^2 +
  i m_{\tilde{\chi}^\pm_j}  \Gamma_{\tilde{\chi}^\pm_j}} \, g \, \overline{u}(q') \,
   \gamma_\rho \left[P_L O^{R \,*}_{ij} + P_R O^{L \, *}_{ij} \right] \nonumber \\&&
  \cdot \left(\slashed{p}-\slashed{k'} + m_{\tilde{\chi}^\pm_j} \right)
  \bigg\{ \left(P_L V^*_{j1} + P_R U_{j1} \right) \gamma^\mu
  [\gamma_\sigma \, , \slashed{k'} ]   \nonumber \\&& +\, 2 g \left(P_L v_2 V^*_{j2} + P_R v_1
  U_{j2} \right) \gamma^\mu \gamma_\sigma \bigg\} \,
  \psi_\mu (p) \, \epsilon^{\rho \, *} (k) \, \epsilon^{\sigma \,*}
  (k')\,, \nonumber
\eeqar

\beqar
  i \mathcal{M}_{\grav \to \tilde{\chi}^0_i  H^{0 \, *}_1 \to \tilde{\chi}^0_i  \, W^+
  W^-} &=&  \frac{i}{2 \mpl} \, \frac{1}{(k+k')^2 - m^2_{H^0_1} + i m_{H^0_1} \Gamma_{H^0_1}} \, g \,
  m_W \cos \left(\beta -\alpha \right) g_{\rho \sigma} \nonumber \\&&
  \times \, \overline{u}(q') \,
  \big[ P_R \left(\sin \alpha N_{i4} + \cos \alpha N_{i3} \right)
  - P_L \left(\sin \alpha N^*_{i4} + \cos \alpha N^*_{i3} \right)
  \big] \gamma^\mu  \nonumber \\&&
  \cdot \,  \left(\slashed{k} +\slashed{k'}\right) \psi_\mu (p) \, \epsilon^{\rho \, *} (k) \, \epsilon^{\sigma \,*}
  (k')\,, \nonumber
\eeqar

\beqar
  i \mathcal{M}_{\grav \to \tilde{\chi}^0_i  H^{0 \, *}_2 \to \tilde{\chi}^0_i  \, W^+
  W^-} &=&  \frac{i}{2 \mpl} \, \frac{1}{(k+k')^2 - m^2_{H^0_2} + i m_{H^0_2} \Gamma_{H^0_2}} \, g \,
  m_W \sin \left(\beta -\alpha \right) g_{\rho \sigma} \nonumber \\&&
  \times \, \overline{u}(q') \,
  \big[ P_R \left(\cos \alpha N_{i4} - \sin \alpha N_{i3} \right)
  - P_L \left(\cos \alpha N^*_{i4} - \sin \alpha N^*_{i3} \right)
  \big] \gamma^\mu  \nonumber \\&&
  \cdot \,  \left(\slashed{k}+\slashed{k'}\right) \psi_\mu (p) \, \epsilon^{\rho \, *} (k) \, \epsilon^{\sigma \,*}
  (k')\,,
\eeqar 
where $\psi_\mu (p)$ represents the decaying gravitino with momentum
$p\,$, $\overline{u}(q')$ represents the produced neutralino with
momentum $q'$, $\epsilon^{\rho \, *} (k)$ and $\epsilon^{\sigma
\,*}(k')$ are the polarization four-vectors for the two $W$ bosons
with momenta $k$ and $k'$, respectively, $g^{\alpha \beta}$ is
the flat-space Lorentz metric tensor, the $\Gamma$ factors in the
propagators are the total widths of the exchanged particles and
$O^L_{ij} = N_{i2} V^*_{j1} - \frac{1}{\sqrt{2}} N_{i4} V^*_{j2}$
and $O^R_{ij} = N^*_{i2} U_{j1} + \frac{1}{\sqrt{2}} N^*_{i3}
U_{j2}$. The Feynman rules for the vertices $W^+ W^- H^0_{1,2}$ and
$\tilde{\chi}^{\pm}_j \, \tilde{\chi}^0_i \, W^{\mp}$ are given in~\cite{GH1986, HK1985}.
The sum of the polarization states for gravitino is 
\beq
  \sum_{s=\pm \frac{3}{2},\pm \frac{1}{2}} \psi^s_\mu(p){ \bar{\psi}}^s_\nu(p) = - \left(\slashed{p} +
  \mgrav \right) \left(g_{\mu \nu} - \frac{p_\mu p_\nu}{\mgrav}
  \right) - \frac{1}{3} \left(\gamma_\mu + \frac{p_\mu}{\mgrav}
  \right) \left(\slashed{p} - \mgrav \right) \left(\gamma_\nu +
  \frac{p_\nu}{\mgrav} \right)\, , \nonumber
\eeq
and the phase space is the same as given in eq.~(\ref{three-B
phase space}).

A remark is in order for the process $\grav \to  \tilde{\chi}^0_i \,
W^+ W^-$. Because of the $1/m_W$ factors in the longitudinal 
polarization states of the $W$ boson, one might worry that this process
might have bad high-energy behavior. Correspondingly, if one restored the
electroweak symmetry to make $m_W$ vanish, then this process would diverge
if there were still some terms proportional to negative powers of
$m_W$ in the final result. However, upon expanding the parameters
in terms of $m_W$, that is, expanding the elements of chargino and
neutralino mixing matrices $N$, $U$, $V$, the masses
$m_{\tilde{\chi}^0_i}$, $m_{\tilde{\chi}^\pm_j}$, and the angle
$\alpha$ etc. (for the $m_W$ dependence of these parameters, see,
for example \cite{GH1988}), in the total amplitude, we have verified that
all terms with negative powers of $m_W$ cancel. As a result, 
this process has good  high energy behavior.
This is a highly non-trivial check on our calculation of the process 
$\grav \to  \tilde{\chi}^0_i \, W^+ W^-$, verifying the relative 
magnitudes, signs and phases of all the individual contributions
to the decay amplitudes.

We conclude by commenting on the differences between the amplitudes
obtained and used this paper and those given~\cite{kmy} and~\cite{0404231},
which are based on the gravitino interaction
Lagrangian given in eq.~(4.58) of~\cite{Moroi:1995fs}. This Lagrangian is
different from that given in eq.~(2.82) of~\cite{Pradlerthesis}, which we
follow here. The latter reference uses the transformation from the
two-component gravitino Lagrangian given in 
Wess and Bagger~\cite{WessBagger} to the four-component
Lagrangian which, as pointed out at the footnote on page 205 of~\cite{HK1985},
requires a factor of $i$ when one defines the Majorana spinors.
This factor results in a sign difference when one writes down the
Feynman rules using the matter part of the Lagrangian.
We also note that our result for gravitino $\to Z + \chi$ differs from~\cite{0404231},
in that it includes the Higgsino contribution.

\end{document}